\documentclass[draft]{agujournal2019}
\usepackage{url} 
\usepackage{lineno}
\usepackage[inline]{trackchanges} 
\usepackage{soul}
\usepackage{amsmath,amssymb}
\usepackage{enumitem}
\draftfalse

\journalname{JAMES}

\begin{document}

%
%

\title{Extratropical low-frequency variability with ENSO forcing:
A reduced-order coupled model study}

%
%




\authors{St\'ephane Vannitsem$^1$, Jonathan Demaeyer$^1$ and Michael Ghil$^{2,3}$}

 \affiliation{1}{Royal Meteorological Institute of Belgium, Avenue Circulaire, 3, 1180 Brussels, Belgium}
 \affiliation{2}{Geosciences Department and Laboratoire de M\'et\'eorologie Dynamique (CNRS and IPSL),\\ Ecole Normale Sup\'erieure and PSL University, Paris, France}
  \affiliation{3}{Department of Atmospheric \& Oceanic Sciences, University of California at Los Angeles,\\ Los Angeles, USA}




\correspondingauthor{St\'ephane Vannitsem}{Stephane.Vannitsem@meteo.be}




\begin{keypoints}
	\item Two PBAs coexist in the model's midlatitudes for both periodic and chaotic ENSO forcing.
	
	\item These local PBAs are nonlinearly unstable, with some trajectories that visit both of them. 
	
	\item The ENSO forcing synchronizes the midlatitude behavior in unexpected ways. 
\end{keypoints}

\begin{abstract}
	The impact of the El Niño--Southern Oscillation (ENSO) on the extratropics is investigated in an idealized, reduced-order model that has a tropical and an extratropical module. Unidirectional ENSO forcing is used to mimick the atmospheric bridge between the tropics and the extratropics. The variability of the coupled ocean--atmosphere extratropical module is then investigated through the analysis of its pullback attractors (PBAs). This analysis focuses on two types of ENSO forcing generated by the tropical module, one periodic and the other aperiodic.
	
	For a substantial range of the ENSO forcing, two chaotic PBAs are found to coexist for the same set of parameter values. Different types of extratropical low-frequency variability are associated with either PBA over the parameter ranges explored. For periodic ENSO forcing, the coexisting PBAs exhibit only weak nonlinear instability. For chaotic forcing, though, they are quite unstable and certain extratropical perturbations induce transitions between the two PBAs. These distinct stability properties may have profound consequences for extratropical climate predictions: in particular, ensemble averaging may no longer help isolate the low-frequency variability signal.
	
\end{abstract}

\section*{Plain Language Summary}
The authors have investigated the variability of a simplified coupled ocean--atmosphere model for the Earth's midlatitudes, subject to the influence of the El Niño--Southern Oscillation (ENSO). This study reveals that multiple climates may coexist, each of which is characterized by distinct types of low-frequency variabiliy (LFV) and predictability properties. When the ENSO forcing is periodic, these climates are fairly robust against perturbations, but when it is chaotic, small perturbations induce transitions between the different climates. These properties could have profound consequences for extratropical climate predictions, since ensemble averaging may no longer be a valid approach to ascertain the LFV signal.

\newpage

\section{Introduction}

The Earth system is a dissipative multiscale system forced by external time-dependent signals of various origins \cite{Ghil.Luc.2020}. Given the system's enormous complexity, one often splits it into subsystems, each of which is affected by the others. Each subsystem can, in turn, be considered as being both forced and dissipative \cite{Lorenz.1963a, Ghil.Chil.1987}. The description of their evolution relies on the use of conservation laws, combined with appropriate simplifications, subject to the influence of the external forcings. The dynamics of such systems can be described using concepts and methods from dynamical systems theory. 

Recent developments in this theory have addressed the effect of time-dependent external forcing, and are now organized in the theory of nonautonomous \cite<NDSs:>[]{Carvalho.ea.2012, Kloed.Rasm.2011} and random \cite<RDSs:>[]{Arnold.1998, Caraballo.Han.2017} dynamical systems. This theory has been applied to the atmospheric, oceanic and climate sciences over the last dozen years or so by several groups of authors 
\cite{Ashwin.ea.2012, Checkrounetal2011, Checkrounetal2018, Ditlev.Ashwin.2018, Drotosetal2015, Drotosetal2016, GCS.2008, Pierinietal2016, Pierinietal2018, Pierini2020, Teletal2020}.    

An important characteristic of systems with time-dependent external forcing or coefficients is that one can no longer assume ergodicity, and thus temporal averages are no longer good approximations to ensemble averages. When periodic forcing is acting, the usual ergodicity property has to be generalized to cycloergodicity, which requires temporal averages to be performed separately at each moment of the cycle. 

When the forcing is erratic --- i.e. either random or chaotic --- there is no equivalence between the two types of averages \cite{Drotosetal2016}. Moreover, as will be seen in the current work, multiple types of qualitatively different solutions may be present for the same parameter values, complicating further the description of the system \cite{Pierinietal2016, Pierinietal2018}. But the tools 
to explore the dynamics of the atmosphere, ocean and coupled climate system in these cases are provided by NDS and RDS theory and are known, alternatively, as pullback attractors \cite<PBAs:>[]{GCS.2008, Checkrounetal2011} or snapshot attractors \cite{Drotosetal2015, Drotosetal2016}.  

For several decades, the influence of the El Niño--Southern Oscillation (ENSO) on the mid- and high latitudes of both hemispheres has been an important area of climate studies \cite{Philander1990, McPhaden.ea.2020}. This area's importance is largely due to the presence of teleconnections between the Tropical Pacific and many regions all over the world \cite{Alexanderetal2002, HoerlingandKumar2002, LopezParages2016, Schemmetal2018}. A particularly important question is exploring the impact of ENSO forcing on the predictability of  extratropical climate \cite{KumarandHoerling1995, Nidheeshetal2017}. 

Most of the above-mentioned studies have relied on comprehensive, high-end models for their analyses. A major difficulty in this case is  the limited number of model runs that can be performed, due to the large computational cost of each run \cite{Ghil.2001, Held.2005}. Moreover, these runs usually start from initial states that are not very far in the past and so the solutions do not necessarily sample the correct asymptotic behavior. This state of affairs also implies that one gets, at best, only a very partial view of a given model's solution space. These obstacles can be overcome by using first simpler models that will provide hints on the possible solutions generated by the use of a vastly larger number of parameter values and initial states, thus providing crucial guidance for future simulations and predictions with larger models.

This paper's aim is to explore the possible existence of multiple types of  extratropical climate trajectories that are compatible with a given type of ENSO forcing. This exploration is performed in the setting of a reduced-order, coupled tropical--extratropical model, and it relies on large numbers of runs and on sophisticated methods for analyzing them. The ENSO forcing studied is both periodic and chaotic, and it covers a wide range of intensities: several PBAs are found for a given type and intensity of the forcing. The stability properties of these PBAs are explored, showing that some are unstable and that model trajectories may transit from one PBA's attractor basin to another. 

Section~\ref{sec:equate} describes the coupled tropical--extratropical model. In Section~\ref{sec:dynamics}, we construct the PBAs and study their properties. Section~\ref{sec:conclude} summarizes the main results and the key messages to keep in mind for ensemble forecasting and climate projections. 


\section{Governing equations for the tropical--extratropical model}
\label{sec:equate}


\subsection{The ENSO module}\label{ssec:ENSO}

The ENSO model used herein was developed in a series of papers by F.-F. Jin, A. Timmermann and colleagues \cite{Jin1996, Jin1997, AnJin2004, Timmermannetal2003, Robertsetal2016}. They  modeled the dynamics of the ocean's upper layer in the Tropical Pacific using a low number of variables. Their two-box ENSO model describes the dynamics of the temperature in the eastern and western Tropical Pacific basins, and it is completed by an equation for the evolution of the thermocline depth. The model represents the horizontal discharge-recharge mechanisms at play in the Tropical Pacific through the heat exchanges between the tropical and subtropical waters, subject to surface wind stress and upwelling of subsurface cold water in the eastern part of the domain \cite{Jin1997}. 

\citeA{Robertsetal2016} introduced a nondimensional model version in which the time, for instance, is normalized by a typical time scale of tropical wave propagation of roughly 3.5 months or 105 days. The latter ENSO model version is coupled here with an extratropical module whose time is nondimentionalized by the Coriolis parameter $f_0$, and the \citeA{Robertsetal2016} equations are slightly modified therefore. 

The nondimensional equations governing our ENSO module are, accordingly:       
\begin{subequations} \label{eq:ENSO}
\begin{align}
\frac{dx}{dt} & =  \rho \delta (x^2-a x) + s x (x+y+c-c \tanh(x+z)), \label{ENSO1} \\
\frac{dy}{dt} & =  -\rho \delta (a y+x^2), \label{ENSO2} \\  
\frac{dz}{dt} & =  \delta (k-z-\frac{x}{2}) \label{ENSO3}.
\end{align}  
\end{subequations}
Here $x, y, z$ and $t$ are dimensionless, $x$ is the temperature difference between the eastern and western basins of the Tropical Pacific, $y$ the western basin's temperature anomaly with respect to a reference value, and $z$ the western basin's thermocline depth anomaly.

The dimensionless parameters are defined as follows
\begin{subequations} \label{eq:dimens}
\begin{align}
a & = \frac{ \alpha b L}{\epsilon h^{\*} \beta}, \qquad  \rho = \frac{\epsilon h^{\*} \beta}{r b L}, 
\qquad   \delta = \frac{r}{f_0}, \label{dimens-1} \\
c & = \frac{C}{S_0}, \qquad k = \frac{K}{S_0},  \qquad  s = \frac{\zeta h^{\*} \beta}{b L f_0}, \label{dimens-2} 
\end{align}  
\end{subequations}
with $f_0$ the Coriolis parameter used to nondimensionalize the time, and the other parameters as defined in \citeA[Table~1]{Robertsetal2016}. For the applications presented in this paper, two sets of parameter values were given special attention. 
These two sets lead to either a periodic or a chaotic solution, and they are listed in Table~\ref{tab:param} below.

\begin{table}[ht!]
\centering
\caption{Dimensionless parameter values of the ENSO module}
\setlength\tabcolsep{6 pt} 
\begin{tabular}{ll} 
\hline
Periodic & Chaotic \\
\hline
\\
$a=6.8927$   &   $a=7.658609809$  \\
$\rho =0.3224$  &  $\rho =0.29016$       \\
$\delta=0.00028058$  &  $\delta=0.0002803$     \\
$c=2.3952$   &   $c=2.3952$   \\
$k=0.4032$   &   $k=0.4032$ \\
$s=0.0010691$  &  $s=0.001069075$ \\
\hline
\end{tabular} \label{tab:param}
\end{table}

The ENSO module is taken here to be unaffected by the midlatitude module, and is thus the driving  (or master) system in our coupled model, in the sense of the skew products of \citeA{Sell.1971}; see also \citeA{Kloed.Rasm.2011}, \citeA{Caraballo.Han.2017} and \citeA[Sec.~III.G]{Ghil.Luc.2020}. Equations~\eqref{eq:ENSO} can thus be integrated independently of the rest of the model, and this is done using a fourth-order Runge-Kutta scheme with a time step $\Delta t = 0.1346$~hours $=0.05$ nondimensional time units. See previous page for nondimensionalization, just below Eq.~\eqref{eq:dimens}.
 
\begin{figure}[ht]
\centering
{\includegraphics[width=0.8\textwidth]{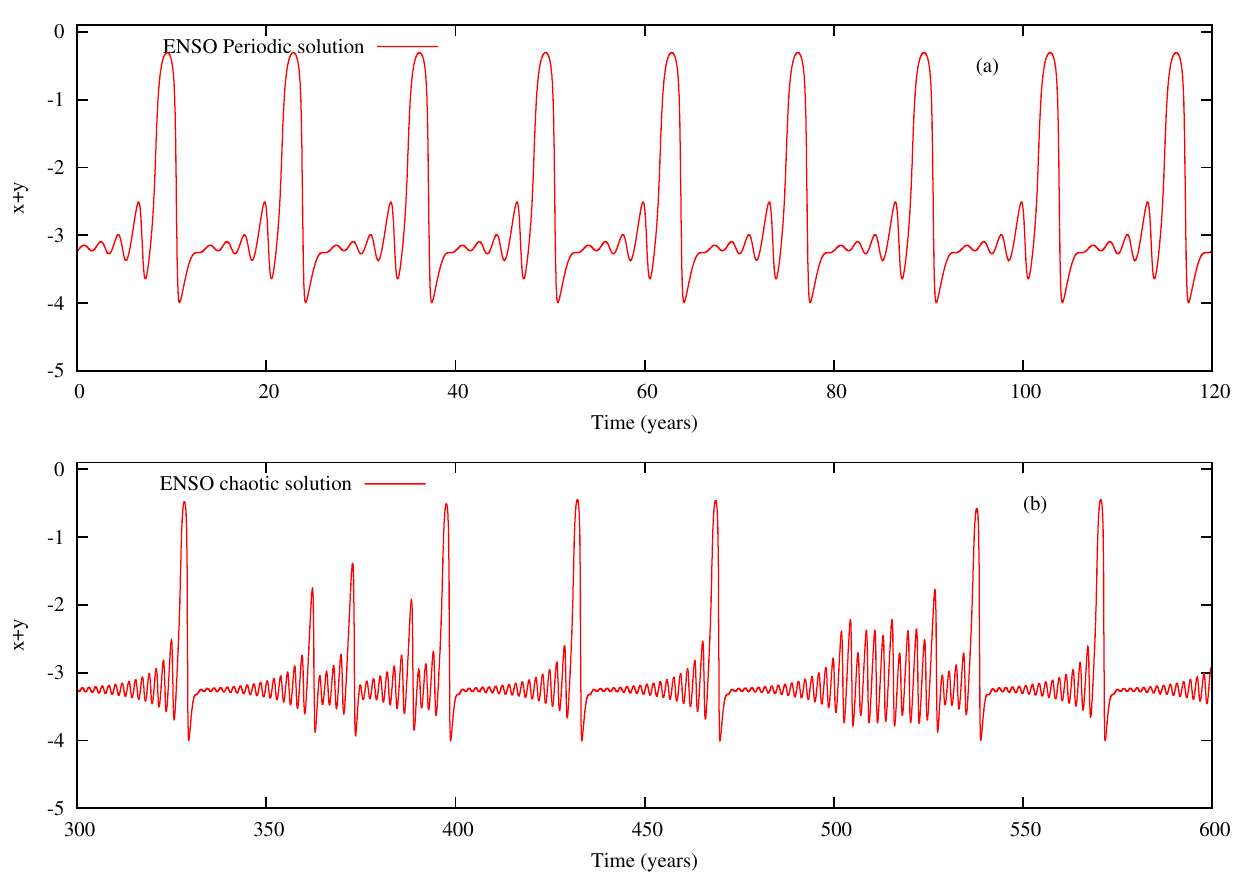}}
\caption{Trajectory segments of the ENSO model for the eastern Tropical Pacific basin's temperature anomalies $x+y$. (a) Periodic case, and (b) chaotic case; see Eqs.~\eqref{eq:ENSO} and Table~\ref{tab:param} for details. The bursting behavior in both cases, with very large excursions towards more positive values, occurs  periodically in panel (a) and irregularly in panel (b). Notice that the total length of the segments is 120~yr in panel (a) and 300~yr in panel (b).}
\label{fig:ENSO_force}
\end{figure}

The results of such integrations are shown in Figs.~\ref{fig:ENSO_force}(a) and \ref{fig:ENSO_force}(b) for the periodic and the chaotic case, respectively. The variable $x + y$ plotted in the figure corresponds in our simple model \eqref{eq:ENSO} to the sea surface temperatures in the eastern Tropical Pacific that are commonly associated with the Ni\~no-3 index.

The periodic solution for the ENSO forcing in Fig.~\ref{fig:ENSO_force}(a) agrees qualitatively with the ``bursting'' behavior emphasized by \citeA{Timmermannetal2003}. Thus, large warm events in the Ni\~no-3 area alternate with several cycles of moderate warm-and-cold events, with a periodicity of roughly 10~yr  \cite[Fig.~11(b)]{Timmermannetal2003}.

A long chaotic solution was started from the initial state $(x = -2.8439, y = -0.62, z = 1.480)$, with an integration length of 20~000~000~time units that corresponds to about 6~145~yr; i.e., 10~000~000 time units $\simeq 3~072 $~yr. In general, such runs were equally divided into a transient portion to reach the asymptotic behavior and an analysis portion; in the present case, both portions were equal to 10~000~000 time units. 

A 300-yr segment of the chaotic solution's analysis interval is plotted in Fig.~\ref{fig:ENSO_force}(b). The large bursts occur much less frequently here, on less realistic time scales of the order of 30--50~yr. On the other hand, the evolution in this case is more erratic --- with much greater irregularity in the timing of the bursts --- as seen in observations and in other chaotic ENSO models \cite<e.g.,>[]{JNG.1994, JNG.1996, Tzip.ea.1994}.


\subsection{The VDDG extratropical model} \label{ssec:VDDG}

The coupled ocean--atmosphere model used herein for the midlatitudes was developed by \citeA{Vannitsemetal2015} and it is called hereafter the VDDG model. S. Vannitsem and colleagues have already used different versions of this VDDG model in order to study low-frequency variabiliy (LFV) within the coupled ocean--atmosphere system \cite{DeCruzetal2016,Vannitsem2017}, the properties of the Lyapunov exponents in such a system \cite{VannitsemLucarini2016, DeCruzetal2018}, the stochastic parametrization of subgrid-scale forcing \cite{Demaeyer2017, Demaeyer2018}, and data assimilation in coupled models \cite{Pennyetal2019, Tondeuretal2020, Carrassietal2020}. 

S. Vannitsem and colleagues have already used different versions of this VDDG model in order to study low-frequency variabiliy (LFV) within the coupled ocean--atmosphere system \cite{DeCruzetal2016,Vannitsem2017}, the properties of the Lyapunov exponents in such a system \cite{VannitsemLucarini2016, DeCruzetal2018}, the stochastic parametrization of subgrid-scale forcing \cite{Demaeyer2018}, and data assimilation in coupled models \cite{Pennyetal2019, Tondeuretal2020, Carrassietal2020}.

Our coupled model's atmospheric module is based on the vorticity equations of a two-layer quasi-geostrophic flow defined on a beta-plane \cite{Gill.1982, Pedlosky.1987}. The equations in pressure coordinates are:
\begin{subequations} \label{eq:VDDG}
\begin{align}
& \frac{\partial}{\partial t} \left( \nabla^2 \psi^1_a \right) + J(\psi^1_a, \nabla^2 \psi^1_a) + \beta \frac{\partial \psi^1_a}{\partial x}
 = - k'_d \nabla^2 (\psi^1-\psi^3) + \frac{f_0}{\Delta p} \omega, \label{VDDG-1} \\
& \frac{\partial}{\partial t} \left( \nabla^2 \psi^3_a \right) + J(\psi^3_a, \nabla^2 \psi^3_a) + \beta \frac{\partial \psi^3_a}{\partial x}
 =  + k'_d \nabla^2 (\psi^1_a-\psi^3_a) - \frac{f_0}{\Delta p}  \omega  
- k_d \nabla^2 (\psi^3_a-\psi_o). \label{VDDG-2} 
\end{align}  
\end{subequations}
Here $\psi^1_a$ and $\psi^3_a$ are the streamfunction fields at 250~hPa and 750~hPa, respectively, while $ \omega =dp/dt$ is the vertical velocity, $f_0$ the Coriolis parameter at $\phi_0 =$~45\textdegree~latitude, and $\beta = df/dy$ at $\phi_0$. The coefficients $k_d$ and $k'_d$ multiply the surface friction term and the internal friction between the layers, respectively. 

An additional term has been introduced in Eq.~\eqref{VDDG-2} in order to account for the presence of a surface boundary velocity $\psi_o$ of the oceanic flow; see Eq.~\eqref{eq:ocean} below. This term corresponds to the Ekman pumping on a moving surface and is the mechanical contribution of the interaction between the ocean and the atmosphere.

The coupled model's ocean dynamics is described by the reduced-gravity, quasi-geostrophic shallow-water model \cite{Gill.1982, Pedlosky.1987}. The forcing is provided by the wind generated by the atmospheric module above. The governing equation is:
\begin{equation} \label{eq:ocean}
\frac{\partial}{\partial t} \left( \nabla^2 \psi_o - \frac{\psi_o}{L_R^2} \right) + J(\psi_o, \nabla^2 \psi_o) + \beta \frac{\partial \psi_o}{\partial x}
= -r \nabla^2 \psi_o + \frac{{\mathrm{curl}}_z \vec{\tau}}{\rho H},
\end{equation}
where $\psi_o$ is the streamfunction, $\rho$ the density of water, $H$ the depth of the fluid layer, $L_R$ the reduced Rossby deformation radius, $r$ a Rayleigh friction coefficient at the bottom of the fluid layer, and curl$_z \vec{\tau}$ is the vertical component of the wind stress curl.

The wind stress in the VDDG model is given by $(\tau_x, \tau_y)=C (u-U,v-V)$, where $(u = -\partial \psi^3_{\rm a}/\partial y, v = \partial \psi^3_{\rm a}/\partial x)$ are the horizontal components of the geostrophic wind, and $(U, V)$ the components of the geostrophic currents in the ocean. One thus gets
\begin{equation}
	\mathrm{curl}_z \tau = C \nabla^2 (\psi^3_{\rm a}-\psi_{\rm o}). 
	\label{eq:stress}
\end{equation}
and the wind stress is proportional to the relative velocity between the flow in the ocean's upper layer and the wind in the lower atmospheric layer. The drag coefficient $d = C/(\rho_0 h)$ gives the strength of the mechanical coupling between the ocean and the atmosphere and it was a key bifurcation parameter in \citeA{Vannitsemetal2015}; see Table~\ref{tab:VDDG}. Here $C$ will also play a crucial role in affecting the VDDG model's behavior subject to ENSO forcing.

The dynamic equations \eqref{eq:VDDG} and \eqref{eq:ocean} are supplemented by temperature equations for the two subsystems. For the atmosphere, 
\begin{equation} \label{eq:temp_a}
\gamma_a \left( \frac{\partial T_a}{\partial t} + J(\psi_a, T_a) -\sigma \omega \frac{p}{R}\right) = -\lambda (T_a-T_o) + E_{a,R},
\end{equation}
with
\begin{equation} \label{eq:rad_a}
E_{a,R} = \epsilon_a \sigma_B T_o^4 - 2 \epsilon_a \sigma_B T_a^4 + R_a.
\end{equation}
Here $R$ is the gas constant, $\epsilon_a$ the emissivity of the atmosphere, $\sigma_B$ the Stefan-Boltzman constant, $R_a$ the shortwave radiation at the top of the atmosphere, $\omega$ the vertical velocity in pressure coordinates, and $\sigma =  -(R/p) (\partial T_a/\partial p - 1/(\rho_a c_p))$ is the static stability, with $p$ the pressure, $\rho_a$ the air density, $c_p$ the specific heat at constant pressure, and $\sigma$ here is taken to be a constant. Note also that, thanks to the hydrostatic relation in pressure coordinates and the
ideal gas relation $p=\rho_a R T_a$, the atmospheric temperature can be written as
$T_a = - (p/R) f_0 (\partial \psi_a/\partial p)$. 
\begin{table}[ht!]
	\centering
	\caption{List of parameters of the extratropical VDDG module}
	\setlength\tabcolsep{6 pt} 
	\begin{tabular}{llll} 
		\hline 
		Parameter (unit) & Value & Parameter (unit) & Value \\
		\hline
		$L_y = \pi L$  (km)         & $5.0 \times 10^3$        & $\gamma_{\rm{o}}$ (J\,m$^{-2}$\, K$^{-1}$) & $4 \times 10^6 \, h$  \\
		$f_0$ (s$^{-1}$)            & $1.032 \times 10^{-4}$   & $C_{\rm{o}}$ (W\,m$^{-2}$)             & 310 \\
		$n = 2 L_y / L_x$           & $1.5$               & $T_{\rm{o}}^0$ (K)                     & $285$ \\
		$R_{\rm{E}}$ (km)           & $6370$                   & $\gamma_{\rm{a}}$ (J\,m$^{-2}$ \, K$^{-1}$) & $1.0 \times 10^7$ \\
		$\phi_0$                    & $\pi/4$            & $C_{\rm{a}}$ (W \, m$^{-2}$)             & $C_{\rm{o}}/4$ \\
		$g^\prime$                  & $3.1 \times 10^{-2}$     & $\epsilon_{\rm{a}}$                    & $0.76$ \\
		$r$ (s$^{-1}$)              & $1.0 \times 10^{-7}$     & $\beta$ (m$^{-1}$ \, s$^{-1}$)           & $1.62 \times 10^{-11}$ \\
		$h$ (m)                     & $100$                  & $T_{\rm{a}}^0$ (K)                     & $270$ \\
		$d$ (s$^{-1}$)              & $C/(\rho_o h)$     & $\lambda$ (W\,m$^{-2}$ \, K$^{-1}$)      &   $1004 \, C$ \\
		$k_d$ (s$^{-1}$)            & $(g C)/(\Delta p)$    & $R$ (J\,kg$^{-1}$\,K$^{-1}$)           & $287$ \\
		$k_d^\prime $ (s$^{-1}$)    & $(g C)/(\Delta p)$     & $\sigma$  (J kg$^{-1}$ Pa$^{-2}$)                & $2.16 \times 10^{-6}$ \\   
		$C$ (kg m$^{-2}$ s$^{-1}$)       & $0.008$ and $0.015$ & & \\ 
		\hline
	\end{tabular} \label{tab:VDDG}
\end{table}

For the ocean, 
\begin{equation} \label{eq:temp_o}
\gamma_o \left( \frac{\partial T_o}{\partial t} + J(\psi_o, T_o)\right) = -\lambda (T_o-T_a) + E_R,
\end{equation}
with
\begin{equation} \label{eq:rad_o}
E_R = -\sigma_B T_o^4 + \epsilon_a \sigma_B T_a^4 + R_o.
\end{equation}
Here $R_o$ is the shortwave radiation entering the ocean,  $\gamma_o$ the heat capacity of the ocean, and $\lambda$ is the inverse of the time scale associated with heat transfer between the ocean and the atmosphere, which includes both the latent and sensible heat fluxes. In fact, we assume that this combined heat transfer is proportional to the temperature difference between the atmosphere and the ocean. 

The temperatures in both modules are linearized around a reference value in order to reduce the quartic terms of the energy balance equations \eqref{eq:rad_a} and \eqref{eq:rad_o} to linear terms, assuming that the temperature fluctuations are small. This modification helps one to reduce the number of terms on the right-hand side of the ordinary differential equations \eqref{eq:temp_a} and \eqref{eq:temp_o} when building the spectral low-order model. 

The model fields in both its atmosphere and its ocean are developed in Fourier series and truncated at a low order.  The number of modes herein is fixed at 10 for the atmosphere and 8 for the ocean, leading to 20 ordinary differential equations for the former and 16 for the latter. This model configuration is the original VDDG one; see also \citeA{Vannitsem2017}. The parameter values used in the present work are listed in Table~\ref{tab:VDDG}.
  

\subsection{Modeling the tropical--extratropical interaction} \label{ssec:interact}

\citeA{Schemmetal2018} investigated the changes of extratropical wintertime cyclogenesis when El Ni\~no or La Ni\~na events are occurring in the Tropical Pacific. These authors showed, in particular, that the background zonal-flow anomaly is more intense over the North Atlantic during La Ni\~na, while it is stronger over the North Pacific during El Ni\~no. This finding tells us that an important effect of the tropical forcing is to change the intensity of the zonal flow in either region, and that the impact of El Ni\~no and La Ni\~na differs from one region to the other. 

In order to mimic this dynamic effect in the extratropical VDDG model used herein, we impose a direct linear forcing of the model's first barotropic atmospheric mode. It is this barotropic streamfunction mode that represents the intensity of the zonal flow within the atmosphere. Its dynamics is written as
\begin{equation}
\frac{d\psi_{a,1}}{dt} = f_1(\psi_{a,1}, \theta_{a,1}) + g \delta (x+y).
\end{equation} 
Here $f_1(\psi_{a,1}, \theta_{a,1})$ is the original right-hand side of the dynamical evolution \eqref{VDDG-1} of $\psi_{a,1}$; $x+y$ represents the eastern Tropical Pacific basin's temperature anomalies, as given by Eq.~\eqref{ENSO2} and plotted in Fig.~\ref{fig:ENSO_force}; and $g$ scales the intensity of the tropical forcing.

Thus $g$ represents the crucial forcing of  the midlatitude VDDG model described in Sec.~\ref{ssec:VDDG} above by the ENSO module of Sec.~\ref{ssec:ENSO}. In our setting, given a positive $g$-value, a positive, warm anomaly will induce an increase of $\psi_{a,1}$,  and hence of the mean zonal flow, $U=- \psi_{a,1} \partial (\sqrt{2} \cos y) /\partial y$. 
This situation corresponds to the intensification of the zonal flow over the North Pacific during an El Ni\~no. If, to the contrary, $g$ is negative, this would correspond to an intensification during La Ni\~na that mimicks the ENSO effect over the North Atlantic. 

The crude analogies with the impact of El Ni\~no and La Ni\~na on the atmosphere overlying the Atlantic and Pacific extratropical basins, along with the formulation of the ENSO forcing in our system, suggest that it is worth exploring both positive and negative $g$-values. Note that a similar analysis based on the western Tropical Pacific's temperature anomaly $y$ has been performed as well; its results were, in fact, quite similar to what is reported in this paper.


\section{Dynamics of the ENSO-forced midlatitudes}
\label{sec:dynamics}

As explained in Sec.~\ref{ssec:interact}, the coupled ocean--atmosphere dynamics in our model's extratropical regions is forced by the Tropical Pacific. The impact of such a time-dependent forcing can be investigated in a self-consistent manner based on the concepts and tools of NDS theory and its PBAs. The latter are asymptotically invariant sets associated with a unique time-dependent forcing of a long-lived system that is started in the distant past; see \citeA[Appendix~A]{GCS.2008} for a didactic presentation in the RDS case and \citeA{Caraballo.Han.2017} for an accessible approach to both the NDS and RDS cases. 

The concept of pullback attraction and the formal definition of a PBA in the deterministic, finite-dimensional case are introduced and motivated succintly in~\ref{app:pba}. Unlike in the better-known, autonomous setting, PBAs are themselves time-dependent objects. Interestingly, a unique global PBA can contain multiple local PBAs, as studied in the case of a double-gyre model of the wind-driven ocean circulation by \citeA{Pierinietal2016, Pierinietal2018}. This novel type of multimodality will be illustrated in the present paper by the coexistence of two chaotic PBAs in our coupled ocean--atmosphere model.

In the present paper, we compute the Lyapunov exponents of the ENSO-forced VDDG model as a a key tool in the systematic investigation of its PBAs.  More precisely, we compute the tangent linearized system of the VDDG model along the forced trajectory. The Lyapunov exponents are computed in this tangent space of 36 variables. 
The methodology follows \citeA{Kuptsov.P.2012} 
but along a forced trajectory.  \citeA{Ruelle1984}, though, is telling us that the Lyapunov exponents can still be computed and are unique, even in a dynamic system with time-dependent forcing, subject to the uniqueness and ergodicity conditions on the latter, which are verified in the present setting; see~\ref{app:pba} for the details of this argument.

We performed a large number of ENSO-forced integrations of the VDDG model starting from different initial states in the extratropics. The length of the integrations was adapted to first explore a substantial number of parameter values with shorter runs and then successively increase the length of the runs to refine the analysis of carefully selected situations. 

A long interval of 10~000~000 time units $\simeq 3~072$~yr was used for specific forcing and coupling parameter values in order to ensure convergence of the Lyapunov spectra. Some very long runs of 50~000~000 time units $\simeq  15~360$~yr were used for two integrations investigating the robustness of the two PBAs found with shorter runs. In general, the transients of all these runs were taken equal to 10~000~000 time units $\simeq 3~072$~yr.  
To check the convergence of trajectories toward the PBAs, we used runs that were 3~072-yr long and had shorter transients of 1~536~yr.


\subsection{Periodic forcing}

\subsubsection{The leading Lyapunov exponents}
\label{sssec:lyapunovper}

\begin{figure}[ht!]
\centering
{\includegraphics[width=140mm]{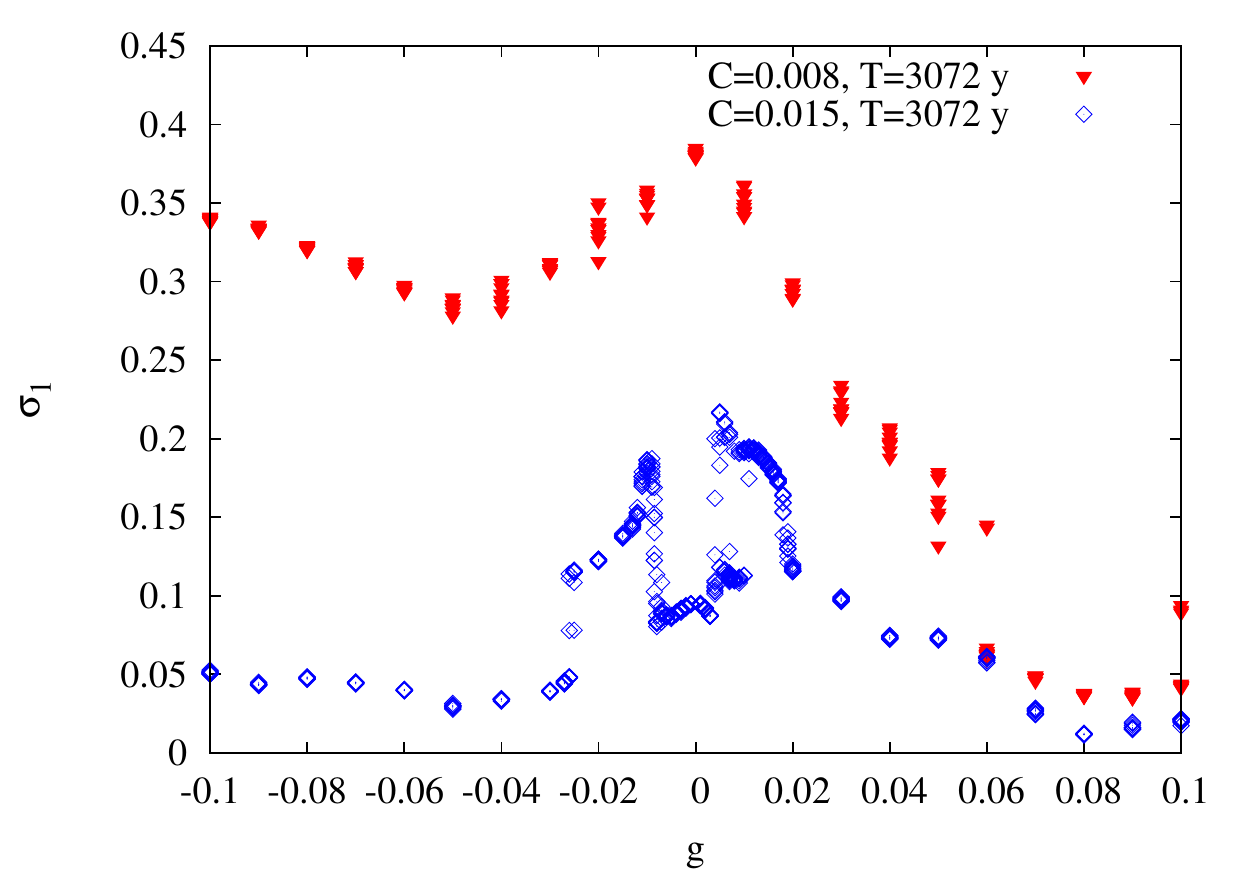}}
\caption{Leading Lyapunov exponent $\sigma_1$ as a function of the ENSO forcing parameter $g$ for periodic ENSO forcing as in Fig.~\ref{fig:ENSO_force}(a). Two values of the air-sea coupling coefficient $C$, 0.008 (red triangles) and 0.015 (blue diamonds), are used,  with an analysis interval of $T = 3~072$~yr. 
For the smaller value of $C$, $\sigma_1$ decreases and hence the predictability increases as a function of the amplitude of $g$. 
For the larger value of $C$, a more complex dependence emerges, with the possibility of multiple pullback attractors (PBAs) coexisting.}
\label{lyapper}
\end{figure}

Figure~\ref{lyapper} displays the value of the leading Lyapunov exponent $\sigma_1$ as a function of the parameter $g$ when the tropical forcing is periodic in time, as in Fig.~\ref{fig:ENSO_force}(a). Plotted are values for the two different configurations of the VDDG model that appear in Table~\ref{tab:VDDG}, namely $C=0.008$ (red inverted triangles) and $C=0.015$ (blue rhomboids). Note that, for certain ranges of $g$, several distinct values of the dominant Lyapunov exponent are obtained for different initial states in the remote past. Since the ENSO forcing acts directly on the atmosphere alone, we expect its effects to be stronger on the coupled VDDG model the larger the air-sea coupling $C$, which passes this forcing on to the ocean as well. 

For $g = 0$, i.e., no ENSO forcing, $\sigma_1$ is of the order of $0.38$ day$^{-1}$ for $C=0.008$ and close to $0.1$ day$^{-1}$ for $C=0.015$. For small $|g|$ and both $C$-values, $\sigma_1 = \sigma_1(g)$ is at its maximum for no ENSO forcing and decreases as the absolute values of $g$ increase away from 0. Since the predictability of the model's extratropical climate is inversely proportional to $\sigma_1$, it appears to increase overall as the tropical forcing becomes stronger. 

There is an important qualitative difference, though, between the behavior of $\sigma_1 = \sigma_1(g)$ for $C = 0.008$ vs. $C = 0.015$; namely, for the latter, there is an abrupt ``splitting'' and ``coalescing'' of $\sigma_1$-values: splitting for $g < 0$ and coalescing for $g > 0$. This coexistence of two branches  $\sigma_1^{\pm}(g)$ of $\sigma_1$-values for the larger air-sea coupling value $C = 0.015$ suggests the coexistence in this case of two distinct PBAs of the coupled VDDG model, as found in the purely oceanic double-gyre problem by \citeA{Pierinietal2016, Pierinietal2018}. 

The potential coexistence of two PBAs seems to extend here for an interval of\\ 
\noindent $- 0.03 \lesssim g \lesssim +0.02$, in the case of $C = 0.015$, while it is entirely absent for $C = 0.008$. Note that the ratio of the $\sigma_1$-values for $C = 0.015$ between the two branches in Fig.~\ref{lyapper} is roughly equal to the ratio between the value for $C=0.008$ and the larger one of the two values for $C=0.015$, i.e., $\sigma_1^{+}(g = 0, C = 0.015) /\sigma_1^{-}(g = 0, C = 0.015) \simeq \sigma_1(g = 0, C = 0.008) /\sigma_1^{+}(g = 0, C = 0.015)$. The difference between the two PBAs is thus quite significant in terms of the potential predictability of midlatitude flow for the same Tropical Pacific surface temperature anomalies. 

The sharp transitions between the upper branch with $\sigma_1^{+}(g)$ to the lower branch with $\sigma_1^{-}(g)$ near $g \simeq - 0.01$ and back near $g \simeq + 0.01$ suggest that the upper branch continues to exist in the interval $ - 0.01 \lesssim g \lesssim + 0.01$ as a “ghost PBA” that is no longer attained by forward integrations of the model. For autonomous systems in the atmospheric sciences, such “ghost equilibria” were defined by \citeA{Legras.Ghil.1985} and “ghost limit cycles” by \citeA{Kimoto.Ghil.1993}. In these two cases, the exact meaning was clear: 
 “A ghost ﬁxed point is a ﬁxed point that has become unstable in one or a few directions in phase space. Still, the system’s trajectories will linger near it for extended time intervals. Likewise, a ghost limit cycle is a closed orbit that has become slightly unstable but is visited, again and again, by the system’s trajectories” \cite{Ghil.SSA.2002}.
 
 In the present, nonautonomous case, it is less clear how a ghost PBA that loses its attractivity in one or more directions might behave and how model trajectories would continue to linger near it. In any case, the results illustrated in Fig.~\ref{lyapper} suggest some form of hysteresis between the upper and lower branches of PBAs in the case of $C = 0.015$. These questions, while quite interesting, will require subsequent work to be solved.
\begin{figure}[ht]
\centering
{\includegraphics[width=140mm]{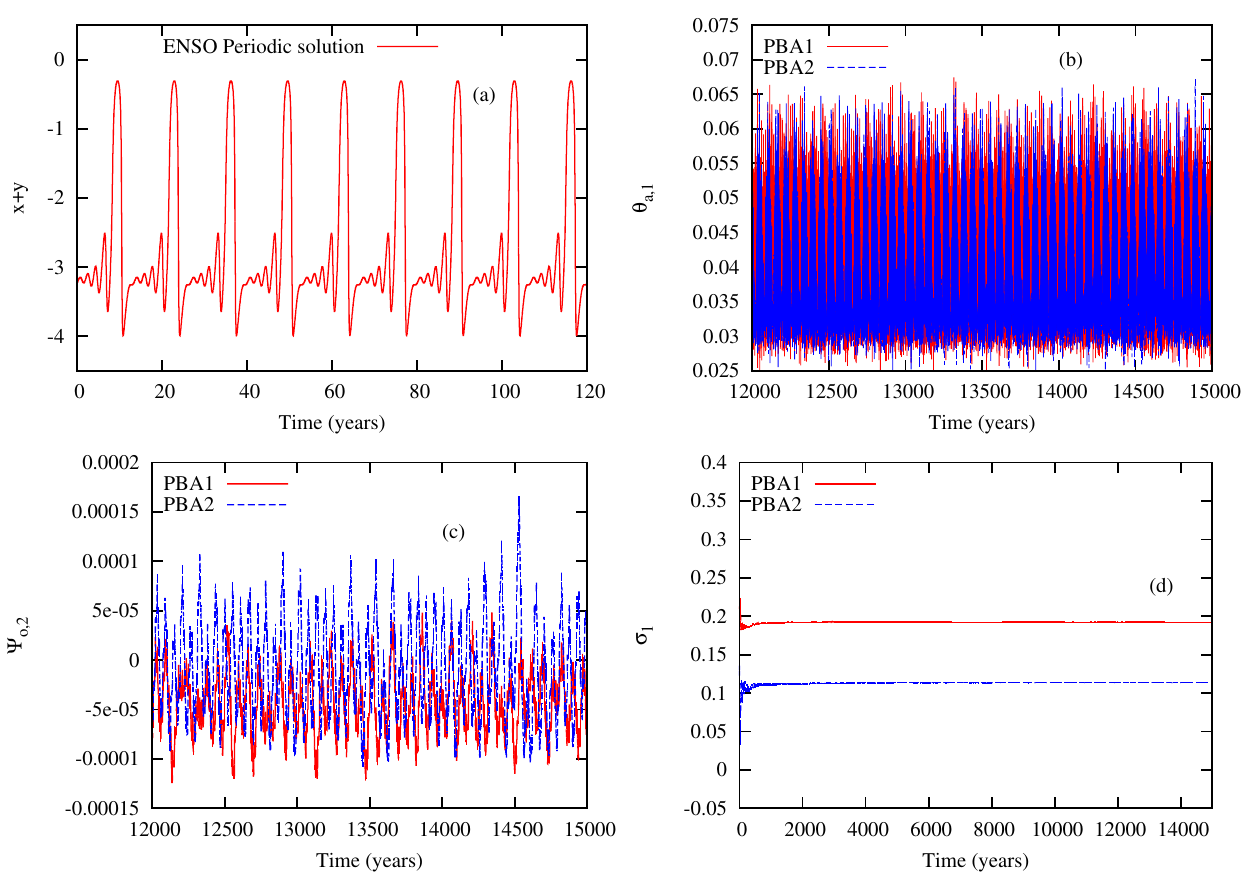}}
\caption{Model trajectory segments from a very long integration of roughly 15~000~yr, using ENSO forcing that is periodic, with $g = 0.01$ and air-sea coupling of $C = 0.015$. (a) A segment of the ENSO forcing, identical to the one displayed in Fig.~\ref{fig:ENSO_force}(a); (b) evolution of the atmospheric temperature variable $\theta_{a,1}$;  (c) evolution of the oceanic streamfunction variable $\Psi_{o,2}$; and (d) convergence of the two leading Lyapunov exponents $\sigma_1^{\pm}$; in panels (b)--(d), the two curves, red and blue, correspond to the two PBAs obtained. 
Cyclostationarity holds, given the periodic forcing, and any PBA segment (not shown) displays features similar to the ones in this figure.}
\label{trajectories}
\end{figure}

\subsubsection{The pullback attractors (PBAs) for $g=0.01$}
\label{ssec:PBAsper}

Let us now focus on the interesting case in which multiple PBAs are present at $g=0.01$. Figure~\ref{trajectories} displays the solutions after a very long time integration of about 15~000~yr, obtained using two distinct initial states that lead to distinct dominant exponents. Given the periodic forcing, asymptotic behavior is attained after a mere 3~072 years.

Panel (a) shows the temporal evolution of the eastern Tropical Pacific's temperature anomaly. The period of this ENSO forcing equals roughly 13~yr. The maxima correspond to warm, El Ni\~no events. In panel (b), the evolution of $\theta_{a,1}(t)$, the first temperature mode in the baroclinic streamfunction \cite{Vannitsemetal2015}, is displayed for the two PBAs. 

The trajectory labeled PBA2 (blue curve) exhibits a clear sequence of low and high values that recur on long time scales of the order of 60~yr. This oscillation is associated with the coupled ocean--atmosphere VDDG model's natural LFV in midlatitudes, which is similar to the one generated in the coupled model without ENSO forcing \cite{Vannitsemetal2015}. The trajectory labeled PBA1 (red curve), on the other hand, looks more erratic. A similar picture is found for the other atmospheric variables (not shown).

For the second mode $\Psi_{o,2}(t)$ of the ocean streamfunction displayed in panel (c), larger intensities of the ocean transport are found for PBA2 (blue curve), while a more erratic, albeit lower-amplitude evolution is found for PBA1 (red curve). Finally, in panel (d), the convergence of the leading Lyapunov exponents is shown, with the two attractors displaying distinct stability properties, as seen in Fig.~\ref{lyapper}. Clearly, $\sigma_1^{\pm}$ correspond to PBAs 1 and 2, respectively, with the coupled PBA2 flow in the midlatitudes being stronger, as per panel (c), but more regular and stable, as per panel (d).     

\begin{figure}[ht!]
\centering
{\includegraphics[width=150mm]{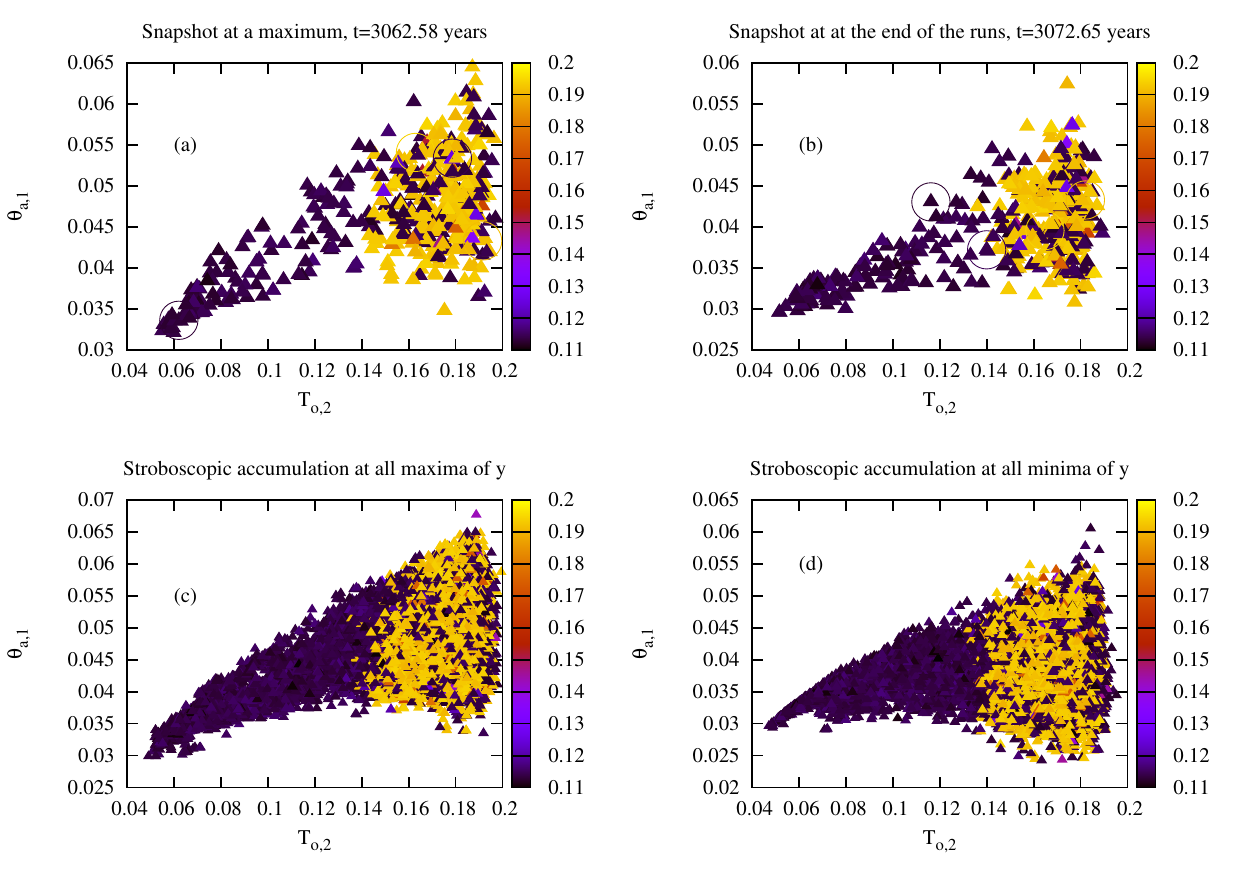}}
\caption{Coupled VDDG model snapshots obtained with 500 trajectories after a transient 
	interval of 3~072~yr. (a,b) Individual snapshots at two different times of the periodic cycle, 
	namely for $3~062.58$~yr and for $t = 3~072.65$~yr, which correspond to 
	$x+y = (x+y)_{\rm{max}}$ and to the end of the run respectively; 
	(c,d) the composites of instantaneous values at (c) the maxima and (d) the minima of the ENSO signal $x+y$, given by Eqs.~\eqref{eq:ENSO}. The warm- and cold-colored solid triangles represent instantaneous maps of solutions belonging to PBA1 and PBA2, respectively, and they are associated with high or low values of $\sigma_1$; the color bar is for $\sigma_1$.The circles displayed in panels (a) and (b) locate the solutions that will be perturbed to check the stability of the PBAs.
}
\label{snapshots}
\end{figure}

To clarify the structure of the PBAs, 500 model integrations were carried out, each starting from a different initial state in the remote past. Snapshots are first displayed in Fig.~\ref{snapshots} at two different times (panels (a) and (b)).  
The snapshots are projections onto the $(\theta_{a,1}, T_{o,2})$ plane and colors indicate the size of the asymptotic Lyapunov exponent. The attractor PBA2 associated with the smaller values $\sigma_1^{-}$ of the leading Lyapunov exponent (cold colors) has a much larger range of variability than PBA1 (warm colors). The projections of the two attractors onto the $(\theta_{a,1}, T_{o,2})$ plane overlap substantially, as seen also in Figs.~\ref{trajectories}(b,c). 

Due to the periodicity of the forcing, the attractors are expected to be cyclostationary and one can select key moments in the ENSO signal to get snapshots that are periodic, too. Panels (c) and (d) show these composite snapshots at the maximum and  minimum of the periodic forcing in the eastern basin's temperature anomaly. These snapshots also illustrate the intricate structure of the two PBAs.

\begin{figure}[ht!]
	\centering
	{\includegraphics[width=140mm]{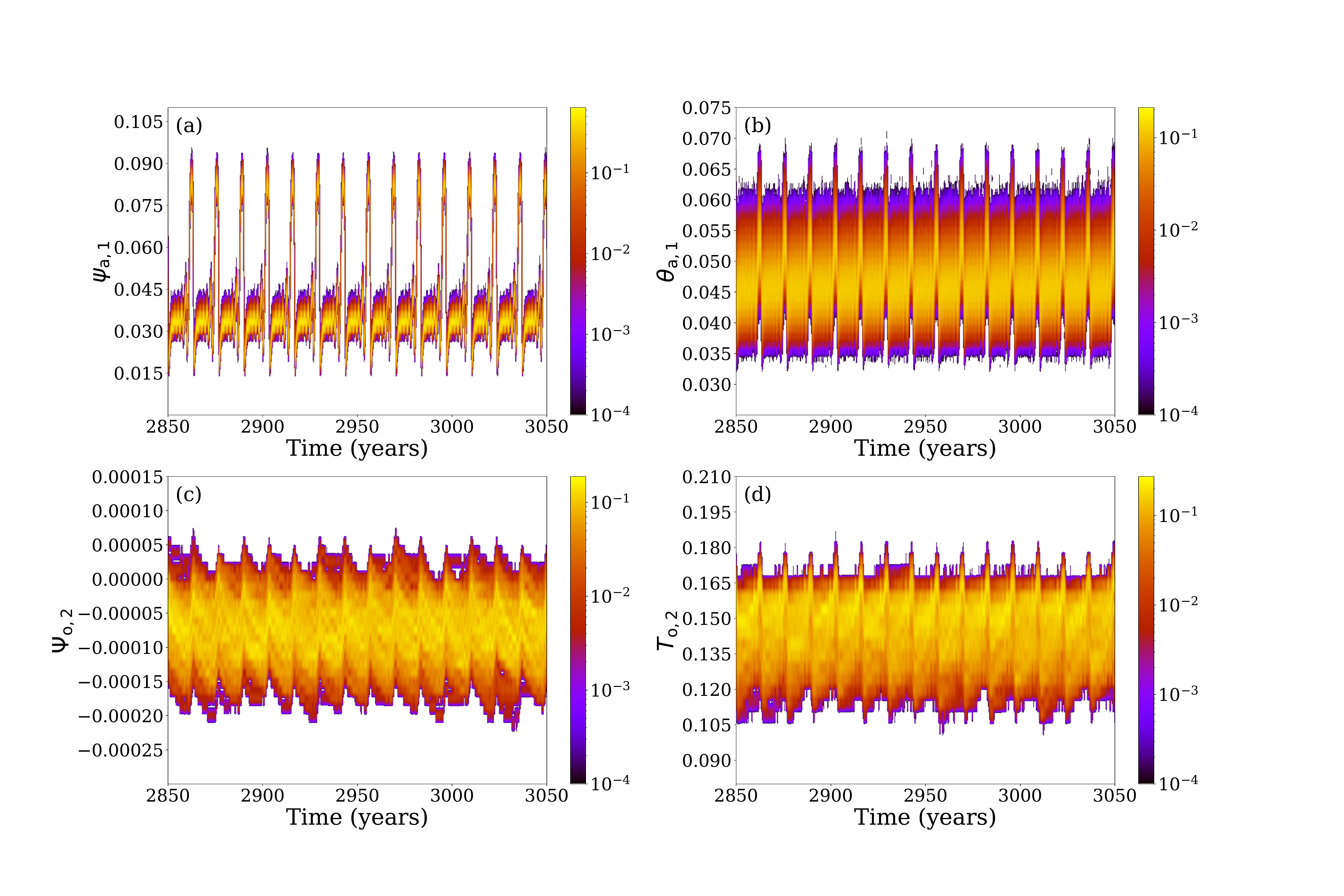}}
	\caption{Solution histograms as a function of time for 500 solutions obtained with periodic forcing and belonging to PBA1. The four histograms are for (a) $\psi_{a,1}$; (b) $\theta_{a,1}$; (c) $\Psi_{o,2}$; and (d) $T_{o,2}$. The color bar indicates the densities $H$ of the values in question, and it is scaled logarithmically, i.e., according to $\log_{10} H$, so as to get a better visual contrast.}
	\label{Histo-PBA1}
\end{figure}

\begin{figure}[ht]
	\centering
	{\includegraphics[width=140mm]{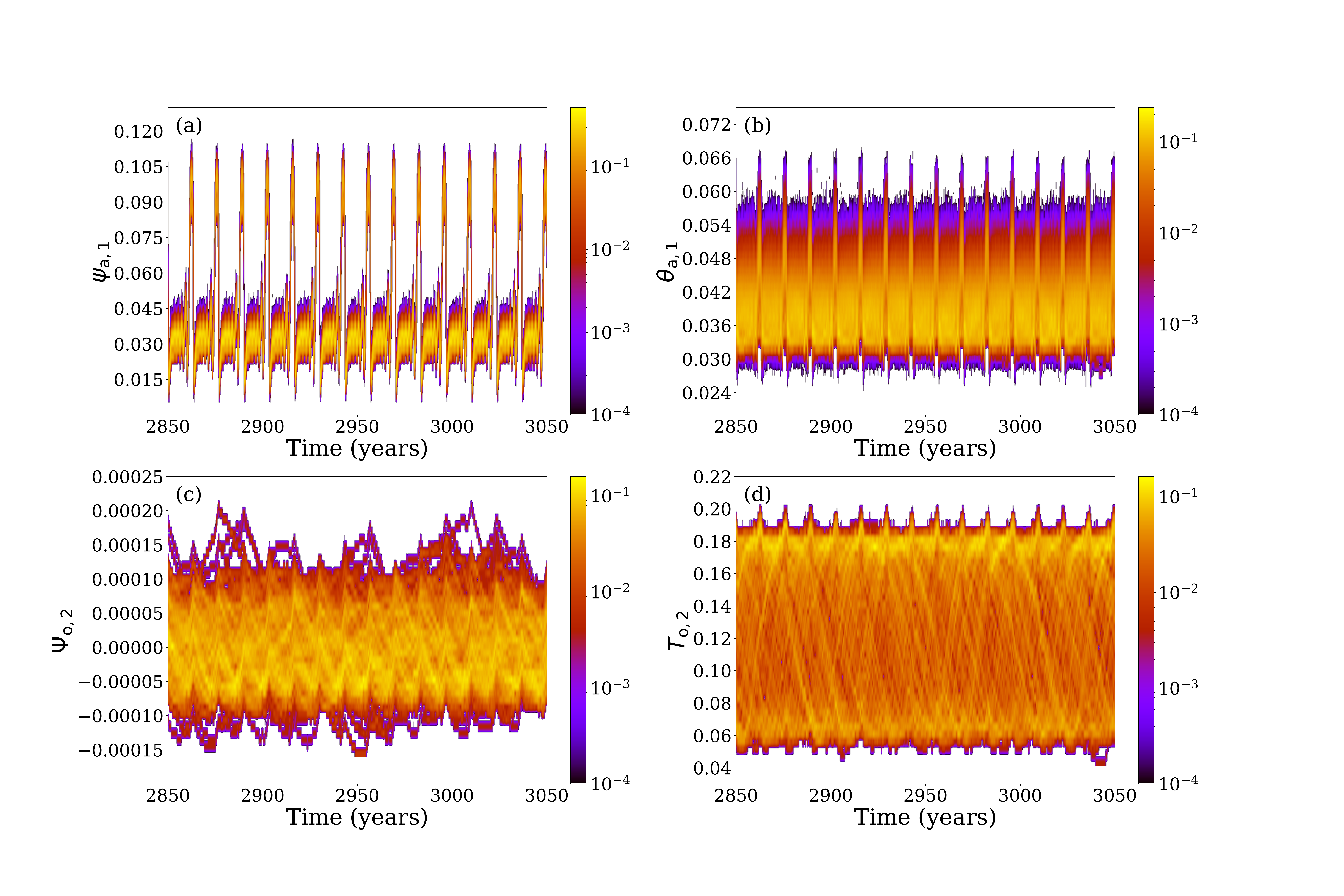}}
	\caption{Same as Fig.~\ref{Histo-PBA1} but for PBA2.}
	\label{Histo-PBA2}
\end{figure}

From the 500 trajectories produced, we also computed the  histograms of the solution values as a function of time. These histograms are displayed in Fig.~\ref{Histo-PBA1} for PBA1, which corresponds to the higher values of $\sigma_1$ and the smaller range of values in Fig.~\ref{snapshots}, and in Fig.~\ref{Histo-PBA2} for PBA2, with its lower values of $\sigma_1$ and the larger range of values in Fig.~\ref{snapshots}.

In Fig.~\ref{Histo-PBA2}(a), the impact of forcing on the distribution of possible values of $\psi_{a,1}$ results in very large and sharp peaks in the zonal flow $\psi_{a,1}$ when El Ni\~no is strong. For the first mode $\theta_{a,1}$ of atmospheric temperature (or of the baroclinic streamfunction), the peaks in panel (b) of the distribution are less pronounced but still significant; they are associated with stronger meridional gradients of the atmospheric temperature when El Ni\~no events are occurring. 

A similar effect of warm events is present for $T_{o,2}$, the second mode of midlatitude ocean temperature, displayed in Fig.~\ref{Histo-PBA2}(d). Note that, for this variable, the nondimensional values have a range of $0.05 \le T_{o,2} \le 0.20$. This range corresponds roughly to dimensional temperature variations of 5 to 20~$\deg$~C, which is quite reasonable. 

For the ocean transport $\Psi_{o,2}$ in Fig.~\ref{Histo-PBA2}(c), the range of values is also large and higher densities $H$ occur near the most negative values. The impact of the forcing on this variable is also visible in the color shading variations of $\Psi_{o,2}$ densities within the range $-0.00010 \le H \le 0.0$. The total nondimensional range of the streamfunction is $- 0.00015 \le \Psi_{o,2} \le + 0.00020$, which corresponds to roughly  $-40$ to $+ 50$~m$^2$s$^{-1}$ or to a velocity of $-0.025$ to $+0.030$~ms$^{-1}$. 

The amplitudes in  Fig.~\ref{Histo-PBA1}(a,b) of the atmospheric zonal flow $\psi_{a,1}$ and temperature gradients  $\theta_{a,1}$ for PBA1 are not that different from those found in Figs.~\ref{Histo-PBA2}(a,b) for PBA2. The main difference between the characteristics of PBA1  and those of PBA2 is the much larger range of high-density values (warm colors) of the ocean variables $\Psi_{o,2}$ and $T_{o,2}$ in Fig.~\ref{Histo-PBA1}(c) and especially in Fig.~\ref{Histo-PBA1}(d). 

In Fig.~\ref{Histo-PBA2}(d), the highest densities $H$ are concentrated very close to the most positive values of the temperature variable $T_{o,2}$, while in Fig.~\ref{Histo-PBA1}(d) they extend pretty much across the entire range of  temperature values; in other words, there is much more mixing in PBA1's ocean than in PBA2. On the other hand, Fig.~\ref{Histo-PBA2}(d) shows higher-density {\em strands} of solutions oscillating regularly between the highests and lowest temperature values. We shall return to this bunching of trajectories into oscillatory strands in the case of chaotic forcing in Sec.~\ref{ssec:chaos} below.

\begin{figure}[htb!]
	\centering
	{\includegraphics[width=65mm]{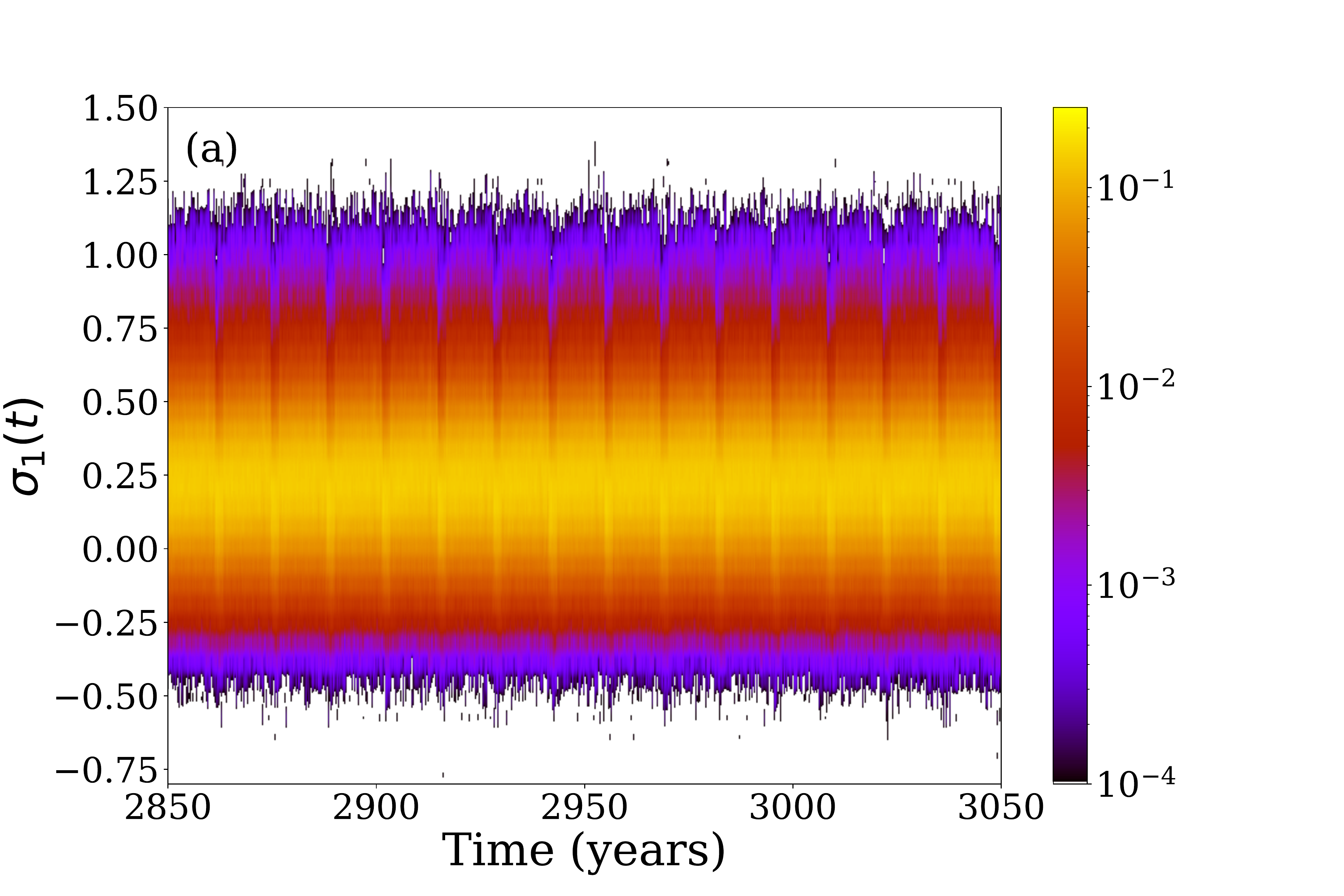}}
	{\includegraphics[width=65mm]{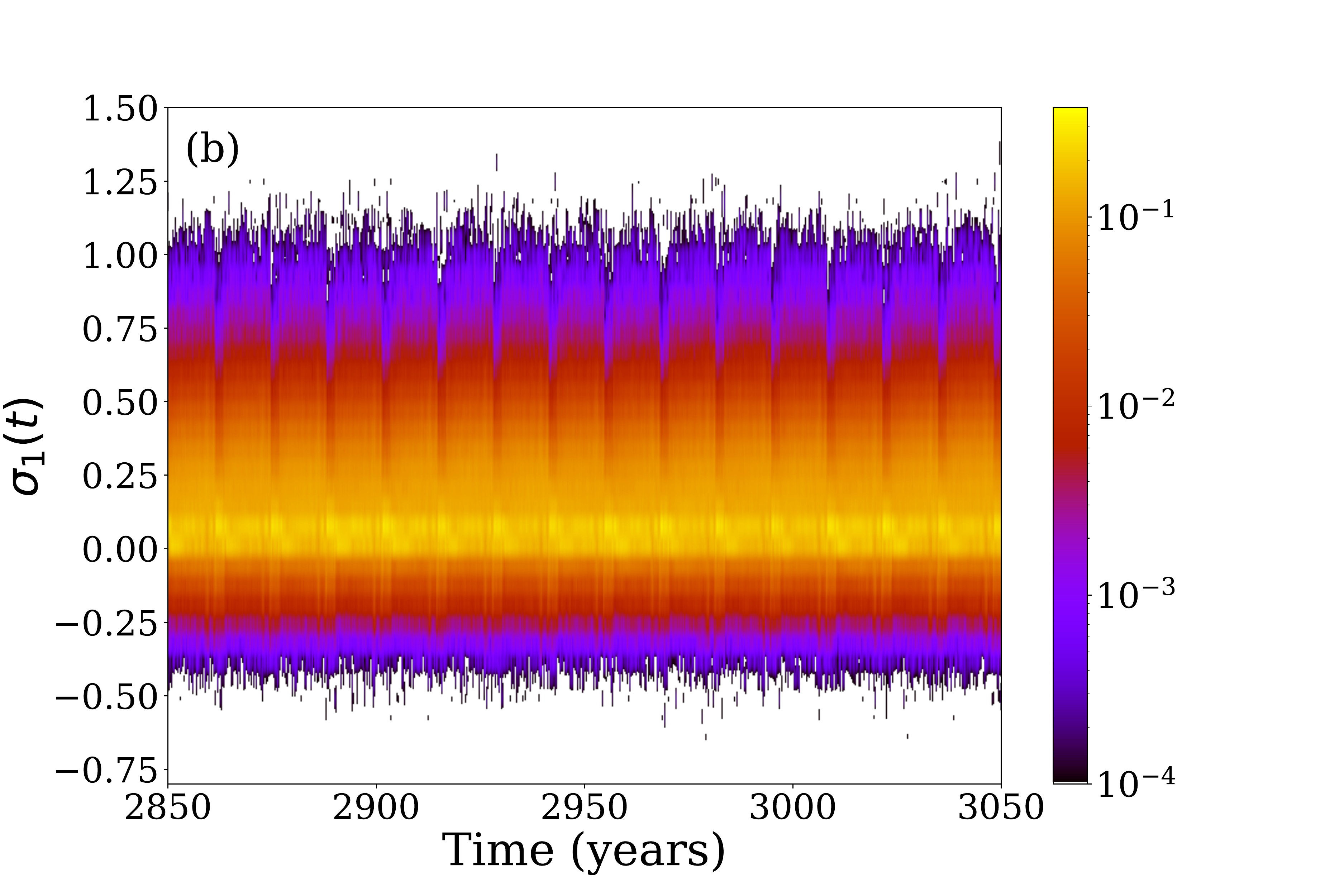}}
	\caption{Time-evolving histogram of the local values of the leading Lyapunov exponent $\sigma_1$ of the two PBAs: (a) PBA1 and (b) PBA2. The local values are defined at each time step $\Delta t=0.05$ time units. 
	Same scaling of the densities as in Figs.~\ref{Histo-PBA1} and \ref{Histo-PBA2}.}
	\label{Histo-loclyap}
\end{figure}

Figure~\ref{Histo-loclyap} is a plot of the histograms of the local values of the leading Lyapunov exponents for both PBAs. For PBA2, the most frequently occurring values are located around 0, and the periodic forcing is imposing a reduction of the higher values when strong El Ni\~nos are occurring. For PBA1, a similar feature is found, except that the most frequently occurring values cluster around $\sigma_1 \simeq 0.25$.

Up to now, the results suggest that two different chaotic PBAs coexist in phase space at the forcing parameter value of $g = 0.01$. But a crucial question when dealing with multiple attractors is their stability, and the possibility of jumping from one PBA to the other when perturbations in the initial states are introduced. If a small perturbation may lead to a transition to the other PBA, intricate boundaries between their respective basins of attractions could be present. If so, 
predictions of a specific solution type may be strongly affected by the presence nearby of solutions with very different properties. 
\begin{figure}[hb!]
	\centering
	{\includegraphics[width=130mm]{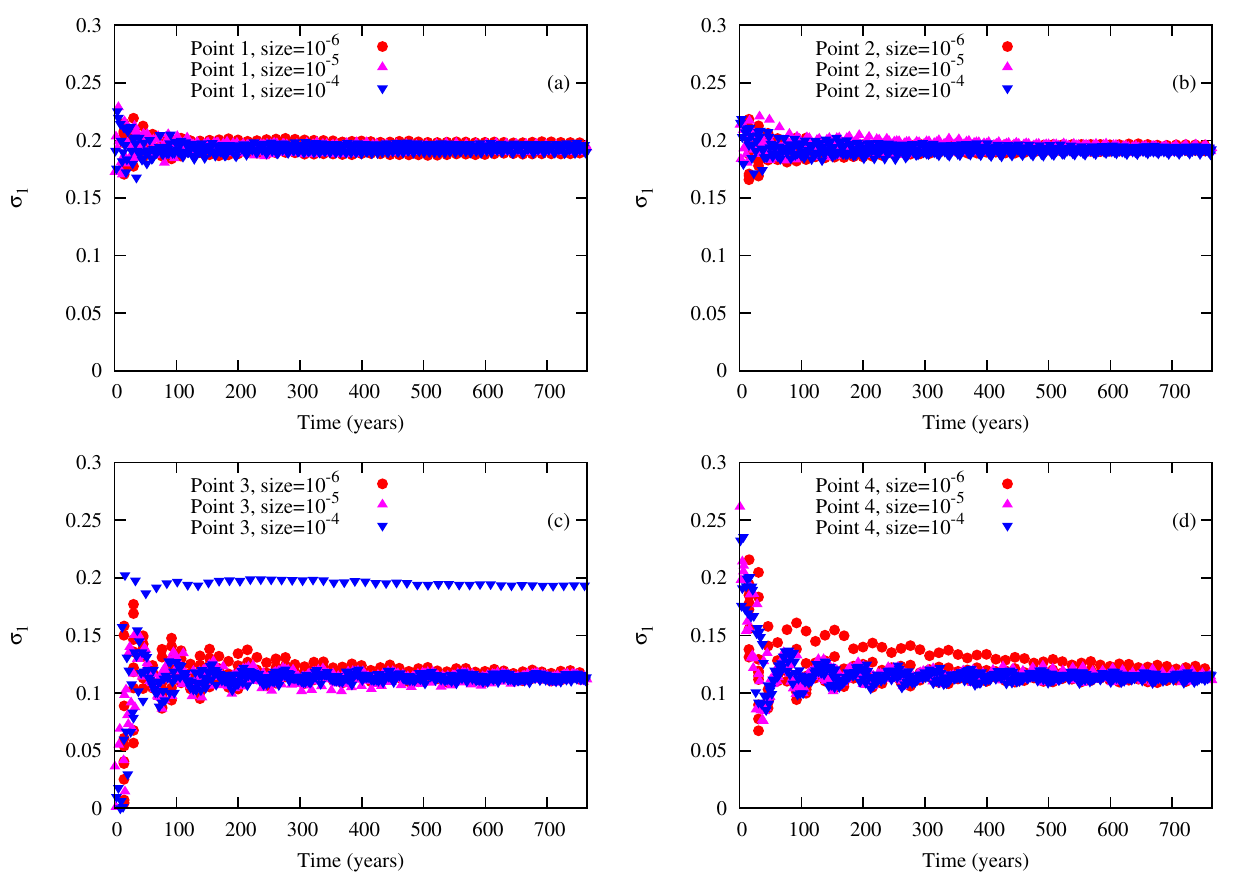}}
	\caption{Convergence of $\sigma_1$ after perturbing the four points selected at random within the snapshots shown in Fig.~\ref{snapshots}(b), at the end of the 3~072-yr--long trajectory analyzed therein. The legend in each of the four panels (a)--(d) indicates the amplitude of the perturbations: $10^{-6}$ (red), $10^{-5}$ (purple) and $10^{-4}$ (blue). For each amplitude, we performed 10 runs with 10 different random perturbations; in other words, 30 runs were done for each point. 
	}
	\label{balllyap}
\end{figure}

Such a possibility renders the problem of prediction --- and in particular the ensemble predictions often carried out in the climate sciences --- much harder, as some perfectly appropriate solutions could rapidly evolve far away from the system's real trajectory. Such rapid divergences could considerably affect the estimation of prediction errors. In the present case of periodic forcing, this situation will lead, in particular, to a loss of cycloergodicity, i.e. temporal averages taken at multiples of the forcing period will no longer equal the ensemble averages.    

To evaluate the stability of the two coexisting PBAs, four points in the snapshots of Fig.~\ref{snapshots}(b) were taken at random and perturbed with uniformly distributed random perturbations of amplitudes in the interval $[10^{-4}, 10^{-6}]$; each of the selected points is marked by a circle. The perturbations are introduced in all the variables, and the amplitude of $10^{-4}$ is of the same order at the natural variability of the ocean dynamics. 

To assess the PBAs' stability, a new set of 1536-yr--long integrations are performed starting from these perturbed initial states and the convergence of the leading Lyapunov exponents $\sigma_1$ is analyzed during the second, 768-yr--long part of these integrations. If a trajectory switches from one PBA to the other, then the corresponding $\sigma_1$ will change in value. As illustrated in Fig. \ref{balllyap}, there is no significant change in $\sigma_1$ with time, and hence no switching between the two PBAs. 
Although this analysis is made for a few points only, it suggests that the two PBAs are robust under changes of initial ensemble states, except for one large perturbation at point 3, in panel (c), with the largest, $10^{-4}$ amplitude. 

\begin{figure}[hb!]
\centering
{\includegraphics[width=130mm]{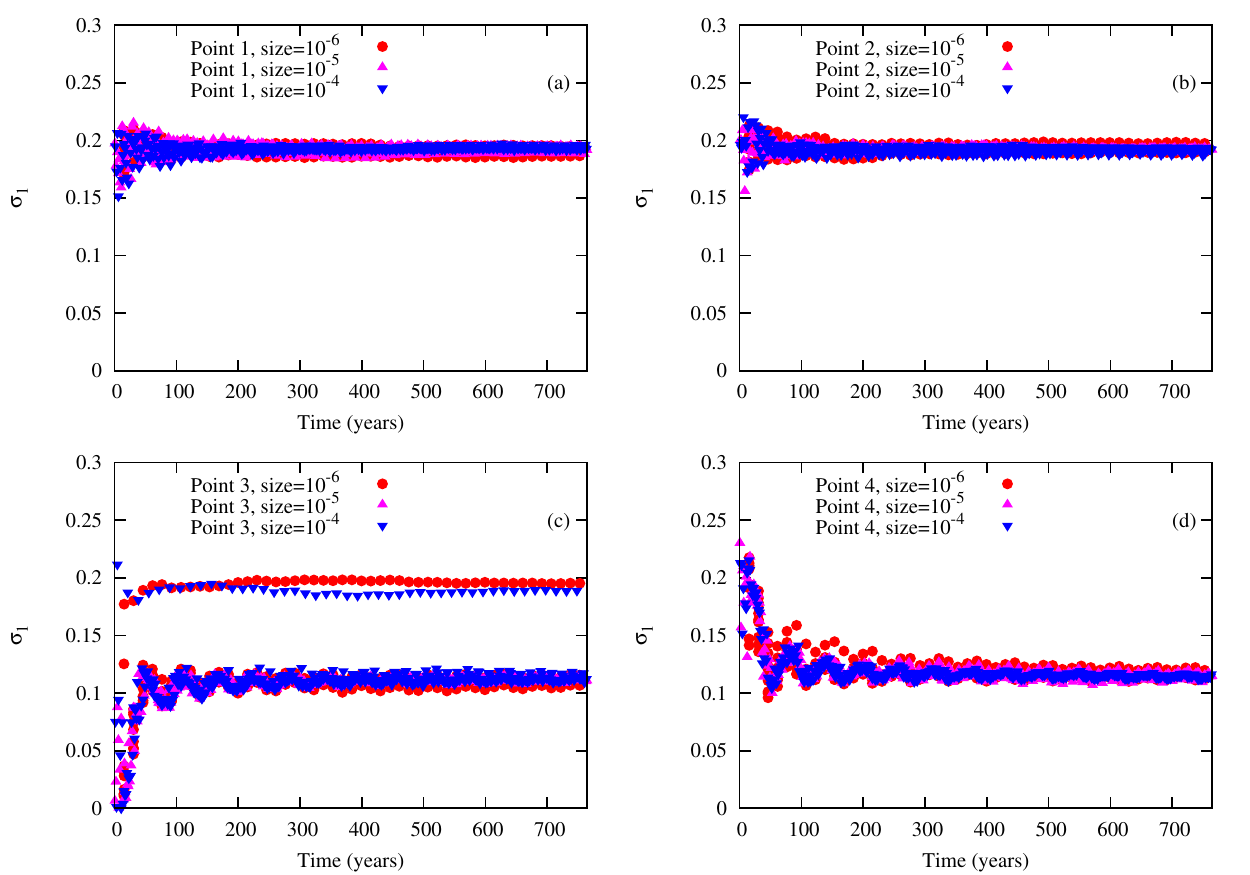}}
\caption{Same as Fig. \ref{balllyap}, but for the four points marked by open circles in Fig.~\ref{snapshots}(a), at a maximum of the ENSO forcing $x+y$.}
\label{ballmax}
\end{figure}

A similar behavior is found in Fig.~\ref{ballmax} when perturbing another four points within the snapshots of Fig.~\ref{snapshots}(a), at a maximum of the ENSO forcing signal. In this case, two trajectories are switching PBAs when perturbing point 3, one with the largest amplitude, as before, and one with the smallest one, of $10^{-6}$ in amplitude. As we shall see in Sec.~\ref{ssec:chaos}, in the presence of  chaotic ENSO forcing, the transitions between the two PBAs are much more frequent. 

These results suggest that two rather robust chaotic PBAs coexist in the presence of periodic ENSO forcing. The two types of solutions are quite different. More specifically, for the forcing parameter value of $g=0.01$, there are key differences in their LFV: For PBA2, with its smaller $\sigma_1$,  the LFV is dominated by multidecadal time scales, as in the LFV found by \citeA{Vannitsemetal2015} for the VDDG model in the absence of ENSO forcing; this PBA also exhibits a high predictability, as expected. For PBA1, with its larger $\sigma_1$, the LFV has a much smaller range, it is closely related to the period of the external forcing, and it is characterized by a lower predictability.


\subsection{Chaotic forcing} \label{ssec:chaos}

\begin{figure}[ht!]
	\centering
	{\includegraphics[width=130mm]{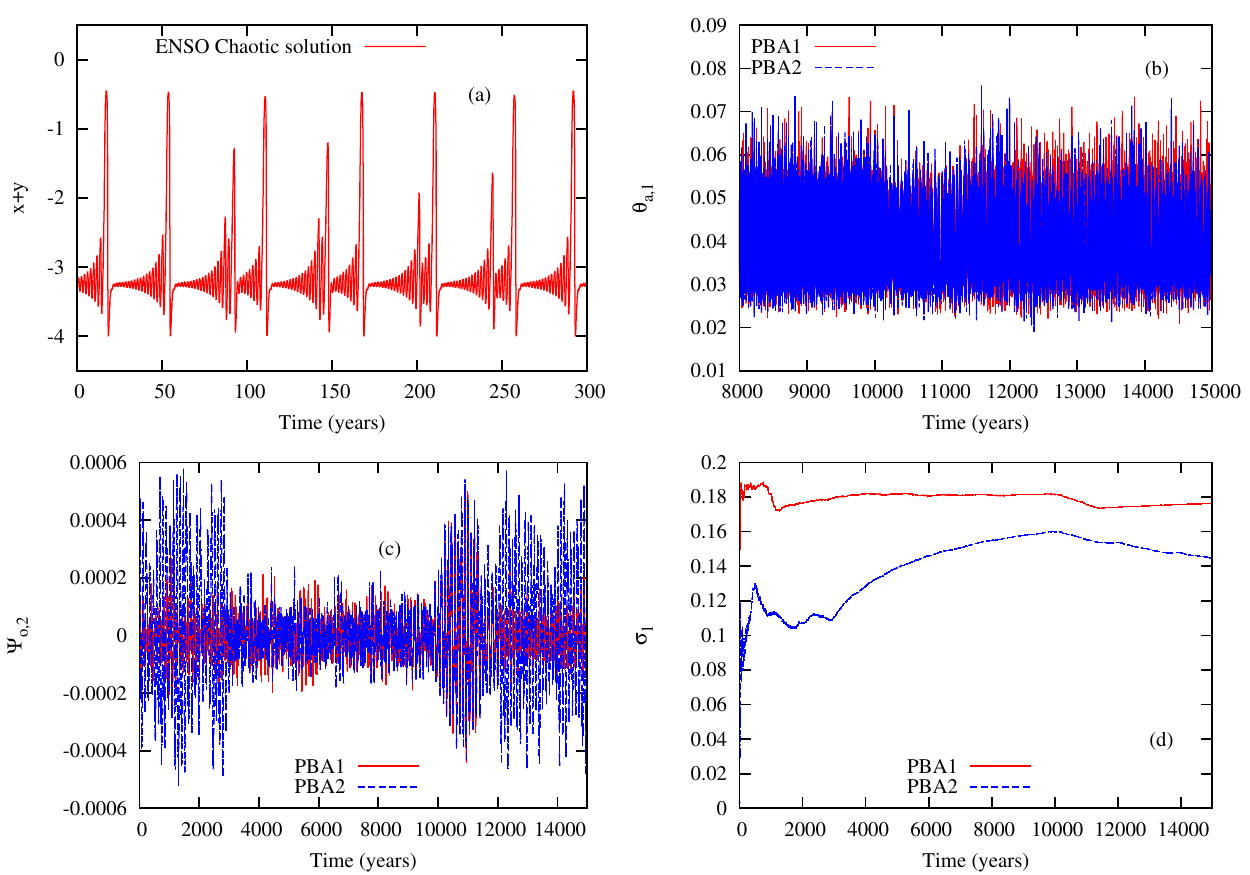}}
	\caption{As in Fig.~\ref{trajectories} but for chaotic ENSO forcing, with $g = 0.03$. (a) Time segment of 300~yr from the chaotic forcing, identical to the one displayed in Fig.~\ref{fig:ENSO_force}(b); (b,c) evolution of the variables $\theta_{a,1}$ and $\psi_{o,2}$ for a long integration of about 15~000~yr, as in Fig.~\ref{lyapper}, with the transient still fixed at 3~072~yr; and (d) convergence of the leading Lyapunov exponent over the same time interval. In panels (b)--(d), the curves for PBA1 are red and those for PBA2 are blue.
	}
	\label{trajcha}
\end{figure}

We consider now the case of chaotic ENSO forcing upon the coupled VDDG midlatitude model obtained when the parameter values in the right column of Table~\ref{tab:param} are used for the tropical module. The typical trajectory of this forcing is illustrated in Fig. \ref{trajcha}(a). Note that, in this case, the recurrence times of large warm events are quite irregular and typically longer than when the  forcing is periodic; compare Figs.~\ref{eq:ENSO}(a) and (b). 

For an ENSO forcing coefficient of $g$=0.03, Fig.~\ref{trajcha} shows the typical evolution of two long integrations of about 15~000~yr: one exhibits a higher value of the dominant Lyapunov exponent  $\sigma_1$ (red curves) and the second one a lower value thereof (blue curves). The inequality $\sigma_1(\rm{PBA1}) > \sigma_1(\rm{PBA2})$, observed already for the periodic case in Fig.~\ref{trajectories}(d), is clearly confirmed in Fig.~\ref{trajcha}(d), where the convergence of $\sigma_1$ as a function of time is displayed for the chaotic forcing in panel (a). It is obvious, though, that the convergence  in panel (d) here takes much longer than in Fig.~\ref{trajectories}(d) and is far from monotone. 

Figures~\ref{trajcha}(b, c) illustrate the corresponding evolution of the atmospheric variable $\theta_{a,1}$ and the oceanic variable $\psi_{o,2}$. In the two panels, the oscillations in both atmospheric and oceanic variables for PBA2 (blue curves) are associated with the slow dynamics on multidecadal time scales present already in the VDDG model with no ENSO forcing \cite{Vannitsemetal2015}. The ENSO forcing, though, seems to dominate the  PBA1's LFV (red curves), in which the very slow intrinsic VDDG dynamics is not as strong. This difference between the two PBAs is particularly evident in the oceanic variable of panel (c).

The atmospheric oscillations in Fig.~\ref{trajcha}(b) have large amplitudes for both PBAs throughout the time interval. For the oceanic oscillations in panel (c), though, the amplitude is quite intermittent, with long episodes of amplitudes that are quite small. Thus, the trajectory from PBA1 (red) has just a single large-amplitude burst, around 11~000~yr. 

An interesting feature is noticeable when taking a closer look at Figs.~\ref{trajcha}(b,c): the mean value of the atmospheric temperatures  $\theta_{a,1}$ is lower during the high-amplitude episodes of the oceanic oscillations than during the low-amplitude episodes of the latter. The entries in Table~\ref{tab:mean+std} clearly show the very large amplitude difference between the ocean's leading variable  $\psi_{o,2}$, with an almost threefold factor between the quiescent middle episode (2) and the two much more active episodes (1) and (3) --- $6.8 \cdot 10^{-5}$ vs. $2.63 \cdot 10^{-4}$ and $2.35 \cdot 10^{-4}$. The corresponding difference in the means of the leading atmospheric variable $\theta_{a,1}$ appears small at first --- $4.11 \cdot 10^{-2}$ in the middle vs. $3.85 \cdot 10^{-2}$ and $3.01 \cdot 10^{-2}$ at the two ends --- but is highly significant given the very small standard deviations, of the order of $6 \cdot 10^{-3}$, for all three episodes.

Thus enhanced oceanic variability suppresses the mean intensity of the meridional temperature gradient, and hence the intensity of the atmospheric westerly jet. From a physical point of view, an intensification of the oceanic circulation leads to an increase of the total heat transport toward the pole. The heat transported is rapidly exchanged with the atmosphere at high latitudes and thus leads to a reduction of the meridional temperature gradient within the atmosphere, which in turn reduces the westerly flow. On the contrary, when oceanic variability is low, poleward heat transport is low and the temperature gradients in both the ocean and the atmosphere are larger in the mean. 

\begin{table}[ht!]
	\centering
	\caption{Means and standard deviations (std) of atmospheric and oceanic VDDG variables for chaotic ENSO forcing. Results for atmospheric winds and oceanic currents in the PBA2 trajectory (blue) of Fig.~\ref{trajcha}. The episodes $(1, 2, 3)$ are based on visual inspection of panel (c).}
	\setlength\tabcolsep{6 pt} 
	\begin{tabular}{llll} 
		\hline 
		Episode & Variable & Mean & Std  \\
		\hline
		(1) & $\simeq 0-3~000$~yr&&\\
		\hline
		& $\psi_{o,2}$ & $2.43 \cdot 10^{-5}$ & $2.63 \cdot 10^{-4}$ \\
		& $\theta_{a,1}$ & $3.85 \cdot 10^{-2}$ & $6.55 \cdot 10^{-3}$ \\
		\hline
		(2) & $\simeq 3~000 - 10~000$~yr &&\\
		\hline
		& $\psi_{o,2}$ & $-1.57 \cdot 10^{-5}$ & $6.8 \cdot 10^{-5}$  \\
		& $\theta_{a,1}$ & $4.11 \cdot 10^{-2}$ & $5.87 \cdot 10^{-3}$ \\
		\hline
		(3) &  $\simeq 10~000 - 15~000$~yr &&\\
		\hline
		& $\psi_{o,2}$ & $1.69 \cdot 10^{-5}$ & $2.35 \cdot 10^{-4}$  \\ 
		& $\theta_{a,1}$ & $3.91 \cdot 10^{-2}$ & $6.01 \cdot 10^{-3}$ \\
		\hline
	\end{tabular} \label{tab:mean+std}
\end{table}

Given the strikingly intermittent behavior in Fig.~\ref{trajcha}(c), it is difficult to  distinguish between the two PBAs, since transitions between the two basins of attractions do,  apparently, occur. The difference in the behavior illustrated for very long trajectories in Figs.~\ref{trajectories} and \ref{trajcha} strongly suggests that the frequent transitions in the second case are induced by the chaotic ENSO forcing.   

\begin{figure}[htb!]
	\centering
	{\includegraphics[width=0.9\textwidth]{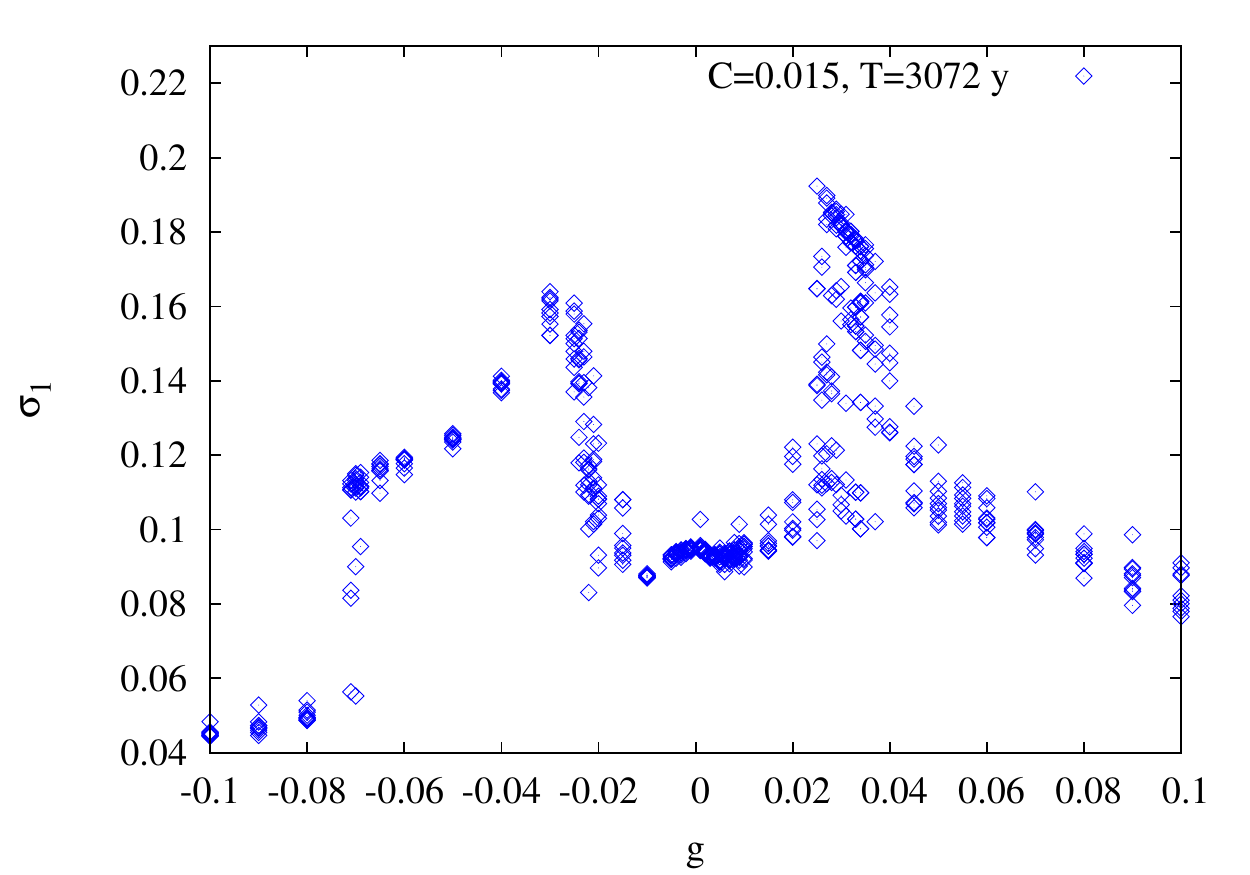}}
	\caption{Same as Fig.~\ref{lyapper} but for the chaotic ENSO forcing, with the friction coefficient $C=0.015$. Ten long trajectories were obtained for each $g$-value, each with a 3~072-yr transient and a 3~072-yr analysis interval of the leading Lyapunov exponents $\sigma_1$. The complex dependence of the $\sigma_1$-values as a function of the ENSO forcing coefficient $g$ strongly suggests the coexistence of two local PBAs, as in the periodic-forcing case.}
	\label{lyapcha}
\end{figure}

The Lyapunov exponent computation used trajectories of $2 \times 3~072$~yr in length, the first part being a transient and the second part being used to actually compute the leading Lyapunov exponents. The results are shown in Fig.~\ref{lyapcha}. Two coexistent PBAs and possible transitions between them are noticeable for a substantial range of forcing parameter values $0.02 \lesssim g \lesssim 0.04$, as we found for the periodic forcing illustrated in Fig.~\ref{lyapper} and discussed in Sec.~\ref{sssec:lyapunovper}.  

In Fig.~\ref{lyapper} and for the case of the air-sea coupling coefficient $C = 0.0015$ (blue curve in the figure), the evidence at hand strongly suggested that, over the interval $- 0.03 \lesssim g \lesssim +0.02$, two local PBAs coexist and are connected by hysteretic jumps among the two at 
$g \simeq - 0.01$ and $g \simeq +0.01$.  Here the equivalent jumps seem to occur at $g \simeq - 0.02$ and $g \simeq +0.02$. In addition, though, the interval $ 0.02 \lesssim g \lesssim 0.04$ appears to also have slow transients between the two PBAs, as we shall see further below.

We turn now to the histograms of the variables as a function of time for an ensemble of 500 trajectories of $2 \times 3~072$~yr. Since transitions appear to be possible between the two attractor basins --- which are characterized by the presence or absence of multidecadal LFV --- it is not easy to isolate the PBAs. We choose, therefore, a simple, scalar criterion: if the asymptotic value of the leading Lyapunov exponent $\sigma_1$ is larger than 0.14, the trajectory is assumed to belong to PBA1, while if $\sigma_1 < 0.14$ we associate it with PBA2. Slight modifications of this threshold value do not modify the conclusions. 
\begin{figure}[ht!]
	{\includegraphics[width=1.0\textwidth]{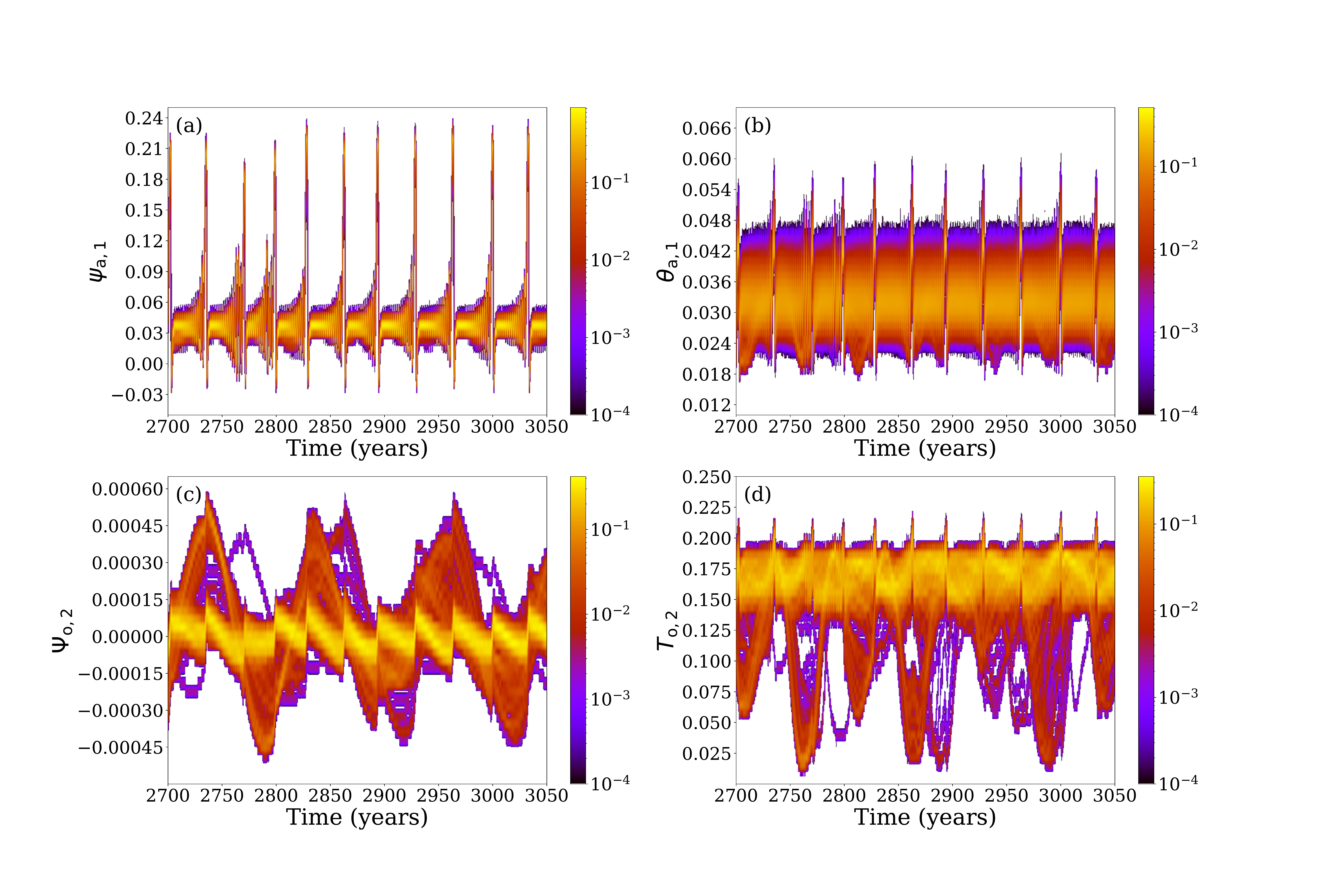}}
	\centering
	\caption{Same solution histograms as in Fig.~\ref{Histo-PBA1} but for chaotic forcing, 
		PBA1 and $g=0.03$. The four histograms displayed are for the variables: (a) $\psi_{a,1}$; (b) $\theta_{a,1}$; (c) $\Psi_{o,2}$; and (d) $T_{o,2}$. The color bar indicates the densities $H$ of the values in question, and it is scaled logarithmically, i.e., according to $\log_{10} H$, so as to get a better visual contrast.}
	\label{histo-high}
\end{figure}

For PBA1, the modulation of the histograms by the forcing is quite striking in all four panels of Fig.~\ref{histo-high}. In the case of the atmospheric variables of panels (a,b), the modulation appears to be cyclostationary, with a periodicity $P_{\rm E}$ of roughly 35~yr. 

Since the ENSO forcing in Figs.~\ref{fig:ENSO_force}(b) and \ref{trajcha}(a) is quite irregular, we carried out a straightfoward check of the ENSO signal produced by Eqs.~\eqref{eq:ENSO} with the parameter values in the second column of Table~\ref{tab:param}. To do so, we first applied a data-adaptive filter provided by the multichannel singular spectrum analysis (M-SSA) of the ENSO model's 3 variables $(x,y,z)$, using $N = 20~000$ data points, sampled every 56.075 days, with a window width of $M = 2~000$ data points; thus $N/3 \ge M \ge 35$~yr, as recommended by \citeA{Ghil.SSA.2002}, \citeA{Alessio.2015}, and references therein. Next, we carried out a maximum entropy spectral analysis of the SSA-filtered signal's $K = 10$ leading reconstructed components \cite<RCs:>[]{Ghil.Vautard.1991, Ghil.SSA.2002}, which capture 12~\% of the total variance.

As expected, the leading oscillatory mode of the signal, carried by the RC pair (1,2) with a variance of 3~\% of the total, has a period of $P_{\rm E} = 36$~yr. Subsequent pairs are associated with approximate harmonics of this basic signal: RCs (3, 4) form the 3rd harmonic with period 11~yr, RCs (5, 6) the 2nd with period 17.7~yr, (7, 8) has  again the full period (34~yr to be precise), and (9, 10) the 4th harmonic with 8.5~yr. 
The highly visible irregularity of the ENSO forcing comes from the sum of all the remaining RCs. Thus, the VDDG model's atmospheric variables pick out the 36-yr mode from the irregular ENSO forcing and amplify it, especially in the zonal velocity histogram of Fig.~\ref{histo-high}(a).

The interaction between the ENSO forcing and the midlatitude LFV is more complex for the oceanic variables in panels (c,d). In particular, the density maxima of the ocean temperatures $T_{o,2}$ in panel (d) are shifted towards high values, as was the case for PBA2 and periodic forcing in Fig.~\ref{Histo-PBA1}(d). Moreover, both evolutive histograms of panels (c,d) are even more strongly affected by the ENSO forcing than the atmospheric histograms in Figs.~\ref{histo-high}(a,b). But the dominant periodicity $P_{\rm V}$--- seen most clearly in the warm colors --- is of roughly 350~yr/$3  \simeq 110$~yr. 

To disentangle the complex effects of ENSO forcing on the midlatitude VDDG model's LFV, we carried out an M-SSA analysis of the latter's atmospheric and oceanic variables, separately. The trajectory used is 20~000~yr long, with a forcing intensity of $g = 0.03$, and it is taken from a trajectory  associated with PBA2. 

Results for the key Fourier modes of the VDDG model's atmospheric and oceanic modules appear in Table~\ref{tab:MSSA}. The periodicities are obtained, as in the case of the ENSO signal, by maximum entropy spectral analysis of the SSA-filtered signal's $K = 10$ leading RCs \cite{Ghil.SSA.2002, Alessio.2015}. As an additional verification of the results in the table, we also tested the individual time series of the atmospheric variable $\theta_{a,1}$ and the oceanic variable $\psi_{o,2}$. For $\theta_{a,1}$ the ten leading RCs capture 12\% of the total variance and the  peaks are at 39, 15.7, 11, and 8.8~yr; for $\psi_{o,2}$, the partial variance captured is of 88\% and the peaks are at 105 and 57~yr.  

\begin{table}[ht!]
	\centering
	\caption{Multichannel singular spectrum analysis (M-SSA) of the midlatitude coupled VDDG model, subject to no forcing ($g = 0$) or to chaotic ENSO forcing, with  $g = 0.03$; for both cases, the internal-coupling parameter value is $C = 0.015$. 
	The length of the time series is $N = 20~000$~yr, the window width is $M = 2~000$~yr, and the time step is $\Delta t = 56.075$~days.
		}
	\setlength\tabcolsep{6 pt} 
	\begin{tabular}{llll} 
		\hline 
		{\bf Forcing} & Variables & RC pair [var\%] & Period (yr)  \\
		\hline
		{\bf No} &&&\\
		\hline
		[Atmos] & $(\psi_{a,1}, \theta_{a,1})$ & (1,2) [70\%] & 58~yr  \\
		& $(\psi_{a,1}, \theta_{a,1})$ & (3,6)~~[9\%] & 115~yr  \\
        & $(\psi_{a,1}, \theta_{a,1})$ & (4,5) [8.5\%] & 38~yr  \\
		\hline
		[Ocean] & $(\psi_{o,2}, T_{o,2})$ & (1,2) [85\%] & 58~yr  \\
		& $(\psi_{o,2}, T_{o,2})$ & (3,4) [8.5\%] & 38 + 115~yr  \\
		& $(\psi_{o,2}, T_{o,2})$ & (5,6) [5.2\%] & 115~yr  \\
	    \hline
		{\bf Yes} &&&\\
	    \hline
		[Atmos]& $(\psi_{a,1}, \theta_{a,1})$ & (1,2) [2.7\%] & 11~yr  \\ 
		& $(\psi_{a,1}, \theta_{a,1})$ & (3,4) [2.5\%] & 35~yr  \\
		& $(\psi_{a,1}, \theta_{a,1})$ & (5,6) [2.4\%] & 17~yr  \\
		& $(\psi_{a,1}, \theta_{a,1})$ & (7,8) [2.3\%] & 34~yr  \\
		& $(\psi_{a,1}, \theta_{a,1})$ & (9,10) [2.1\%] & 8.8~yr  \\
		\hline
	    [Ocean] & $(\psi_{o,2}, T_{o,2})$ & (1,2) [55\%] & 54~yr  \\ 
	    & $(\psi_{o,2}, T_{o,2})$ & (3,4) 11\%] & 53 + 115~yr  \\ 
	    & $(\psi_{o,2}, T_{o,2})$ & (5,8) [9.5\%] & long trends \\
	    & $(\psi_{o,2}, T_{o,2})$ & (6,7) [9.2\%] & 115~yr  \\ 
		\hline
	\end{tabular} \label{tab:MSSA}
\end{table}

\citeA{Vannitsemetal2015} found the dominant periodicity of the VDDG model in the absence of the time-dependent ENSO forcing to be roughly 60~yr. Table~\ref{tab:MSSA} here confirms that, in this case, the dominant oscillatory pair (1,2) has a period of $P_{\rm V} \simeq 58$~yr; the associated variance is of 70\% in the atmosphere and of 85\% in the ocean. A shorter periodicity of 38~yr and a longer one of 115~yr are also present, in both the VDDG model's atmospheric and oceanic modules. The long, 115-yr period is clearly a subharmonic of the main 58-yr periodicity, since $58 \times 2 = 116$~yr, and such small differences are negligible in the presence of complex nonlinear dynamics with a substantial fraction of continuous spectrum \cite{Alessio.2015, Ghil.SSA.2002}.

For chaotic ENSO forcing with scaling parameter $g=0.03$, as in  Figs.~\ref{trajcha}--\ref{histo-high}, M-SSA yields a leading RC pair (1,2) with the period of 54~yr in the ocean, where its variance is of 55\%. This period appears to be a slight modification --- due to some rectification effect that remains to be clarified --- of the  VDDG model's intrinsic periodicity. The 54~yr period is no longer present in the atmosphere, nor is this periodicity's long, 115-yr subharmonic present in the latter. 

As previously noted in Sec.~\ref{sssec:lyapunovper}, the ENSO forcing acts directly on the VDDG model's atmosphere and only indirectly on its ocean. Table~\ref{tab:MSSA} confirms that the periodicity $P_{\rm E} \simeq 36$~yr that dominates the chaotic ENSO forcing and was visually detected in Figs.~\ref{histo-high}(a,b) is indeed present in the M-SSA results. It appears in the table, as 35~yr or 34~yr in RC pairs (3,4) and (7.8), along with its 2nd harmonic of 17~yr in RCs (5,6), its 3rd harmonic of 11~yr in RCs (1,2), and as its 4th harmonic 8.8~yr in RCs (9,10). Given the highly anharmonic, spiky appearance of this signal in Fig.~\ref{histo-high}(a), the heavy loading of the harmonics is not surprising at all, with the variances of the dominant $P_{\rm E}$ and its 4 harmonics capturing variances between 2.7\% and 2.1\%.

The M-SSA results for the oceanic variables include only the intrinsic periodicity $P_{\rm V}$, in RC pairs (1,2) and (3,4), along with its subharmonic of 115~yr in pairs (3,4) and (6,7), and even longer trends in RCs 5 and 8. The highly anharmonic and interwoven character of the oscillations makes their varimax separation still rather incomplete. The distinct spectral signatures of the atmospheric and oceanic variables in Table~\ref{tab:MSSA} was found also by \citeA{VannitsemGhil2017} in observational data for coupled ocean--atmosphere dynamics over the North Atlantic basin.

 The highly visible $P_{\rm V} \simeq 115$-yr periodicity in the oceanic histograms of Figs.~\ref{histo-high}(c,d) is a degree-3 subharmonic of the ENSO forcing at roughly 36~yr. It is the presence of this 110-yr periodicity in the oceanic spectrum of the unforced VDDG model that causes the nonlinear resonance with the ENSO forcing periodicity of $P_{\rm E} \simeq 36$~yr to occur, rather than a simpler resonance with the forcing periodicity itself. Such phenomena have been found in paleoclimate studies \cite[and references therein]{Ghil.1994} and will be discussed further in Sec.~\ref{ssec:discuss}.\\

Figures~\ref{histo-high}(c,d) reveal another interesting phenomenon, namely the splitting of tracks between groups of trajectories that we called strands in discussing Fig.~\ref{Histo-PBA2}(d). This splitting is even more obvious in Fig.~\ref{histo-low} for PBA2, and we'll describe it in discussing the latter figure below.

For PBA2, the picture in Fig.~\ref{histo-low} is quite different: a strong LFV signal on time scales of $P_{\rm V} \simeq 110$~yr is present in all the midlatitude variables, while the mean periodicity of the forcing is much shorter, of $P_{\rm E} \simeq 35$~yr. This feature is particularly visible when following the density maxima for the ocean variables in panels (c) and (d). It is quite intriguing that, in this evolutionary histogram representation, most of the trajectories fall into one of three distinct strands that seem to have the same very large periodicity but are phase shifted with respect to each other. The mutual phase shifts are by approximately 35~yr, i.e., they roughly equal the periodicity of the forcing, and one clearly sees in panel (c) that $P_{\rm V} \simeq 3 P_{\rm E}$.
\begin{figure}[ht!]
	{\includegraphics[width=1.0\textwidth]{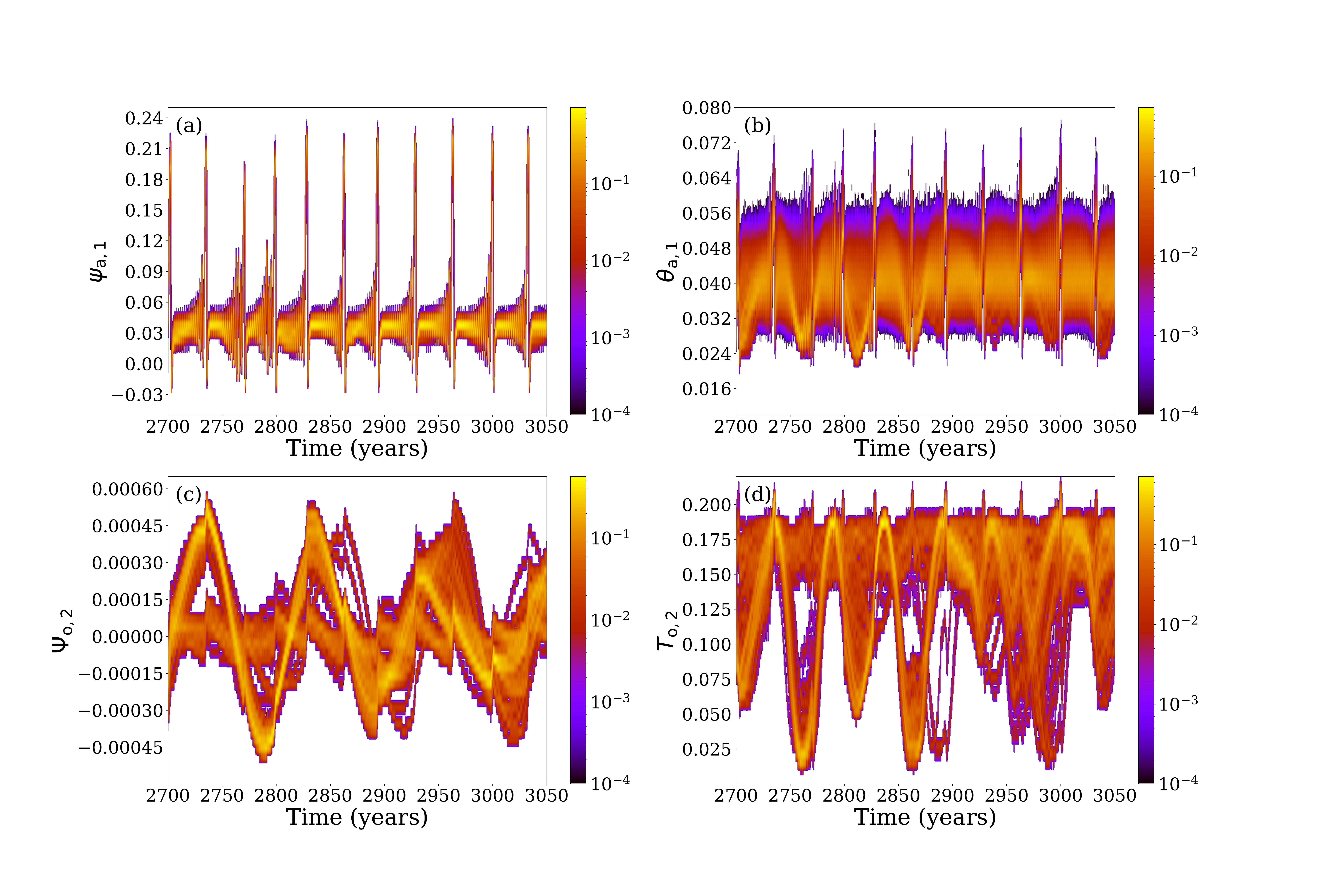}}
	\centering
	\caption{As in Fig.~\ref{histo-high}, but for PBA2.} 
	\label{histo-low}
\end{figure}

\begin{figure}
\centering
{\includegraphics[width=65mm]{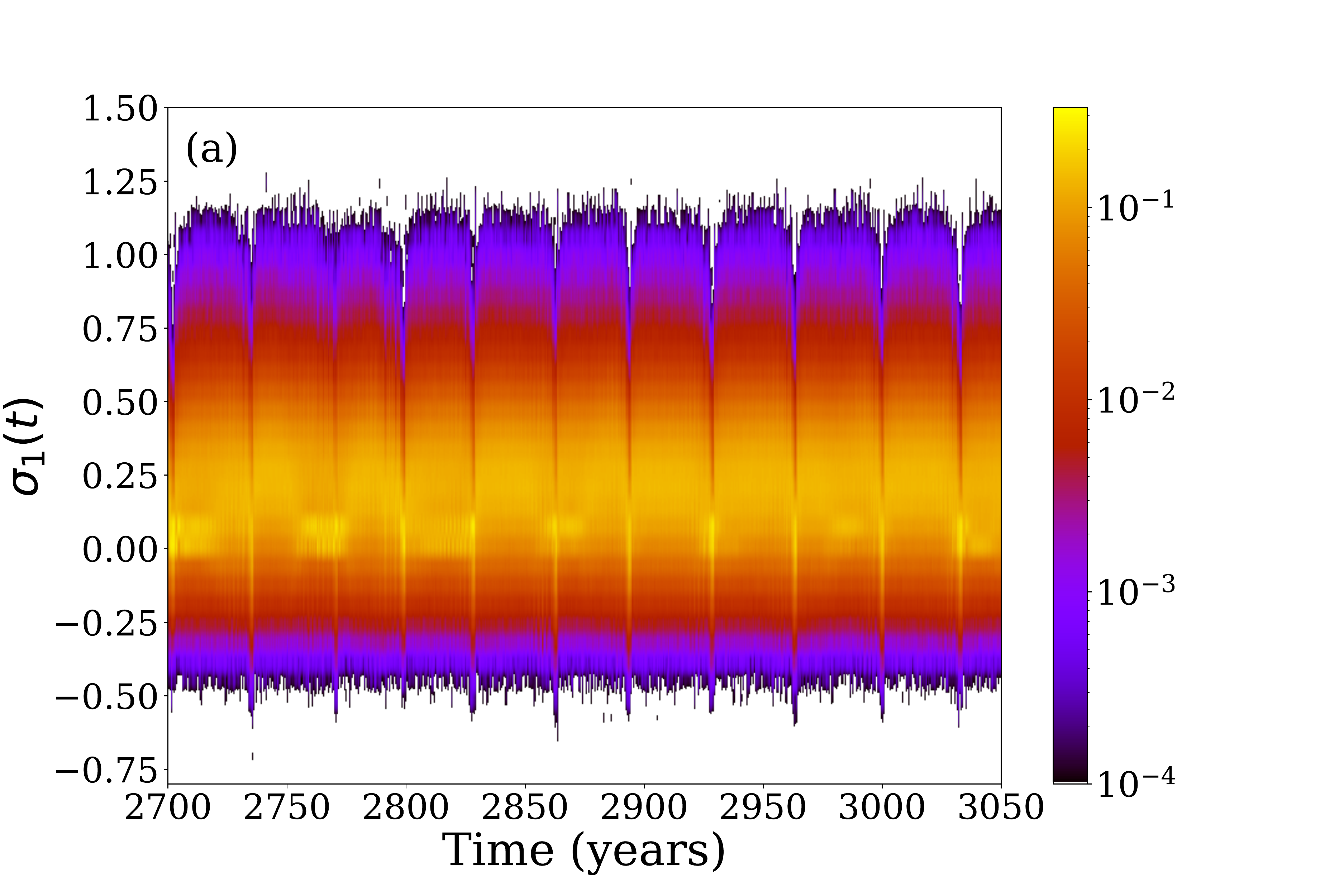}}
{\includegraphics[width=65mm]{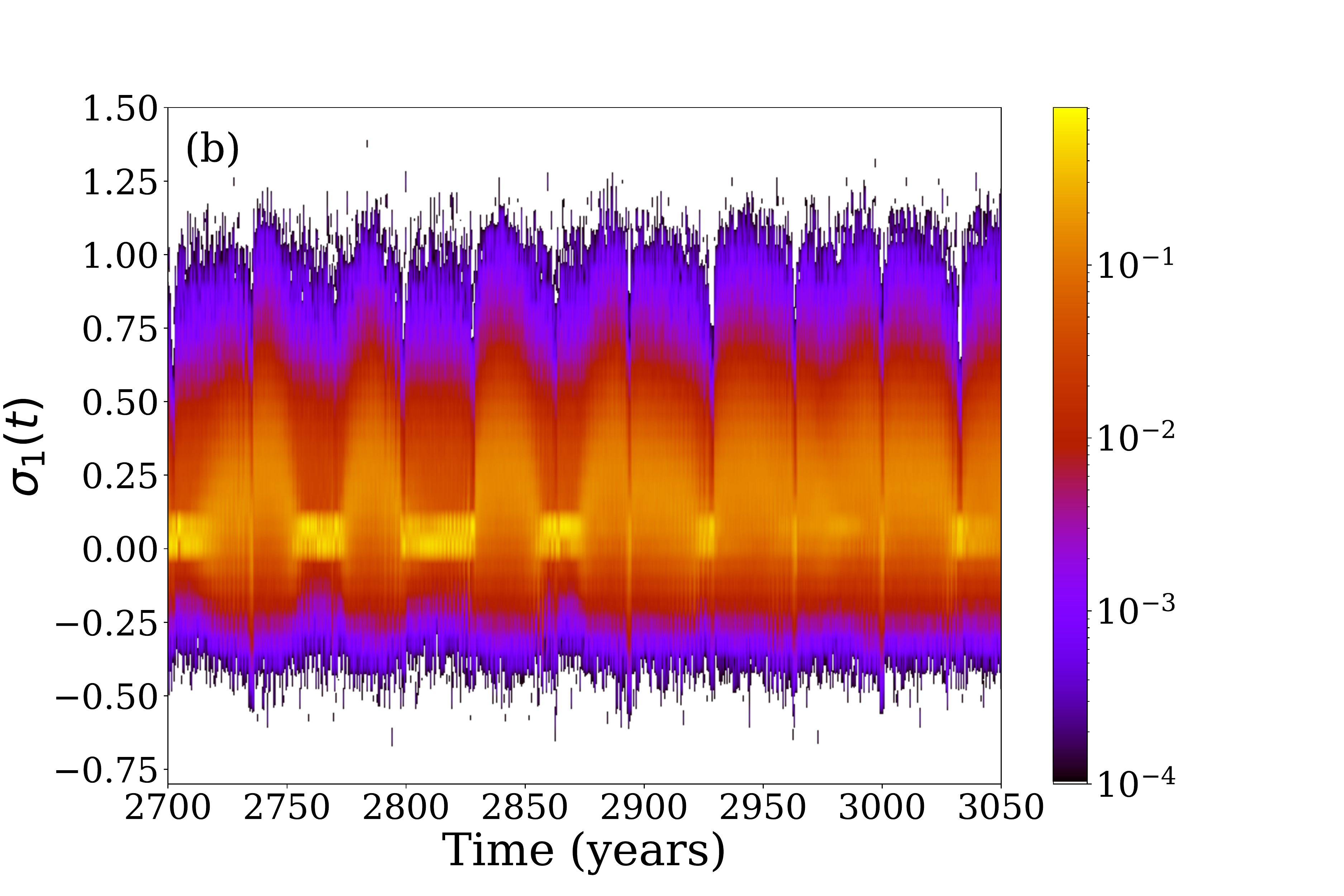}}
\caption{Same time-evolving histogram of $\sigma_1$ as in Fig.~\ref{Histo-loclyap} but for chaotic ENSO forcing with $g=0.03$: (a) PBA1 and (b) PBA2.
}
\label{histo-loclyap-cha}
\end{figure}

This phenomenon indicates a coherent dynamics induced by the presence
of the forcing, even if the latter acts on a distinct, and much shorter, characteristic time scale. The striking interaction of intrinsic and forced behavior is also present in the variations of the local predictability for PBA2, as illustrated in Fig.~\ref{histo-loclyap-cha}b: comparison with Fig.~\ref{histo-low}(d) shows that high predictability occurs preferentially when low values of $T_{o,2}$ predominate. This conditioning of high predictability on low $T_{o,2}$ values for PBA2 contrasts with the increase of predictability for PBA1 in Fig.~\ref{histo-loclyap-cha}(a) being synchronized with high amplitudes of the external forcing in Fig.~\ref{histo-low}(a).

A key question is whether the statistic properties of the ENSO-forced VDDG dynamics illustrated in Figs.~ \ref{histo-high}--\ref{histo-loclyap-cha} are robust and, in particular, independent of the time of initialization of the trajectories. To check the latter issue, we started another set of 500 trajectories from  a new set of initial states of the VDDG model, 1~536~yr in the past. 

\begin{figure}[ht!]
{\includegraphics[width=130mm]{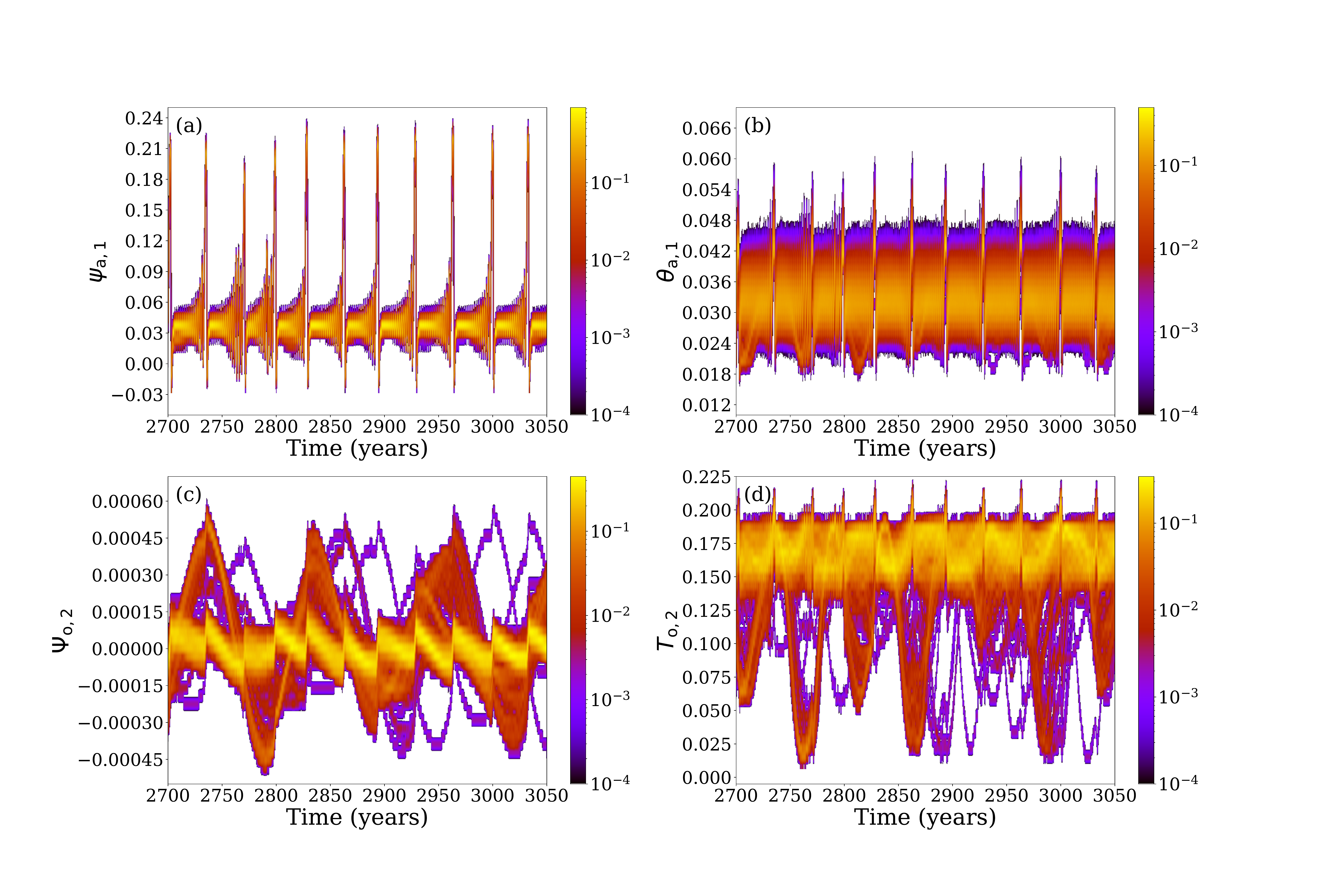}}
\caption{Same as Fig.~\ref{histo-high}, for PBA1 and 500 random initial states starting from 1~536~yr in the past, instead of 3~072~yr.}
\label{histo-high-trsh}
\end{figure}

\begin{figure}[hb!]
{\includegraphics[width=130mm]{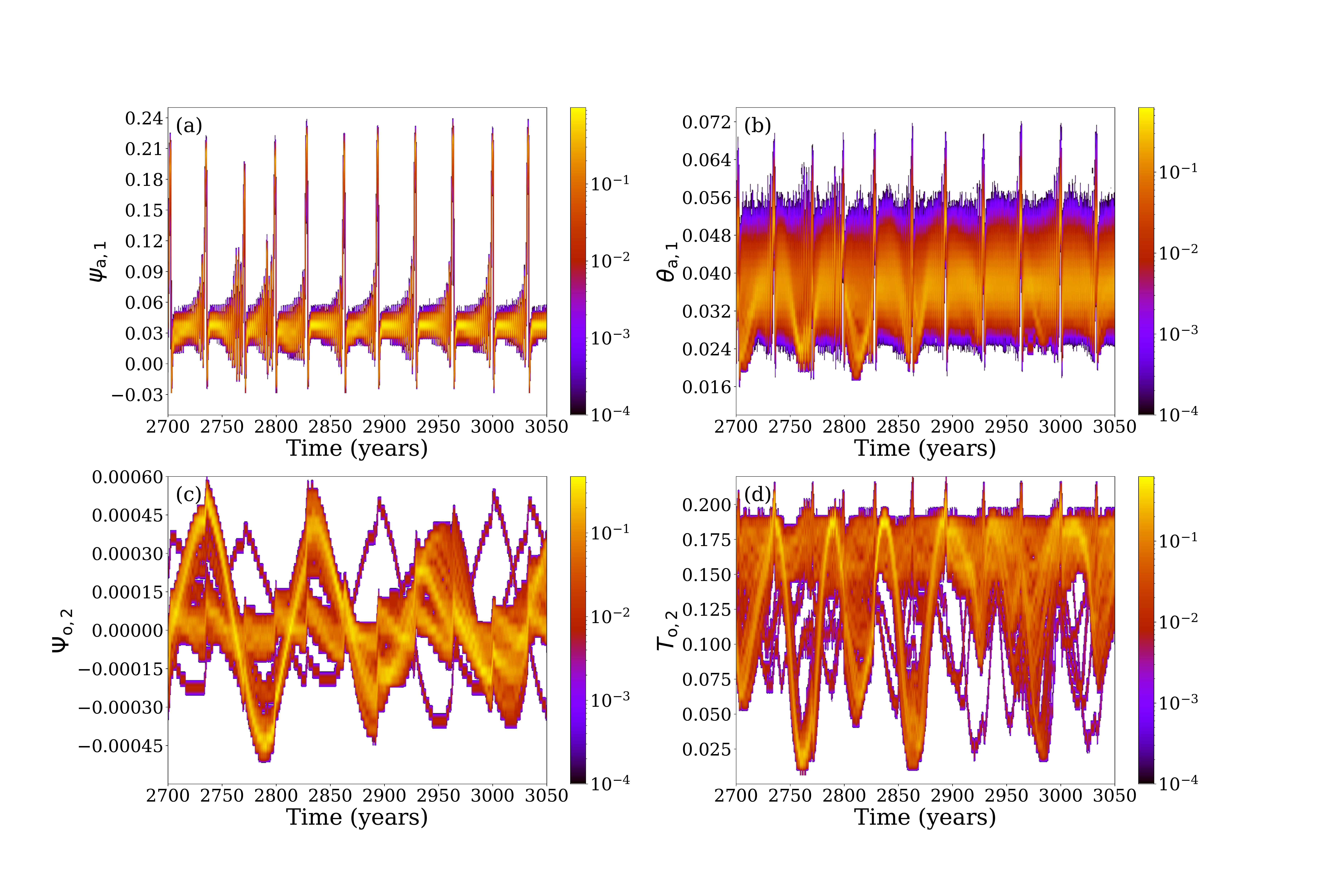}}
\caption{Same as Fig.~\ref{histo-high-trsh}, but for PBA2.}
\label{histo-low-trsh}
\end{figure}

The histograms are now displayed in Figs.~\ref{histo-high-trsh} and \ref{histo-low-trsh}, for PBA1 and PBA2, respectively. Comparison with Figs.~\ref{histo-high} and \ref{histo-low} clearly shows that (a) the shorter transient of 1~536~yr already provides sufficient convergence to either of the two PBAs' asymptotic behavior, respectively; and (b) that the evolution of the solutions within this asymptotic regime still separates into the same two PBAs and does not depend on the details of the sample set used for initialization. 

The characteristics of each PBA are pairwise the same, and the only differences are clearly attributable to sampling issues. Hence the distributions displayed in Figs.~\ref{histo-high}--\ref{histo-loclyap-cha} are genuine asymptotic features of our coupled VDDG model's solution set for the specific forcing chosen. 

\begin{figure}[ht!]
{\includegraphics[width=130mm]{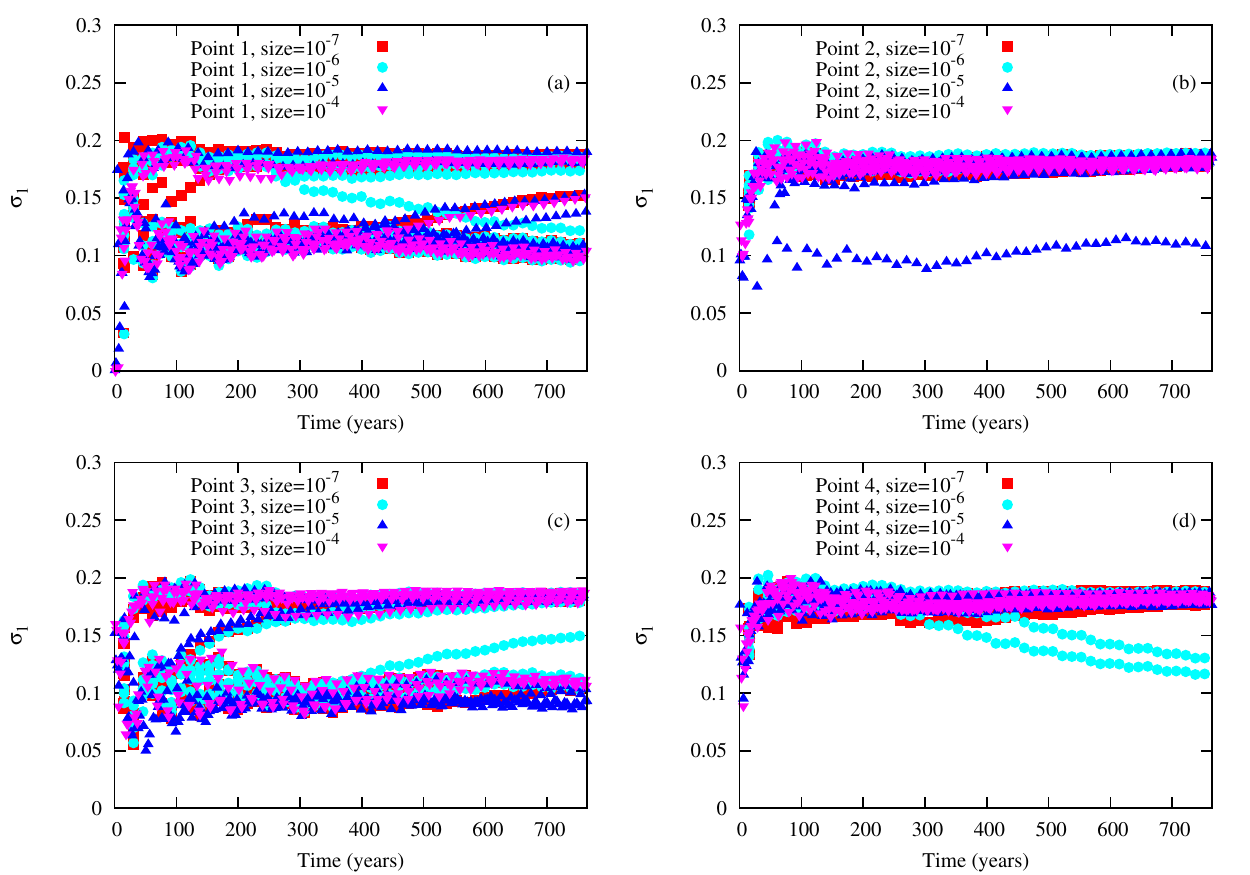}}
\caption{
Convergence of the leading Lyapunov exponent $\sigma_1$ after perturbing the initial states of four different points selected at random on the two PBAs in the middle of the total analysis interval of $2 \times 3~072$~yr. The convergence is analyzed over 768~yr. For several points on PBA2, transitions to PBA1 are found. 
}
\label{chaosball1}
\end{figure}

As for the periodic case, we may wonder whether the asymptotic dynamics we found is robust to perturbations of the state of the extratropical system. Several experiments have been performed with different perturbation amplitudes, as illustrated in Fig.~\ref{chaosball1}. The perturbations are introduced around four different points at two different instants in the forcing signal's evolution. Two of these points are associated with PBA1's high values of the dominant Lyapunov exponent $\sigma_1$ and illustrated in panels (a) and (c) of the figure, while the two others, illustrated in panels (b) and (d), are associated with PBA2 and a low value of $\sigma_1$. 

Transitions are noticeable in both directions, from PBA1 to PBA2 and vice-versa, but PBA1 is clearly more robust than PBA2. Even for very small perturbations of the solutions, with amplitudes as low as $10^{-7}$, some solutions initially belonging to PBA2 switch to PBA1. The transitions  from PBA1 to PBA2 seem to occur only for relatively large perturbations. It thus appears that, in spite of its larger $\sigma_1$, it is PBA1 that has the larger basin of attraction.


\section{Concluding remarks} \label{sec:conclude}

\subsection{Summary} \label{ssec:summary}

As stated in the Introduction, the theory of nonautonomous dynamical systems (NDSs) and of their pullback attractors (PBAs) provides key concepts and tools to study the dynamics of systems with time-dependent forcing and coefficients. In recent years, this theory has provided several important advances in the climate sciences, as reviewed, for instance, by \citeA{Ghil.Luc.2020} and \citeA{Teletal2020}.

An interesting result of this line of work is that, in the presence of either periodic or chaotic forcing, multiple PBAs can coexist within a unique global attractor. In particular,  \citeA{Pierinietal2016, Pierinietal2018} have considered a severely truncated model of wind-driven, midlatitude ocean dynamics in which two PBAs coexist, one with rather quiescent, small-amplitude behavior and the other with much more energetic and irregular behavior for the same prescribed time-dependent forcing. 

In the present paper, we have studied a considerably more detailed climate model, due to \citeA{Vannitsemetal2015}; it was dubbed herein the VDDG model and summarized in Sec.~\ref{sec:equate}. This coupled ocean--atmosphere model represents extratropical basin-wide dynamics and it comprises two reduced-order modules, atmospheric and oceanic. \citeA{Vannitsemetal2015} found in their model low-frequency variability (LFV) that develops coherently in both modules, while \citeA{VannitsemGhil2017} verified that its multidecadal coupled mode shares common features with reanalysis data sets. The VDDG model was forced herein by an ENSO model based on the work of \citeA{Timmermannetal2003}.

For both periodic and chaotic ENSO forcing, the extratropical model displays multiple PBAs for a substantial range of the scaling parameter $g$. This result confirms that multiplicity of PBAs is not an artefact of low-order dynamics. For the two PBAs coexisting here, the LFV behavior can be drastically different: PBA2 features a strong multidecadal signal, intrinsic to the VDDG model, while for PBA1 a much weaker signal is dominated by the recurrence of warm El Niño and cold La Niña events; see Figs.~\ref{trajectories} and \ref{trajcha} and Secs.~\ref{ssec:PBAsper}, and \ref{ssec:chaos}, respectively. 

This difference in the nature of the two PBAs' variability is accompanied by drastically different predictability properties, as shown by the respective leading Lyapunov exponents $\sigma_1$, with $\sigma_1^{+}$ for PBA1 and $\sigma_1^{-}$ for PBA2; roughly speaking, $\sigma_1^{+} \simeq 2 \times \sigma_1^{-}$, as seen in Figs.~\ref{trajectories}d and \ref{trajcha}d. This finding is consistent with the fact that, in the unforced VDDG model, predictability was higher when multidecadal LFV was present \cite{Vannitsemetal2015}. When the midlatitude model is forced by the chaotic ENSO solution, an even richer dynamics is found for PBA2, characterized by LFV on centennial time scales: as seen in Fig.~\ref{histo-low}c, phase locking of all the trajectories along one of three paths in phase space, or strands, occurs in this case, but not for PBA1.

In the present work, we have also explored the stability of these attractors by perturbing the initial states of the extratropical model and following the trajectories starting from the perturbed states. In the case of periodic ENSO forcing, the extratropical model looks quite robust even when the initial perturbations are of the same order of magnitude as the oceanic streamfunction field's variability; see Figs. \ref{balllyap}, \ref{ballmax} and \ref{chaosball1}. 

When the system is forced by the chaotic ENSO signal, though, PBA1 is much more robust than PBA2. The latter displays trajectories that can easily escape to the other attractor once perturbed, even for perturbations that are quite small in amplitude; see Figs.~\ref{chaosball1}(a--c). Thus PBA2 could be qualified as metastable. For larger initial-perturbation amplitudes, the perturbed solutions can visit one or the other attractor, as in Fig.~\ref{chaosball1}, suggesting that holes  or {\em melancholia states} \cite{Luc.Bodai.2017, Ghil.Luc.2020} may develop in the basin boundaries of the local PBAs. 

An additional matter of both theoretical and practical interest in the present results is the complex interplay between time-dependent forcing and intrinsic climate variability in the spectral domain. Visual examination of Figs.~\ref{Histo-PBA2}(c,d), \ref{histo-high} and \ref{histo-low} already suggested a complex interaction between the ENSO forcing and the VDDG model's intrinsic variability. In Sec.~\ref{ssec:chaos}, we took a closer look at the spectral characteristics of the chaotic ENSO forcing, on the one hand, and on the VDDG's model's behavior with and without chaotic ENSO, on the other. Table~\ref{tab:MSSA}, in particular, confirmed the results of \citeA{Vannitsemetal2015} as to the VDDG model's dominant intrinsic periodicity of $P_{\rm V} \simeq 58$~yr and pointed to an important subharmonic thereof, at $2 \times P_{\rm V} \simeq 115$~yr.

In the presence of chaotic ENSO forcing, the VDDG atmospheric module does show the dominant periodicity of the forcing, at $P_{\rm E} \simeq 36$~yr, and several harmonics thereof. The latter are due to the highly anharmonic, spiky character of the ENSO  signal itself, cf. Fig.~\ref{fig:ENSO_force}(b),
as well as to the model's rather noisy atmosphere, cf.~\citeA{Vannitsemetal2015}.

The ENSO-forced VDDG model's ocean LFV, though, is still dominated by the intrinsic periodicity $P_{\rm V} \simeq 58$~yr, and by its subharmonic $2 \times P_{\rm V} \simeq 115$~yr. The latter is, at the same time, a subharmonic of the forcing frequency, $3 \times P_{\rm E} \simeq 115$~yr, and it is precisely this nonlinear resonance, $2 \times P_{\rm V} \simeq 3 \times P_{\rm E}$, that characterizes the interaction of the chaotic forcing with the model's chaotic LFV.

Such rational, as opposed to 1:1, resonances are well known in other subfields of the climate sciences. Thus the Great Inequality of Jupiter and Saturn --- i.e., the 2: 5 near-resonance between the revolution periods of Jupiter and Saturn around the Sun \cite{Wilson.1985} --- constitutes a major perturbation of the regular, quasi-periodic evolution of the solar system \cite{Varadi.ea.1999}. And the interaction of quasi-periodic, orbitally induced insolation forcing and of a still periodic, but nonlinear climate oscillator
can lead to multiple rational resonances and combination tones, as well as to chaotic Quaternary glaciation cycles \cite{Ghil.1994}. In the latter case, the intrinsic climate oscillations were due to coupling of an energy balance model with an ice mass balance model into a very low-order model \cite{Ghil.Chil.1987, LeTreut.ea.1988, Ghil.1994}.

\subsection{Discussion} \label{ssec:discuss}

In the present analysis, we have focused merely on two different types of ENSO forcing, periodic and chaotic, and on two specific air-sea coupling parameter values that lead to the presence of multiple PBAs. At this stage, we have thus only lifted the veil on a small number of possible solutions and types of behavior in the current model. The striking results, though, warrant further exploration. Proceeding more systematically, even for the current model version, requires intensive computer calculations as the number of trajectories needed is large and the necessary run times are quite long.

This being said, the possibility of alternating between two PBAs with very different behavior and predictability — even in the presence of purely deterministic forcing, whether periodic or chaotic --- has profound implications for interannual climate predictions and longer-term climate projections \cite<e.g.,>[]{IPCC.2013}. \citeA{Lorenz.1990} already formulated a low-order, midlatitude  atmospheric model with seasonal forcing, in which the bistability of a perpetual-summer climate, combined with the chaotic character of the unique perpetual-winter climate, led to low interannual predictability. In fact, the \citeA{Lorenz.1990} model's two summer climates, which were both stable at constant forcing, were characterized, respectively, by a strong vs. a weak oscillation of the model's westerly flow, somewhat like our PBA2 and PBA1 in the case of periodic ENSO forcing.

The present results thus extend those of  \citeA{Lorenz.1990} in two ways: (i) to longer, multidecadal time scales; and (ii) to a model that couples the atmosphere and ocean, as well as the Tropics and extratropics. These results confirm, therewith, the limitations on predictability due to the interactions of time-dependent forcing and intrinsic climate variability for a model that sits on a substantially higher rung of the climate modeling hierarchy; for the importance and usefulness of such a hierarchy, see, for instance, \citeA{Ghil.2001}, \citeA{Held.2005} and \citeA{Ghil.Luc.2020}, and references therein.

It thus appears fairly likely that the climate system could possess --- on the time scales of interest herein --- multiple PBAs that are affected by chaotic forcing. In the kind of situation depicted in Fig.~\ref{chaosball1}, individual members of a typical prediction ensemble will persist in a single local PBA or visit the two (or more) coexisting PBAs; hence, the first and second moments of the distribution generated by the ensemble are of very limited use for actual prediction. Since switches between PBAs may be highly intermittent, with very long intervals of dwelling in either one of them, more reliable ensemble predictions would require both much longer runs and many more of them. 

This state of affairs also implies that analyzing the dynamics of a more detailed and, presumably, more realistic climate model requires a very large ensemble starting from initial states in the distant past. Achieving this computational feat would allow for as long a spin-up interval as necessary to reach the model's global attractor, as well as for a very long time to analyze possibly multiple local PBAs, as done herein. Such an achievement would also shed further light on the so-called signal-to-noise paradox in IPCC-class climate models. 

For quite a while, there has been a suspicion that the LFV signal in such models is much weaker than observed \cite{Kravtsov.ea.2018, Smithetal2020}. This underestimation of interannual-to-multidecadal LFV  is usually attributed to omission of mechanisms and errors in the parameter values used in model development. \citeA{Oreillyetal2018}, though, have raised the possibility that the choice of stratospheric initial states might play a role in the limited amplitude of a detailed climate model's North Atlantic Oscillation. The results herein indicate that the limited, and possibly quite suboptimal, choice of initial states might play a key role in missing much of IPCC-class models' range of long-term behavior in general, and of vigorous LFV in particular.

The  spectacular channeling of highly irregular trajectories into distinct strands in phase space, as found in panels (c) and (d) of Figs.~\ref{histo-low} and \ref{histo-low-trsh}, illustrates the complex dynamics that the interaction of chaotic forcing with intrinsic LFV can give rise to, namely the phase locking by the former of the latter's set of trajectories along very specific paths. Other forms of striking response of chaotic variability to external forcing had been reported already in more highly idealized settings \cite{Checkrounetal2011, Checkrounetal2018, Pierinietal2016, Pierinietal2018, Pierini2020}. As already mentioned in a previous paragraph of this section, the findings herein raise the likelihood of such interesting and challenging behavior occurring in the climate system itself and require, therefore, further confirmation by the study of increasingly detailed and sophisticated models.

The Modular Arbitrary-Order Ocean-Atmosphere Model \cite<MAOOAM:>[]{DeCruzetal2016} is well adapted to studying the above-mentioned striking results across a hierarchy of increasingly well-resolved models, given its modular structure that facilitates the use of an arbitrary number of basis functions. Of course, this does not automatically include additional physical or chemical mechanisms that are present in high-end models. Still, performing multiannual climate predictions and multidecadal climate projections in the presence of large uncertainties in the external forcing, on the one side, and the limited amount of computer resources to run comprehensive climate models, on the other, pose great challenges. These challenges compel us to explore the wealth of nonlinear climate behavior in the presence of time-dependent forcing, natural and anthropogenic, across the intermediate rungs that MAOOAM can occupy within a full hierarchy of climate models \cite{Ghil.2001, Ghil.2019, Held.2005, Ghil.Luc.2020}.

\begin{acknowledgments}
The work of S. Vannitsem is partially supported by the Belgian Federal Science Policy Office under contract B2/20E/P1/ROADMAP, financed in the context of the European JPI-Climate/JPI-Oceans initiative. 
The present paper is TiPES contribution \# 74; this project has received funding from the EU’s Horizon 2020 research and innovation programme under grant agreement No. 820970, and it helps support the work of M. Ghil. Work on this paper has also been supported by the EIT Climate-KIC (grant no. 190733); EIT Climate-KIC is supported by the European Institute of Innovation \& Technology (EIT), a body of the European Union.
\end{acknowledgments}


\appendix 

\section{NDSs and PBAs} \label{app:pba}

The brief presentation here follows \citeA{Caraballo.Han.2017}. One must first distinguish between autonomous and nonautonomous systems of ordinary differential equations (ODEs), as the paradigmatic examples of models that generate the two corresponding types of dynamical systems. We have, respectively, the initial-value problems
\begin{subequations} \label{eq:DS}
\begin{align}
\frac{dx}{dt} & = f(x), \quad x(t_0) = x_0, \;\; {\mathrm{and}} \label{DDS} \\
\frac{dx}{dt} & = g(t, x), \quad x(t_0) = x_0; \label{NDS} 
\end{align}  
\end{subequations}
here $t \in {\mathbb R}$, $x \in {\mathbb R}^d$, $f: {\mathbb R}^d \to {\mathbb R}^d$ in Eq.~\eqref{DDS}, while $g: {\mathbb R} \times {\mathbb R}^d \to {\mathbb R}^d$ in Eq.~\eqref{NDS}. 

One assumes that $f$ and $g$ have ``nice'' properties that guarantee the existence, uniqueness and continuous dependence on initial states and on parameters for the solutions of Eqs.~\eqref{DDS} and \eqref{NDS}, respectively. Furthermore, \citeA{Caraballo.Han.2017} show that, provided the vector field $g(t,x)$ is dissipative, solutions of Eq.~\eqref{NDS} exist and satisfy the two other properties globally, i.e., for all $t \in \mathbb{R}$. We call such a global solution $\varphi(t, t_0, x_0)$.

There are two key distinctions between the two cases:
\begin{enumerate}[label=(\alph*)]
\item In the autonomous setting, solutions cannot intersect, since there is only one trajectory through a given point $x_0 \in {\mathbb R}^d$, due to uniqueness. Hence, for $d = 2$, the only possible (forward) attracting sets are fixed points and limit cycles, i.e., chaotic behavior and strange attractors can only occur for $d \ge 3$. The NDS setting is different in these respects, i.e., intersections are possible at two times $t_1$ and $t_2 \neq t_1$, and thus chaos can occur for $d = 2$ and periodic forcing, as is the case, for instance, in the Van der Pol oscillator \cite<e.g.,>[]{Guck.Holm.1983}.
\item In the autonomous setting, solutions depend only on the time $t - t_0$ elapsed since initial time, while in the NDS setting, they depend separately on the initial time $t_0$ and the current time $t$, at which we observe the system. In the former setting, it suffices to consider forward-in-time attraction, which results in attractors that are fixed, time-independent objects, such as  fixed points, limit cycles, tori and  strange attractors, In the latter case, we need to define pullback attraction and the PBAs that it leads to.
\end{enumerate}

Given the uniqueness and the continuous dependence of the global solutions to Eq.~\eqref{NDS} on initial states and on parameters, it is straightforward to verify that a global solution $\varphi$ of \eqref{NDS} satisﬁes:
\begin{enumerate}[label=(\roman*)]
\item the {\em initial value property} at $t=t_0$, namely $\varphi(t_0, t_0, x_0) = x_0$; and 

\item the {\em two-parameter semigroup} evolution property,
$$ \varphi(t_2, t_0, x_0) = \varphi(t_2, t_1, \varphi(t_1, t_0, x_0)) \quad {\mathrm{for}} \quad t_0 \le t \le t_2,$$
which coresponds to the concatenation of solutions; i.e., to go from $t_0$ to $t_2$ one can go first  from $t_0$ to $t_1$ and then from $t_1$ to $t_2$.
\end{enumerate}

One can then provide the following definition of a {\em process}.\\
\noindent {\bf Definition A.1}. Let $\mathbb{R}_{\ge}^2 = \{(t,t_0) \in \mathbb{R}^2 : t \ge t_0\}$. A process on $\mathbb{R}^d$ is a family of mappings
$$\varphi(t, t_0, \cdot)) : \mathbb{R}^d \to \mathbb{R}^d, \quad (t, t_0) \in \mathbb{R}_{\ge}^2,$$
which satisfy 
\begin{enumerate}[label=(\roman*)]
\item the initial value property $\varphi(t_0, t_0, x) = x$ for all $x \in \mathbb{R}^d$ and any $t_0 \in \mathbb{R}$;

\item the two-parameter semigroup property for all $x \in \mathbb{R}^d$ and both $(t_2, t_1) \in \mathbb{R}_{\ge}^2$ and $(t_1, t_0) \in \mathbb{R}_{\ge}^2$; and

\item the continuity property that the mapping $(t, t_0, x) \mapsto \varphi(t, t_0, x)$ be continuous on $\mathbb{R}_{\ge}^2 \times \mathbb{R}^d$.
\end{enumerate}
This is the so-called process formulation of an NDS. An alternative formulation is the so-called  {\em skew-product} formulation, which goes back to the work of G.~R. Sell, as reviewed in \citeA{Sell.1971}. A process as defined above is also called a two-parameter semigroup on $\mathbb{R}^d$, in contrast with the one-parameter semigroup of an autonomous dynamical systems, since the former depends not just on the initial time $t_0$, as in the latter case, but also on the current time $t$.

This difference matters, in particular, in determining the asymptotic behavior of the solutions. In the autonomous case, a global solution is invariant with respect to translation, $\varphi(t, t_0, x_0) = \varphi(t - t_0, 0, x_0)$. Hence, the usual forward asymptotic behavior for $t \to + \infty$ and $t_0$ fixed is the same as the behavior for $t$ fixed and $t_0 \to - \infty$. This equivalence may no longer hold when the translation invariance is lost, in the NDS case. 

A simple case in which analytical computations can be carried out explicitly is given in \citeA[Sec.~3.2.1]{Caraballo.Han.2017}, namely
\begin{equation} \label{eq:sin}
\frac{dx}{dt} = - ax + b\sin t, \quad x(t_0) = x_0, \quad t \ge t_0.
\end{equation}
Individual solutions do not have a forward limit as $t \to + \infty$ for $t_0$ fixed, but the difference between any two solutions vanishes in this limit. The particular solution
\begin{equation} \label{fwd:sin} 
A(t) = \frac{b(a \sin t - \cos t)}{a^2 + 1}
\end{equation}
provides the long-term information on the behavior of all the solutions of Eq.~\eqref{eq:sin}. This result is best captured by recognizing that the {\em pullback} limit
\begin{equation} \label{pba:sin}
\lim \limits_{t_0 \to - \infty} \varphi(t, t_0, x_0) = A(t) \quad \mathrm{for ~all}\; t \;\mathrm{and}\; x_0 \in {\mathbb R}
\end{equation}
yields $A(t)$ as the {\em pullback attractor} of all the solutions of Eq.~\eqref{eq:sin}. 

One is thus led to the following rigorous definition of a PBA  for a forced dissipative dynamical system subject to a time-dependent forcing, where we have generalized $\mathbb{R}^d$ to a  finite-dimensional metric space $\mathcal{X}$ and replaced $t_0$ by $s$, for greater symmetry. 

\noindent {\bf Definition A.2}. A PBA is a collection $\bigcup_{t \in \Re} \mathcal{A}(t)$ of invariant sets that depend on time and satisfy the following conditions: 
\begin{enumerate}
\item For all $t$, $\mathcal{A}(t)$ is a compact subset in $\mathcal{X}$ that is invariant with respect to the two-parameter semi-group $\mathcal{F}(t,s)$,
\begin{equation}
\mathcal{F}(t,s) \mathcal{A}(s) = \mathcal{A}(t) \quad \mathrm{for~every} \; s \; \leq t ; \quad \mathrm{and}
\end{equation} 

\item for all $t$, pullback attraction is reached when 
\begin{equation}
\lim_{s \to -\infty} D_{\rm H}(\mathcal{F}(t,s) \mathcal{B},\mathcal{A}(t)) = 0 \,\,\, \rm{for} \, \rm{all} \, \mathcal{B} \in \mathcal{C} 
\end{equation}
where $D_{\rm H}(E,D)$ is the Hausdorff semi-distance between two sets, and $\mathcal{C}$ is a collection of bounded sets in $\mathcal{X}$. 
\end{enumerate} 

More general definitions for PBAs exist in infinite-dimensional spaces, like those required by the solutions of partial differential or delay-differential equations, but mathematical rigor in these cases requires more technical details \cite<e.g.,>[]{Carvalho.ea.2012, Kloed.Rasm.2011}.

The finite-dimensional definition above follows \citeA[Appendix~A and references therein]{Charo.ea.2021}. In fact, both deterministic and stochastic versions of forcing have been applied, for instance, by \citeA{Checkrounetal2018} in the study of an infinite-dimensional, delay-differential equation model of ENSO. The deterministic forcing corresponded to the purely periodic, seasonal changes in insolation, while the stochastic component represented the westerly windbursts appearing in other models by F.-F. Jin and A. Timmermann \cite{Timm.Jin.2002}; see also \citeA[Sec.~4.3]{Checkrounetal2011}.

In the present paper, we used the forced VDDG model's Lyapunov exponents as a key tool in the systematic investigation of its PBAs obtained herein. In the PBA framework, the Lyapunov exponents of a system subject to time-dependent forcing exist and are well defined \cite{Ruelle1984}, provided this forcing is ergodic and unique. Since our forcing histories originate from a nonlinear oscillator with an ergodic attractor, namely the ENSO module of Sec.~\ref{ssec:ENSO}, the ergodicity of the time-dependent forcing is ensured. 

As to the uniqueness, we are also assuming that there are no initial errors in the tropical module's trajectory, which is independent of the extratropical module. One can also think of the Lyapunov exponent calculation as being in a $36 + 3$--dimensional space, with the 3-dimensional ENSO component of the trajectory being exact and not subject to divergence or convergence.

\bibliography{Pullback-MG_v2}

\begin{thebibliography}{}

\bibitem [\protect \citeauthoryear {%
Alessio%
}{%
Alessio%
}{%
{\protect \APACyear {2015}}%
}]{%
Alessio.2015}
\APACinsertmetastar {%
Alessio.2015}%
\begin{APACrefauthors}%
Alessio, S\BPBI M.%
\end{APACrefauthors}%
\unskip\
\newblock
\APACrefYear{2015}.
\newblock
\APACrefbtitle {{Digital Signal Processing and Spectral Analysis for
  Scientists: Concepts and Applications}} {{Digital Signal Processing and
  Spectral Analysis for Scientists: Concepts and Applications}}.
\newblock
\APACaddressPublisher{}{Springer Science \& Business Media}.
\PrintBackRefs{\CurrentBib}

\bibitem [\protect \citeauthoryear {%
Alexander%
\ \protect \BOthers {.}}{%
Alexander%
\ \protect \BOthers {.}}{%
{\protect \APACyear {2002}}%
}]{%
Alexanderetal2002}
\APACinsertmetastar {%
Alexanderetal2002}%
\begin{APACrefauthors}%
Alexander, M\BPBI A.%
, Bladé, I.%
, Newman, M.%
, Lanzante, J\BPBI R.%
, Lau, N\BHBI C.%
\BCBL {}\ \BBA {} Scott, J\BPBI D.%
\end{APACrefauthors}%
\unskip\
\newblock
\APACrefYearMonthDay{2002}{}{}.
\newblock
{\BBOQ}\APACrefatitle {The atmospheric bridge: {The influence of ENSO}
  teleconnections on air–sea interaction over the global oceans} {The
  atmospheric bridge: {The influence of ENSO} teleconnections on air–sea
  interaction over the global oceans}.{\BBCQ}
\newblock
\APACjournalVolNumPages{{J. Climate}}{15}{}{2205-2231}.
\newblock
\begin{APACrefDOI} \doi{10.1175/1520-0442(2002)015<2205:TABTIO>2.0.CO;2}
  \end{APACrefDOI}
\PrintBackRefs{\CurrentBib}

\bibitem [\protect \citeauthoryear {%
An%
\ \BBA {} Jin%
}{%
An%
\ \BBA {} Jin%
}{%
{\protect \APACyear {2004}}%
}]{%
AnJin2004}
\APACinsertmetastar {%
AnJin2004}%
\begin{APACrefauthors}%
An, S\BHBI I.%
\BCBT {}\ \BBA {} Jin, F\BHBI F.%
\end{APACrefauthors}%
\unskip\
\newblock
\APACrefYearMonthDay{2004}{}{}.
\newblock
{\BBOQ}\APACrefatitle {Nonlinearity and asymmetry of {ENSO}} {Nonlinearity and
  asymmetry of {ENSO}}.{\BBCQ}
\newblock
\APACjournalVolNumPages{{J. Climate}}{17}{}{2399--2412}.
\PrintBackRefs{\CurrentBib}

\bibitem [\protect \citeauthoryear {%
Arnold%
}{%
Arnold%
}{%
{\protect \APACyear {1998}}%
}]{%
Arnold.1998}
\APACinsertmetastar {%
Arnold.1998}%
\begin{APACrefauthors}%
Arnold, L.%
\end{APACrefauthors}%
\unskip\
\newblock
\APACrefYear{1998}.
\newblock
\APACrefbtitle {{Random Dynamical Systems}} {{Random Dynamical Systems}}.
\newblock
\APACaddressPublisher{New York/Berlin}{Springer-Verlag}.
\PrintBackRefs{\CurrentBib}

\bibitem [\protect \citeauthoryear {%
Ashwin%
, Wieczorek%
, Vitolo%
\BCBL {}\ \BBA {} Cox%
}{%
Ashwin%
\ \protect \BOthers {.}}{%
{\protect \APACyear {2012}}%
}]{%
Ashwin.ea.2012}
\APACinsertmetastar {%
Ashwin.ea.2012}%
\begin{APACrefauthors}%
Ashwin, P.%
, Wieczorek, S.%
, Vitolo, R.%
\BCBL {}\ \BBA {} Cox, P.%
\end{APACrefauthors}%
\unskip\
\newblock
\APACrefYearMonthDay{2012}{}{}.
\newblock
{\BBOQ}\APACrefatitle {Tipping points in open systems: bifurcation,
  noise-induced and rate-dependent examples in the climate system} {Tipping
  points in open systems: bifurcation, noise-induced and rate-dependent
  examples in the climate system}.{\BBCQ}
\newblock
\APACjournalVolNumPages{Philosophical Transactions of the Royal Society A:
  Mathematical, Physical and Engineering Sciences}{370}{1962}{1166--1184}.
\PrintBackRefs{\CurrentBib}

\bibitem [\protect \citeauthoryear {%
Caraballo%
\ \BBA {} Han%
}{%
Caraballo%
\ \BBA {} Han%
}{%
{\protect \APACyear {2017}}%
}]{%
Caraballo.Han.2017}
\APACinsertmetastar {%
Caraballo.Han.2017}%
\begin{APACrefauthors}%
Caraballo, T.%
\BCBT {}\ \BBA {} Han, X.%
\end{APACrefauthors}%
\unskip\
\newblock
\APACrefYear{2017}.
\newblock
\APACrefbtitle {{Applied Nonautonomous and Random Dynamical Systems: Applied
  Dynamical Systems}} {{Applied Nonautonomous and Random Dynamical Systems:
  Applied Dynamical Systems}}.
\newblock
\APACaddressPublisher{}{Springer Science + Business Media}.
\PrintBackRefs{\CurrentBib}

\bibitem [\protect \citeauthoryear {%
Carrassi%
\ \protect \BOthers {.}}{%
Carrassi%
\ \protect \BOthers {.}}{%
{\protect \APACyear {2020}}%
}]{%
Carrassietal2020}
\APACinsertmetastar {%
Carrassietal2020}%
\begin{APACrefauthors}%
Carrassi, A.%
, Grudzien, C.%
, Bocquet, M.%
, Demaeyer, J.%
, Raanes, P.%
\BCBL {}\ \BBA {} Vannitsem, S.%
\end{APACrefauthors}%
\unskip\
\newblock
\APACrefYearMonthDay{2020}{}{}.
\newblock
{\BBOQ}\APACrefatitle {Data assimilation for chaotic systems} {Data
  assimilation for chaotic systems}.{\BBCQ}
\newblock
\BIn{} S\BHBI K.~Park\ \BBA {} X.~Liang\ (\BEDS), \APACrefbtitle {{Data
  Assimilation for Atmospheric, Oceanic and Hydrological Applications}.} {{Data
  Assimilation for Atmospheric, Oceanic and Hydrological Applications}.}
\newblock
\APACaddressPublisher{}{Springer Science \& Business Media}.
\PrintBackRefs{\CurrentBib}

\bibitem [\protect \citeauthoryear {%
Carvalho%
, Langa%
\BCBL {}\ \BBA {} Robinson%
}{%
Carvalho%
\ \protect \BOthers {.}}{%
{\protect \APACyear {2012}}%
}]{%
Carvalho.ea.2012}
\APACinsertmetastar {%
Carvalho.ea.2012}%
\begin{APACrefauthors}%
Carvalho, A.%
, Langa, J\BPBI A.%
\BCBL {}\ \BBA {} Robinson, J.%
\end{APACrefauthors}%
\unskip\
\newblock
\APACrefYear{2012}.
\newblock
\APACrefbtitle {{Attractors for Infinite-Dimensional Non-Autonomous Dynamical
  Systems}} {{Attractors for Infinite-Dimensional Non-Autonomous Dynamical
  Systems}}.
\newblock
\APACaddressPublisher{}{Springer Science \& Business Media}.
\PrintBackRefs{\CurrentBib}

\bibitem [\protect \citeauthoryear {%
Char{\'o}%
, Chekroun%
, Sciamarella%
\BCBL {}\ \BBA {} Ghil%
}{%
Char{\'o}%
\ \protect \BOthers {.}}{%
{\protect \APACyear {2021}}%
}]{%
Charo.ea.2021}
\APACinsertmetastar {%
Charo.ea.2021}%
\begin{APACrefauthors}%
Char{\'o}, G\BPBI D.%
, Chekroun, M\BPBI D.%
, Sciamarella, D.%
\BCBL {}\ \BBA {} Ghil, M.%
\end{APACrefauthors}%
\unskip\
\newblock
\APACrefYearMonthDay{2021}{}{}.
\newblock
{\BBOQ}\APACrefatitle {Topological effects of noise on nonlinear dynamics}
  {Topological effects of noise on nonlinear dynamics}.{\BBCQ}
\newblock
\APACjournalVolNumPages{arXiv preprint arXiv:2010.09611v5 [nlin.CD]}{}{}{}.
\PrintBackRefs{\CurrentBib}

\bibitem [\protect \citeauthoryear {%
Checkroun%
, Ghil%
\BCBL {}\ \BBA {} Neelin%
}{%
Checkroun%
\ \protect \BOthers {.}}{%
{\protect \APACyear {2018}}%
}]{%
Checkrounetal2018}
\APACinsertmetastar {%
Checkrounetal2018}%
\begin{APACrefauthors}%
Checkroun, M\BPBI D.%
, Ghil, M.%
\BCBL {}\ \BBA {} Neelin, J\BPBI D.%
\end{APACrefauthors}%
\unskip\
\newblock
\APACrefYearMonthDay{2018}{}{}.
\newblock
{\BBOQ}\APACrefatitle {Pullback attractor crisis in a delay differential {ENSO}
  model} {Pullback attractor crisis in a delay differential {ENSO}
  model}.{\BBCQ}
\newblock
\BIn{} A\BPBI A.~Tsonis\ (\BED), \APACrefbtitle {{Advances in Nonlinear
  Geosciences}} {{Advances in Nonlinear Geosciences}}\ (\BPGS\ 1--33).
\newblock
\APACaddressPublisher{Cham, Switzerland}{Springer}.
\newblock
\begin{APACrefDOI} \doi{10.1007/978-3-319-58895-7_1} \end{APACrefDOI}
\PrintBackRefs{\CurrentBib}

\bibitem [\protect \citeauthoryear {%
Checkroun%
, Simmonet%
\BCBL {}\ \BBA {} Ghil%
}{%
Checkroun%
\ \protect \BOthers {.}}{%
{\protect \APACyear {2011}}%
}]{%
Checkrounetal2011}
\APACinsertmetastar {%
Checkrounetal2011}%
\begin{APACrefauthors}%
Checkroun, M\BPBI D.%
, Simmonet, E.%
\BCBL {}\ \BBA {} Ghil, M.%
\end{APACrefauthors}%
\unskip\
\newblock
\APACrefYearMonthDay{2011}{}{}.
\newblock
{\BBOQ}\APACrefatitle {Stochastic climate dynamics: Random attractors and
  time-dependent invariant measures} {Stochastic climate dynamics: Random
  attractors and time-dependent invariant measures}.{\BBCQ}
\newblock
\APACjournalVolNumPages{Physica D: Nonlinear Phenomena}{240}{}{1685-1700}.
\PrintBackRefs{\CurrentBib}

\bibitem [\protect \citeauthoryear {%
{De Cruz}%
, Demaeyer%
\BCBL {}\ \BBA {} Vannitsem%
}{%
{De Cruz}%
\ \protect \BOthers {.}}{%
{\protect \APACyear {2016}}%
}]{%
DeCruzetal2016}
\APACinsertmetastar {%
DeCruzetal2016}%
\begin{APACrefauthors}%
{De Cruz}, L.%
, Demaeyer, J.%
\BCBL {}\ \BBA {} Vannitsem, S.%
\end{APACrefauthors}%
\unskip\
\newblock
\APACrefYearMonthDay{2016}{}{}.
\newblock
{\BBOQ}\APACrefatitle {{The Modular Arbitrary-Order Ocean-Atmosphere Model:
  MAOOAM v1.0}} {{The Modular Arbitrary-Order Ocean-Atmosphere Model: MAOOAM
  v1.0}}.{\BBCQ}
\newblock
\APACjournalVolNumPages{Geosci. Mod. Dev.}{9}{8}{2793--2808}.
\PrintBackRefs{\CurrentBib}

\bibitem [\protect \citeauthoryear {%
{De Cruz}%
, Schubert%
, Demaeyer%
, Lucarini%
\BCBL {}\ \BBA {} Vannitsem%
}{%
{De Cruz}%
\ \protect \BOthers {.}}{%
{\protect \APACyear {2018}}%
}]{%
DeCruzetal2018}
\APACinsertmetastar {%
DeCruzetal2018}%
\begin{APACrefauthors}%
{De Cruz}, L.%
, Schubert, S.%
, Demaeyer, J.%
, Lucarini, V.%
\BCBL {}\ \BBA {} Vannitsem, S.%
\end{APACrefauthors}%
\unskip\
\newblock
\APACrefYearMonthDay{2018}{}{}.
\newblock
{\BBOQ}\APACrefatitle {{Exploring the Lyapunov instability properties of
  high-dimensional atmospheric and climate models}} {{Exploring the Lyapunov
  instability properties of high-dimensional atmospheric and climate
  models}}.{\BBCQ}
\newblock
\APACjournalVolNumPages{Nonlin. Proc. Geophys.}{25}{2}{387--412}.
\newblock
\begin{APACrefDOI} \doi{10.5194/npg-25-387-2018} \end{APACrefDOI}
\PrintBackRefs{\CurrentBib}

\bibitem [\protect \citeauthoryear {%
Demaeyer%
\ \BBA {} Vannitsem%
}{%
Demaeyer%
\ \BBA {} Vannitsem%
}{%
{\protect \APACyear {2017}}%
}]{%
Demaeyer2017}
\APACinsertmetastar {%
Demaeyer2017}%
\begin{APACrefauthors}%
Demaeyer, J.%
\BCBT {}\ \BBA {} Vannitsem, S.%
\end{APACrefauthors}%
\unskip\
\newblock
\APACrefYearMonthDay{2017}{}{}.
\newblock
{\BBOQ}\APACrefatitle {Stochastic parametrization of subgrid-scale processes in
  coupled ocean–atmosphere systems: benefits and limitations of response
  theory} {Stochastic parametrization of subgrid-scale processes in coupled
  ocean–atmosphere systems: benefits and limitations of response
  theory}.{\BBCQ}
\newblock
\APACjournalVolNumPages{Quarterly Journal of the Royal Meteorological
  Society}{143}{703}{881-896}.
\newblock
\begin{APACrefDOI} \doi{https://doi.org/10.1002/qj.2973} \end{APACrefDOI}
\PrintBackRefs{\CurrentBib}

\bibitem [\protect \citeauthoryear {%
Demaeyer%
\ \BBA {} Vannitsem%
}{%
Demaeyer%
\ \BBA {} Vannitsem%
}{%
{\protect \APACyear {2018}}%
}]{%
Demaeyer2018}
\APACinsertmetastar {%
Demaeyer2018}%
\begin{APACrefauthors}%
Demaeyer, J.%
\BCBT {}\ \BBA {} Vannitsem, S.%
\end{APACrefauthors}%
\unskip\
\newblock
\APACrefYearMonthDay{2018}{}{}.
\newblock
{\BBOQ}\APACrefatitle {Comparison of stochastic parameterizations in the
  framework of a coupled ocean-atmosphere model} {Comparison of stochastic
  parameterizations in the framework of a coupled ocean-atmosphere
  model}.{\BBCQ}
\newblock
\APACjournalVolNumPages{Nonl. Proc. Geophys.}{25}{3}{605-631}.
\newblock
\begin{APACrefDOI} \doi{10.5194/npg-25-605-2018} \end{APACrefDOI}
\PrintBackRefs{\CurrentBib}

\bibitem [\protect \citeauthoryear {%
Ditlevsen%
\ \BBA {} Ashwin%
}{%
Ditlevsen%
\ \BBA {} Ashwin%
}{%
{\protect \APACyear {2018}}%
}]{%
Ditlev.Ashwin.2018}
\APACinsertmetastar {%
Ditlev.Ashwin.2018}%
\begin{APACrefauthors}%
Ditlevsen, P\BPBI D.%
\BCBT {}\ \BBA {} Ashwin, P.%
\end{APACrefauthors}%
\unskip\
\newblock
\APACrefYearMonthDay{2018}{jun}{}.
\newblock
{\BBOQ}\APACrefatitle {Complex climate response to astronomical forcing: {The
  Middle-Pleistocene Transition} in glacial cycles and changes in frequency
  locking} {Complex climate response to astronomical forcing: {The
  Middle-Pleistocene Transition} in glacial cycles and changes in frequency
  locking}.{\BBCQ}
\newblock
\APACjournalVolNumPages{Frontiers in Physics}{6}{}{}.
\newblock
\begin{APACrefDOI} \doi{10.3389/fphy.2018.00062} \end{APACrefDOI}
\PrintBackRefs{\CurrentBib}

\bibitem [\protect \citeauthoryear {%
Dr\'otos%
, B\'odai%
\BCBL {}\ \BBA {} T\'el%
}{%
Dr\'otos%
\ \protect \BOthers {.}}{%
{\protect \APACyear {2015}}%
}]{%
Drotosetal2015}
\APACinsertmetastar {%
Drotosetal2015}%
\begin{APACrefauthors}%
Dr\'otos, G.%
, B\'odai, T.%
\BCBL {}\ \BBA {} T\'el, T.%
\end{APACrefauthors}%
\unskip\
\newblock
\APACrefYearMonthDay{2015}{}{}.
\newblock
{\BBOQ}\APACrefatitle {Probabilistic Concepts in a Changing Climate: A Snapshot
  Attractor Picture} {Probabilistic concepts in a changing climate: A snapshot
  attractor picture}.{\BBCQ}
\newblock
\APACjournalVolNumPages{J. Climate}{28}{}{3275--3288}.
\PrintBackRefs{\CurrentBib}

\bibitem [\protect \citeauthoryear {%
Dr\'otos%
, B\'odai%
\BCBL {}\ \BBA {} T\'el%
}{%
Dr\'otos%
\ \protect \BOthers {.}}{%
{\protect \APACyear {2016}}%
}]{%
Drotosetal2016}
\APACinsertmetastar {%
Drotosetal2016}%
\begin{APACrefauthors}%
Dr\'otos, G.%
, B\'odai, T.%
\BCBL {}\ \BBA {} T\'el, T.%
\end{APACrefauthors}%
\unskip\
\newblock
\APACrefYearMonthDay{2016}{}{}.
\newblock
{\BBOQ}\APACrefatitle {Quantifying nonergodicity in nonautonomous dissipative
  dynamical systems: An application to climate change} {Quantifying
  nonergodicity in nonautonomous dissipative dynamical systems: An application
  to climate change}.{\BBCQ}
\newblock
\APACjournalVolNumPages{Phys. Rev. E}{94}{}{022214}.
\newblock
\begin{APACrefDOI} \doi{10.1103/PhysRevE.94.022214} \end{APACrefDOI}
\PrintBackRefs{\CurrentBib}

\bibitem [\protect \citeauthoryear {%
Ghil%
}{%
Ghil%
}{%
{\protect \APACyear {1994}}%
}]{%
Ghil.1994}
\APACinsertmetastar {%
Ghil.1994}%
\begin{APACrefauthors}%
Ghil, M.%
\end{APACrefauthors}%
\unskip\
\newblock
\APACrefYearMonthDay{1994}{}{}.
\newblock
{\BBOQ}\APACrefatitle {Cryothermodynamics: the chaotic dynamics of
  paleoclimate} {Cryothermodynamics: the chaotic dynamics of
  paleoclimate}.{\BBCQ}
\newblock
\APACjournalVolNumPages{{Physica D}}{77}{}{130--159}.
\PrintBackRefs{\CurrentBib}

\bibitem [\protect \citeauthoryear {%
Ghil%
}{%
Ghil%
}{%
{\protect \APACyear {2001}}%
}]{%
Ghil.2001}
\APACinsertmetastar {%
Ghil.2001}%
\begin{APACrefauthors}%
Ghil, M.%
\end{APACrefauthors}%
\unskip\
\newblock
\APACrefYearMonthDay{2001}{}{}.
\newblock
{\BBOQ}\APACrefatitle {Hilbert problems for the geosciences in the 21st
  century} {Hilbert problems for the geosciences in the 21st century}.{\BBCQ}
\newblock
\APACjournalVolNumPages{Nonlinear Processes in Geophysics}{8}{4/5}{211--211}.
\newblock
\begin{APACrefDOI} \doi{10.5194/npg-8-211-2001} \end{APACrefDOI}
\PrintBackRefs{\CurrentBib}

\bibitem [\protect \citeauthoryear {%
Ghil%
}{%
Ghil%
}{%
{\protect \APACyear {2019}}%
}]{%
Ghil.2019}
\APACinsertmetastar {%
Ghil.2019}%
\begin{APACrefauthors}%
Ghil, M.%
\end{APACrefauthors}%
\unskip\
\newblock
\APACrefYearMonthDay{2019}{}{}.
\newblock
{\BBOQ}\APACrefatitle {A century of nonlinearity in the geosciences} {A century
  of nonlinearity in the geosciences}.{\BBCQ}
\newblock
\APACjournalVolNumPages{Earth and Space Science}{6}{}{1007--1042}.
\newblock
\begin{APACrefDOI} \doi{10.1029/2019EA000599} \end{APACrefDOI}
\PrintBackRefs{\CurrentBib}

\bibitem [\protect \citeauthoryear {%
Ghil%
\ \protect \BOthers {.}}{%
Ghil%
\ \protect \BOthers {.}}{%
{\protect \APACyear {2002}}%
}]{%
Ghil.SSA.2002}
\APACinsertmetastar {%
Ghil.SSA.2002}%
\begin{APACrefauthors}%
Ghil, M.%
, Allen, M\BPBI R.%
, Dettinger, M\BPBI D.%
, Ide, K.%
, Kondrashov, D.%
, Mann, M\BPBI E.%
\BDBL {}Yiou, P.%
\end{APACrefauthors}%
\unskip\
\newblock
\APACrefYearMonthDay{2002}{}{}.
\newblock
{\BBOQ}\APACrefatitle {{Advanced spectral methods for climatic time series}}
  {{Advanced spectral methods for climatic time series}}.{\BBCQ}
\newblock
\APACjournalVolNumPages{Reviews of Geophysics}{40}{1}{{41 pages}}.
\newblock
\begin{APACrefDOI} \doi{10.1029/2000RG000092} \end{APACrefDOI}
\PrintBackRefs{\CurrentBib}

\bibitem [\protect \citeauthoryear {%
Ghil%
, Chekroun%
\BCBL {}\ \BBA {} Simonnet%
}{%
Ghil%
\ \protect \BOthers {.}}{%
{\protect \APACyear {2008}}%
}]{%
GCS.2008}
\APACinsertmetastar {%
GCS.2008}%
\begin{APACrefauthors}%
Ghil, M.%
, Chekroun, M\BPBI D.%
\BCBL {}\ \BBA {} Simonnet, E.%
\end{APACrefauthors}%
\unskip\
\newblock
\APACrefYearMonthDay{2008}{aug}{}.
\newblock
{\BBOQ}\APACrefatitle {Climate dynamics and fluid mechanics: Natural
  variability and related uncertainties} {Climate dynamics and fluid mechanics:
  Natural variability and related uncertainties}.{\BBCQ}
\newblock
\APACjournalVolNumPages{Physica D: Nonlinear
  Phenomena}{237}{14-17}{2111--2126}.
\newblock
\begin{APACrefDOI} \doi{10.1016/j.physd.2008.03.036} \end{APACrefDOI}
\PrintBackRefs{\CurrentBib}

\bibitem [\protect \citeauthoryear {%
Ghil%
\ \BBA {} Childress%
}{%
Ghil%
\ \BBA {} Childress%
}{%
{\protect \APACyear {1987}}%
}]{%
Ghil.Chil.1987}
\APACinsertmetastar {%
Ghil.Chil.1987}%
\begin{APACrefauthors}%
Ghil, M.%
\BCBT {}\ \BBA {} Childress, S.%
\end{APACrefauthors}%
\unskip\
\newblock
\APACrefYear{1987}.
\newblock
\APACrefbtitle {{Topics in Geophysical Fluid Dynamics: Atmospheric Dynamics,
  Dynamo Theory, and Climate Dynamics}} {{Topics in Geophysical Fluid Dynamics:
  Atmospheric Dynamics, Dynamo Theory, and Climate Dynamics}}.
\newblock
\APACaddressPublisher{Berlin/Heidelberg}{Springer Science+Business Media}.
\newblock
\APACrefnote{{Reissued in pdf, 2012.}}
\PrintBackRefs{\CurrentBib}

\bibitem [\protect \citeauthoryear {%
Ghil%
\ \BBA {} Lucarini%
}{%
Ghil%
\ \BBA {} Lucarini%
}{%
{\protect \APACyear {2020}}%
}]{%
Ghil.Luc.2020}
\APACinsertmetastar {%
Ghil.Luc.2020}%
\begin{APACrefauthors}%
Ghil, M.%
\BCBT {}\ \BBA {} Lucarini, V.%
\end{APACrefauthors}%
\unskip\
\newblock
\APACrefYearMonthDay{2020}{}{}.
\newblock
{\BBOQ}\APACrefatitle {The physics of climate variability and climate change}
  {The physics of climate variability and climate change}.{\BBCQ}
\newblock
\APACjournalVolNumPages{Reviews of Modern Physics}{92}{3}{{035002}}.
\newblock
\begin{APACrefDOI} \doi{10.1103/revmodphys.92.035002} \end{APACrefDOI}
\PrintBackRefs{\CurrentBib}

\bibitem [\protect \citeauthoryear {%
Ghil%
\ \BBA {} Vautard%
}{%
Ghil%
\ \BBA {} Vautard%
}{%
{\protect \APACyear {1991}}%
}]{%
Ghil.Vautard.1991}
\APACinsertmetastar {%
Ghil.Vautard.1991}%
\begin{APACrefauthors}%
Ghil, M.%
\BCBT {}\ \BBA {} Vautard, R.%
\end{APACrefauthors}%
\unskip\
\newblock
\APACrefYearMonthDay{1991}{}{}.
\newblock
{\BBOQ}\APACrefatitle {Interdecadal oscillations and the warming trend in
  global temperature time series} {Interdecadal oscillations and the warming
  trend in global temperature time series}.{\BBCQ}
\newblock
\APACjournalVolNumPages{Nature}{350}{}{324--327}.
\PrintBackRefs{\CurrentBib}

\bibitem [\protect \citeauthoryear {%
Gill%
}{%
Gill%
}{%
{\protect \APACyear {1982}}%
}]{%
Gill.1982}
\APACinsertmetastar {%
Gill.1982}%
\begin{APACrefauthors}%
Gill, A\BPBI E.%
\end{APACrefauthors}%
\unskip\
\newblock
\APACrefYear{1982}.
\newblock
\APACrefbtitle {{Atmosphere-Ocean Dynamics}} {{Atmosphere-Ocean Dynamics}}.
\newblock
\APACaddressPublisher{New York, U.S.A.}{Academic Press}.
\PrintBackRefs{\CurrentBib}

\bibitem [\protect \citeauthoryear {%
Guckenheimer%
\ \BBA {} Holmes%
}{%
Guckenheimer%
\ \BBA {} Holmes%
}{%
{\protect \APACyear {1983}}%
}]{%
Guck.Holm.1983}
\APACinsertmetastar {%
Guck.Holm.1983}%
\begin{APACrefauthors}%
Guckenheimer, J.%
\BCBT {}\ \BBA {} Holmes, P\BPBI J.%
\end{APACrefauthors}%
\unskip\
\newblock
\APACrefYear{1983}.
\newblock
\APACrefbtitle {{Nonlinear Oscillations, Dynamical Systems, and Bifurcations of
  Vector Fields}} {{Nonlinear Oscillations, Dynamical Systems, and Bifurcations
  of Vector Fields}}.
\newblock
\APACaddressPublisher{}{Springer Science \& Business Media}.
\PrintBackRefs{\CurrentBib}

\bibitem [\protect \citeauthoryear {%
Held%
}{%
Held%
}{%
{\protect \APACyear {2005}}%
}]{%
Held.2005}
\APACinsertmetastar {%
Held.2005}%
\begin{APACrefauthors}%
Held, I\BPBI M.%
\end{APACrefauthors}%
\unskip\
\newblock
\APACrefYearMonthDay{2005}{}{}.
\newblock
{\BBOQ}\APACrefatitle {The gap between simulation and understanding in climate
  modeling} {The gap between simulation and understanding in climate
  modeling}.{\BBCQ}
\newblock
\APACjournalVolNumPages{Bulletin of the American Meteorological
  Society}{86}{11}{1609--1614}.
\newblock
\begin{APACrefDOI} \doi{10.1175/bams-86-11-1609} \end{APACrefDOI}
\PrintBackRefs{\CurrentBib}

\bibitem [\protect \citeauthoryear {%
Hoerling%
\ \BBA {} Kumar%
}{%
Hoerling%
\ \BBA {} Kumar%
}{%
{\protect \APACyear {2002}}%
}]{%
HoerlingandKumar2002}
\APACinsertmetastar {%
HoerlingandKumar2002}%
\begin{APACrefauthors}%
Hoerling, M\BPBI P.%
\BCBT {}\ \BBA {} Kumar, A.%
\end{APACrefauthors}%
\unskip\
\newblock
\APACrefYearMonthDay{2002}{}{}.
\newblock
{\BBOQ}\APACrefatitle {Atmospheric Response Patterns Associated with Tropical
  Forcing} {Atmospheric response patterns associated with tropical
  forcing}.{\BBCQ}
\newblock
\APACjournalVolNumPages{J. Climate}{15}{}{2184-2203}.
\newblock
\begin{APACrefDOI} \doi{10.1175/1520-0442(2002)015<2184:ARPAWT>2.0.CO;2}
  \end{APACrefDOI}
\PrintBackRefs{\CurrentBib}

\bibitem [\protect \citeauthoryear {%
{IPCC}%
}{%
{IPCC}%
}{%
{\protect \APACyear {2014}}%
}]{%
IPCC.2013}
\APACinsertmetastar {%
IPCC.2013}%
\begin{APACrefauthors}%
{IPCC}.%
\end{APACrefauthors}%
\unskip\
\newblock
\APACrefYear{2014}.
\newblock
\APACrefbtitle {{Climate Change 2013: The Physical Science Basis. Contribution
  of Working Group I to the Fifth Assessment Report of the Intergovernmental
  Panel on Climate Change}} {{Climate Change 2013: The Physical Science Basis.
  Contribution of Working Group I to the Fifth Assessment Report of the
  Intergovernmental Panel on Climate Change}}\ ({T. Stocker et al.}, \BED{}).
\newblock
\APACaddressPublisher{Cambridge, UK}{Cambridge University Press}.
\newblock
\begin{APACrefDOI} \doi{10.1017/cbo9781107415324} \end{APACrefDOI}
\PrintBackRefs{\CurrentBib}

\bibitem [\protect \citeauthoryear {%
Jin%
}{%
Jin%
}{%
{\protect \APACyear {1996}}%
}]{%
Jin1996}
\APACinsertmetastar {%
Jin1996}%
\begin{APACrefauthors}%
Jin, F\BHBI F.%
\end{APACrefauthors}%
\unskip\
\newblock
\APACrefYearMonthDay{1996}{}{}.
\newblock
{\BBOQ}\APACrefatitle {Tropical ocean-atmosphere interaction, the {Pacific cold
  tongue, and the El-Ni\~no-Southern Oscillation}} {Tropical ocean-atmosphere
  interaction, the {Pacific cold tongue, and the El-Ni\~no-Southern
  Oscillation}}.{\BBCQ}
\newblock
\APACjournalVolNumPages{Science}{274}{}{76-78}.
\PrintBackRefs{\CurrentBib}

\bibitem [\protect \citeauthoryear {%
Jin%
}{%
Jin%
}{%
{\protect \APACyear {1997}}%
}]{%
Jin1997}
\APACinsertmetastar {%
Jin1997}%
\begin{APACrefauthors}%
Jin, F\BHBI F.%
\end{APACrefauthors}%
\unskip\
\newblock
\APACrefYearMonthDay{1997}{}{}.
\newblock
{\BBOQ}\APACrefatitle {An equatorial ocean recharge paradigm for {ENSO. Part I:
  Conceptual model}} {An equatorial ocean recharge paradigm for {ENSO. Part I:
  Conceptual model}}.{\BBCQ}
\newblock
\APACjournalVolNumPages{{J. Atmos. Sci.}}{54}{}{811-829}.
\PrintBackRefs{\CurrentBib}

\bibitem [\protect \citeauthoryear {%
Jin%
, Neelin%
\BCBL {}\ \BBA {} Ghil%
}{%
Jin%
\ \protect \BOthers {.}}{%
{\protect \APACyear {1994}}%
}]{%
JNG.1994}
\APACinsertmetastar {%
JNG.1994}%
\begin{APACrefauthors}%
Jin, F\BHBI F.%
, Neelin, J\BPBI D.%
\BCBL {}\ \BBA {} Ghil, M.%
\end{APACrefauthors}%
\unskip\
\newblock
\APACrefYearMonthDay{1994}{}{}.
\newblock
{\BBOQ}\APACrefatitle {{El Ni\~no on the devil's staircase: Annual subharmonic
  steps to chaos}} {{El Ni\~no on the devil's staircase: Annual subharmonic
  steps to chaos}}.{\BBCQ}
\newblock
\APACjournalVolNumPages{{Science}}{264}{}{70--72}.
\PrintBackRefs{\CurrentBib}

\bibitem [\protect \citeauthoryear {%
Jin%
, Neelin%
\BCBL {}\ \BBA {} Ghil%
}{%
Jin%
\ \protect \BOthers {.}}{%
{\protect \APACyear {1996}}%
}]{%
JNG.1996}
\APACinsertmetastar {%
JNG.1996}%
\begin{APACrefauthors}%
Jin, F\BHBI F.%
, Neelin, J\BPBI D.%
\BCBL {}\ \BBA {} Ghil, M.%
\end{APACrefauthors}%
\unskip\
\newblock
\APACrefYearMonthDay{1996}{}{}.
\newblock
{\BBOQ}\APACrefatitle {{El Ni\~no/Southern Oscillation and the annual cycle:
  Subharmonic frequency-locking and aperiodicity}} {{El Ni\~no/Southern
  Oscillation and the annual cycle: Subharmonic frequency-locking and
  aperiodicity}}.{\BBCQ}
\newblock
\APACjournalVolNumPages{Physica D: Nonlinear Phenomena}{98}{}{442--465}.
\PrintBackRefs{\CurrentBib}

\bibitem [\protect \citeauthoryear {%
Kimoto%
\ \BBA {} Ghil%
}{%
Kimoto%
\ \BBA {} Ghil%
}{%
{\protect \APACyear {1993}}%
}]{%
Kimoto.Ghil.1993}
\APACinsertmetastar {%
Kimoto.Ghil.1993}%
\begin{APACrefauthors}%
Kimoto, M.%
\BCBT {}\ \BBA {} Ghil, M.%
\end{APACrefauthors}%
\unskip\
\newblock
\APACrefYearMonthDay{1993}{}{}.
\newblock
{\BBOQ}\APACrefatitle {{Multiple flow regimes in the Northern Hemisphere
  winter. Part II: Sectorial regimes and preferred transitions}} {{Multiple
  flow regimes in the Northern Hemisphere winter. Part II: Sectorial regimes
  and preferred transitions}}.{\BBCQ}
\newblock
\APACjournalVolNumPages{Journal of the Atmospheric Sciences}{50}{}{2645--2673}.
\PrintBackRefs{\CurrentBib}

\bibitem [\protect \citeauthoryear {%
Kloeden%
\ \BBA {} Rasmussen%
}{%
Kloeden%
\ \BBA {} Rasmussen%
}{%
{\protect \APACyear {2011}}%
}]{%
Kloed.Rasm.2011}
\APACinsertmetastar {%
Kloed.Rasm.2011}%
\begin{APACrefauthors}%
Kloeden, P\BPBI E.%
\BCBT {}\ \BBA {} Rasmussen, M.%
\end{APACrefauthors}%
\unskip\
\newblock
\APACrefYear{2011}.
\newblock
\APACrefbtitle {{Nonautonomous Dynamical Systems}} {{Nonautonomous Dynamical
  Systems}}.
\newblock
\APACaddressPublisher{}{American Mathematical Society}.
\PrintBackRefs{\CurrentBib}

\bibitem [\protect \citeauthoryear {%
Kravtsov%
, Grimm%
\BCBL {}\ \BBA {} Gu%
}{%
Kravtsov%
\ \protect \BOthers {.}}{%
{\protect \APACyear {2018}}%
}]{%
Kravtsov.ea.2018}
\APACinsertmetastar {%
Kravtsov.ea.2018}%
\begin{APACrefauthors}%
Kravtsov, S.%
, Grimm, C.%
\BCBL {}\ \BBA {} Gu, S.%
\end{APACrefauthors}%
\unskip\
\newblock
\APACrefYearMonthDay{2018}{}{}.
\newblock
{\BBOQ}\APACrefatitle {Global-scale multidecadal variability missing in
  state-of-the-art climate models} {Global-scale multidecadal variability
  missing in state-of-the-art climate models}.{\BBCQ}
\newblock
\APACjournalVolNumPages{npj Climate and Atmospheric Science}{1}{1}{34}.
\newblock
\begin{APACrefDOI} \doi{10.1038/s41612-018-0044-6} \end{APACrefDOI}
\PrintBackRefs{\CurrentBib}

\bibitem [\protect \citeauthoryear {%
Kumar%
\ \BBA {} Hoerling%
}{%
Kumar%
\ \BBA {} Hoerling%
}{%
{\protect \APACyear {1995}}%
}]{%
KumarandHoerling1995}
\APACinsertmetastar {%
KumarandHoerling1995}%
\begin{APACrefauthors}%
Kumar, A.%
\BCBT {}\ \BBA {} Hoerling, M\BPBI P.%
\end{APACrefauthors}%
\unskip\
\newblock
\APACrefYearMonthDay{1995}{}{}.
\newblock
{\BBOQ}\APACrefatitle {Prospects and Limitations of Seasonal Atmospheric GCM
  Predictions} {Prospects and limitations of seasonal atmospheric gcm
  predictions}.{\BBCQ}
\newblock
\APACjournalVolNumPages{Bull. Amer. Meteor. Soc.}{76}{}{335--345}.
\newblock
\begin{APACrefDOI} \doi{10.1175/1520-0477(1995)076<0335:PALOSA>2.0.CO;2}
  \end{APACrefDOI}
\PrintBackRefs{\CurrentBib}

\bibitem [\protect \citeauthoryear {%
Kuptsov%
\ \BBA {} Parlitz%
}{%
Kuptsov%
\ \BBA {} Parlitz%
}{%
{\protect \APACyear {2012}}%
}]{%
Kuptsov.P.2012}
\APACinsertmetastar {%
Kuptsov.P.2012}%
\begin{APACrefauthors}%
Kuptsov, P\BPBI V.%
\BCBT {}\ \BBA {} Parlitz, U.%
\end{APACrefauthors}%
\unskip\
\newblock
\APACrefYearMonthDay{2012}{}{}.
\newblock
{\BBOQ}\APACrefatitle {Theory and computation of covariant {Lyapunov vectors}}
  {Theory and computation of covariant {Lyapunov vectors}}.{\BBCQ}
\newblock
\APACjournalVolNumPages{Journal of Nonlinear Science}{22}{5}{727--762}.
\PrintBackRefs{\CurrentBib}

\bibitem [\protect \citeauthoryear {%
Legras%
\ \BBA {} Ghil%
}{%
Legras%
\ \BBA {} Ghil%
}{%
{\protect \APACyear {1985}}%
}]{%
Legras.Ghil.1985}
\APACinsertmetastar {%
Legras.Ghil.1985}%
\begin{APACrefauthors}%
Legras, B.%
\BCBT {}\ \BBA {} Ghil, M.%
\end{APACrefauthors}%
\unskip\
\newblock
\APACrefYearMonthDay{1985}{}{}.
\newblock
{\BBOQ}\APACrefatitle {Persistent anomalies, blocking, and variations in
  atmospheric predictability} {Persistent anomalies, blocking, and variations
  in atmospheric predictability}.{\BBCQ}
\newblock
\APACjournalVolNumPages{Journal of the Atmospheric Sciences}{42}{}{433--471}.
\PrintBackRefs{\CurrentBib}

\bibitem [\protect \citeauthoryear {%
Le~Treut%
, Portes%
, Jouzel%
\BCBL {}\ \BBA {} Ghil%
}{%
Le~Treut%
\ \protect \BOthers {.}}{%
{\protect \APACyear {1988}}%
}]{%
LeTreut.ea.1988}
\APACinsertmetastar {%
LeTreut.ea.1988}%
\begin{APACrefauthors}%
Le~Treut, H.%
, Portes, J.%
, Jouzel, J.%
\BCBL {}\ \BBA {} Ghil, M.%
\end{APACrefauthors}%
\unskip\
\newblock
\APACrefYearMonthDay{1988}{}{}.
\newblock
{\BBOQ}\APACrefatitle {{Isotopic modeling of climatic oscillations:
  implications for a comparative study of marine and ice-core records}}
  {{Isotopic modeling of climatic oscillations: implications for a comparative
  study of marine and ice-core records}}.{\BBCQ}
\newblock
\APACjournalVolNumPages{J. Geophys. Res.}{93}{}{9365--9383}.
\PrintBackRefs{\CurrentBib}

\bibitem [\protect \citeauthoryear {%
L\'opez-Parages%
, Rodríguez-Fonseca%
, Dommenget%
\BCBL {}\ \BBA {} Frauen%
}{%
L\'opez-Parages%
\ \protect \BOthers {.}}{%
{\protect \APACyear {2016}}%
}]{%
LopezParages2016}
\APACinsertmetastar {%
LopezParages2016}%
\begin{APACrefauthors}%
L\'opez-Parages, J.%
, Rodríguez-Fonseca, B.%
, Dommenget, D.%
\BCBL {}\ \BBA {} Frauen, C.%
\end{APACrefauthors}%
\unskip\
\newblock
\APACrefYearMonthDay{2016}{}{}.
\newblock
{\BBOQ}\APACrefatitle {{ENSO influence on the North Atlantic European climate:}
  a non-linear and non-stationary approach} {{ENSO influence on the North
  Atlantic European climate:} a non-linear and non-stationary approach}.{\BBCQ}
\newblock
\APACjournalVolNumPages{Climate Dynamics}{47}{}{2071-2084}.
\newblock
\begin{APACrefDOI} \doi{10.1007/s00382-015-2951-0} \end{APACrefDOI}
\PrintBackRefs{\CurrentBib}

\bibitem [\protect \citeauthoryear {%
Lorenz%
}{%
Lorenz%
}{%
{\protect \APACyear {1963}}%
}]{%
Lorenz.1963a}
\APACinsertmetastar {%
Lorenz.1963a}%
\begin{APACrefauthors}%
Lorenz, E\BPBI N.%
\end{APACrefauthors}%
\unskip\
\newblock
\APACrefYearMonthDay{1963}{}{}.
\newblock
{\BBOQ}\APACrefatitle {Deterministic nonperiodic flow} {Deterministic
  nonperiodic flow}.{\BBCQ}
\newblock
\APACjournalVolNumPages{Journal of the Atmospheric Sciences}{20}{}{130--141}.
\PrintBackRefs{\CurrentBib}

\bibitem [\protect \citeauthoryear {%
Lorenz%
}{%
Lorenz%
}{%
{\protect \APACyear {1990}}%
}]{%
Lorenz.1990}
\APACinsertmetastar {%
Lorenz.1990}%
\begin{APACrefauthors}%
Lorenz, E\BPBI N.%
\end{APACrefauthors}%
\unskip\
\newblock
\APACrefYearMonthDay{1990}{}{}.
\newblock
{\BBOQ}\APACrefatitle {Can chaos and intransitivity lead to interannual
  variability?} {Can chaos and intransitivity lead to interannual
  variability?}{\BBCQ}
\newblock
\APACjournalVolNumPages{Tellus A}{42}{3}{378--389}.
\PrintBackRefs{\CurrentBib}

\bibitem [\protect \citeauthoryear {%
Lucarini%
\ \BBA {} B{\'{o}}dai%
}{%
Lucarini%
\ \BBA {} B{\'{o}}dai%
}{%
{\protect \APACyear {2017}}%
}]{%
Luc.Bodai.2017}
\APACinsertmetastar {%
Luc.Bodai.2017}%
\begin{APACrefauthors}%
Lucarini, V.%
\BCBT {}\ \BBA {} B{\'{o}}dai, T.%
\end{APACrefauthors}%
\unskip\
\newblock
\APACrefYearMonthDay{2017}{}{}.
\newblock
{\BBOQ}\APACrefatitle {Edge states in the climate system: exploring global
  instabilities and critical transitions} {Edge states in the climate system:
  exploring global instabilities and critical transitions}.{\BBCQ}
\newblock
\APACjournalVolNumPages{Nonlinearity}{30}{7}{R32--R66}.
\newblock
\begin{APACrefDOI} \doi{10.1088/1361-6544/aa6b11} \end{APACrefDOI}
\PrintBackRefs{\CurrentBib}

\bibitem [\protect \citeauthoryear {%
McPhaden%
, Santoso%
\BCBL {}\ \BBA {} Cai%
}{%
McPhaden%
\ \protect \BOthers {.}}{%
{\protect \APACyear {2020}}%
}]{%
McPhaden.ea.2020}
\APACinsertmetastar {%
McPhaden.ea.2020}%
\begin{APACrefauthors}%
McPhaden, M\BPBI J.%
, Santoso, A.%
\BCBL {}\ \BBA {} Cai, W.%
\end{APACrefauthors}%
\ (\BEDS).
\unskip\
\newblock
\APACrefYear{2020}.
\newblock
\APACrefbtitle {{El Ni{\~n}o Southern Oscillation in a Changing Climate}} {{El
  Ni{\~n}o Southern Oscillation in a Changing Climate}}\ (\BVOL~253).
\newblock
\APACaddressPublisher{}{John Wiley \& Sons}.
\PrintBackRefs{\CurrentBib}

\bibitem [\protect \citeauthoryear {%
Nidheesh%
\ \protect \BOthers {.}}{%
Nidheesh%
\ \protect \BOthers {.}}{%
{\protect \APACyear {2017}}%
}]{%
Nidheeshetal2017}
\APACinsertmetastar {%
Nidheeshetal2017}%
\begin{APACrefauthors}%
Nidheesh, A\BPBI G.%
, Lengaigne, M.%
, Vialard, J.%
, Izumo, T.%
, Unnikrishnan, A\BPBI S.%
\BCBL {}\ \BBA {} Cassou, C.%
\end{APACrefauthors}%
\unskip\
\newblock
\APACrefYearMonthDay{2017}{}{}.
\newblock
{\BBOQ}\APACrefatitle {Influence of {ENSO on the Pacific decadal oscillation in
  CMIP models}} {Influence of {ENSO on the Pacific decadal oscillation in CMIP
  models}}.{\BBCQ}
\newblock
\APACjournalVolNumPages{Climate Dynamics}{49}{}{3309-3326}.
\newblock
\begin{APACrefDOI} \doi{10.1007/s00382-016-3514-8} \end{APACrefDOI}
\PrintBackRefs{\CurrentBib}

\bibitem [\protect \citeauthoryear {%
O'Reilly%
, Weisheimer%
, Woollings%
, Gray%
\BCBL {}\ \BBA {} MacLeod%
}{%
O'Reilly%
\ \protect \BOthers {.}}{%
{\protect \APACyear {2019}}%
}]{%
Oreillyetal2018}
\APACinsertmetastar {%
Oreillyetal2018}%
\begin{APACrefauthors}%
O'Reilly, C\BPBI H.%
, Weisheimer, A.%
, Woollings, T.%
, Gray, L\BPBI J.%
\BCBL {}\ \BBA {} MacLeod, D.%
\end{APACrefauthors}%
\unskip\
\newblock
\APACrefYearMonthDay{2019}{}{}.
\newblock
{\BBOQ}\APACrefatitle {The importance of stratospheric initial conditions for
  winter {North Atlantic Oscillation} predictability and implications for the
  signal-to-noise paradox} {The importance of stratospheric initial conditions
  for winter {North Atlantic Oscillation} predictability and implications for
  the signal-to-noise paradox}.{\BBCQ}
\newblock
\APACjournalVolNumPages{Quart. J. Royal Met. Soc.}{145}{718}{131-146}.
\newblock
\begin{APACrefDOI} \doi{10.1002/qj.3413} \end{APACrefDOI}
\PrintBackRefs{\CurrentBib}

\bibitem [\protect \citeauthoryear {%
Pedlosky%
}{%
Pedlosky%
}{%
{\protect \APACyear {1987}}%
}]{%
Pedlosky.1987}
\APACinsertmetastar {%
Pedlosky.1987}%
\begin{APACrefauthors}%
Pedlosky, J.%
\end{APACrefauthors}%
\unskip\
\newblock
\APACrefYear{1987}.
\newblock
\APACrefbtitle {{Geophysical Fluid Dynamics}} {{Geophysical Fluid Dynamics}}\
  (\PrintOrdinal{2nd}\ \BEd).
\newblock
\APACaddressPublisher{New York}{Springer-Verlag}.
\PrintBackRefs{\CurrentBib}

\bibitem [\protect \citeauthoryear {%
Penny%
\ \protect \BOthers {.}}{%
Penny%
\ \protect \BOthers {.}}{%
{\protect \APACyear {2019}}%
}]{%
Pennyetal2019}
\APACinsertmetastar {%
Pennyetal2019}%
\begin{APACrefauthors}%
Penny, S.%
, Bach, E.%
, Bhargava, K.%
, Chang, C\BHBI C.%
, Da, C.%
, Sun, L.%
\BCBL {}\ \BBA {} Yoshida, T.%
\end{APACrefauthors}%
\unskip\
\newblock
\APACrefYearMonthDay{2019}{}{}.
\newblock
{\BBOQ}\APACrefatitle {Strongly Coupled Data Assimilation in Multiscale Media:
  Experiments Using a Quasi-Geostrophic Coupled Model} {Strongly coupled data
  assimilation in multiscale media: Experiments using a quasi-geostrophic
  coupled model}.{\BBCQ}
\newblock
\APACjournalVolNumPages{J. Adv. Model. Earth Syst}{6}{}{1803-1829}.
\PrintBackRefs{\CurrentBib}

\bibitem [\protect \citeauthoryear {%
Philander%
}{%
Philander%
}{%
{\protect \APACyear {1990}}%
}]{%
Philander1990}
\APACinsertmetastar {%
Philander1990}%
\begin{APACrefauthors}%
Philander, S.%
\end{APACrefauthors}%
\unskip\
\newblock
\APACrefYear{1990}.
\newblock
\APACrefbtitle {El Niño and the Southern Oscillation} {El niño and the
  southern oscillation}.
\newblock
\APACaddressPublisher{}{Academic Press}.
\PrintBackRefs{\CurrentBib}

\bibitem [\protect \citeauthoryear {%
Pierini%
}{%
Pierini%
}{%
{\protect \APACyear {2020}}%
}]{%
Pierini2020}
\APACinsertmetastar {%
Pierini2020}%
\begin{APACrefauthors}%
Pierini, S.%
\end{APACrefauthors}%
\unskip\
\newblock
\APACrefYearMonthDay{2020}{}{}.
\newblock
{\BBOQ}\APACrefatitle {Statistical Significance of Small Ensembles of
  Simulations and Detection of the Internal Climate Variability: An Excitable
  Ocean System Case Study} {Statistical significance of small ensembles of
  simulations and detection of the internal climate variability: An excitable
  ocean system case study}.{\BBCQ}
\newblock
\APACjournalVolNumPages{J. Stat. Phys.}{179}{}{1475--1495}.
\newblock
\begin{APACrefDOI} \doi{10.1007/s10955-019-02409-x} \end{APACrefDOI}
\PrintBackRefs{\CurrentBib}

\bibitem [\protect \citeauthoryear {%
Pierini%
, Chekroun%
\BCBL {}\ \BBA {} Ghil%
}{%
Pierini%
\ \protect \BOthers {.}}{%
{\protect \APACyear {2018}}%
}]{%
Pierinietal2018}
\APACinsertmetastar {%
Pierinietal2018}%
\begin{APACrefauthors}%
Pierini, S.%
, Chekroun, M\BPBI D.%
\BCBL {}\ \BBA {} Ghil, M.%
\end{APACrefauthors}%
\unskip\
\newblock
\APACrefYearMonthDay{2018}{}{}.
\newblock
{\BBOQ}\APACrefatitle {The onset of chaos in nonautonomous dissipative
  dynamical systems: a low-order ocean-model case} {The onset of chaos in
  nonautonomous dissipative dynamical systems: a low-order ocean-model
  case}.{\BBCQ}
\newblock
\APACjournalVolNumPages{Nonlin. Proc. Geophys.}{25}{}{671--692}.
\newblock
\begin{APACrefDOI} \doi{10.5194/npg-25-671-2018} \end{APACrefDOI}
\PrintBackRefs{\CurrentBib}

\bibitem [\protect \citeauthoryear {%
Pierini%
, Ghil%
\BCBL {}\ \BBA {} Chekroun%
}{%
Pierini%
\ \protect \BOthers {.}}{%
{\protect \APACyear {2016}}%
}]{%
Pierinietal2016}
\APACinsertmetastar {%
Pierinietal2016}%
\begin{APACrefauthors}%
Pierini, S.%
, Ghil, M.%
\BCBL {}\ \BBA {} Chekroun, M\BPBI D.%
\end{APACrefauthors}%
\unskip\
\newblock
\APACrefYearMonthDay{2016}{}{}.
\newblock
{\BBOQ}\APACrefatitle {Exploring the pullback attractors of a low-order
  quasigeostrophic ocean model: The deterministic case} {Exploring the pullback
  attractors of a low-order quasigeostrophic ocean model: The deterministic
  case}.{\BBCQ}
\newblock
\APACjournalVolNumPages{{J. Climate}}{29}{}{4185-4202}.
\PrintBackRefs{\CurrentBib}

\bibitem [\protect \citeauthoryear {%
Roberts%
, Guckenheimer%
, Widiasih%
, Timmermann%
\BCBL {}\ \BBA {} Jones%
}{%
Roberts%
\ \protect \BOthers {.}}{%
{\protect \APACyear {2016}}%
}]{%
Robertsetal2016}
\APACinsertmetastar {%
Robertsetal2016}%
\begin{APACrefauthors}%
Roberts, A.%
, Guckenheimer, J.%
, Widiasih, E.%
, Timmermann, A.%
\BCBL {}\ \BBA {} Jones, C\BPBI K\BPBI R\BPBI T.%
\end{APACrefauthors}%
\unskip\
\newblock
\APACrefYearMonthDay{2016}{}{}.
\newblock
{\BBOQ}\APACrefatitle {Mixed-mode oscillations of {El Ni\~no-Southern
  Oscillation}} {Mixed-mode oscillations of {El Ni\~no-Southern
  Oscillation}}.{\BBCQ}
\newblock
\APACjournalVolNumPages{J Atmos Sci}{73}{}{1755-1766}.
\PrintBackRefs{\CurrentBib}

\bibitem [\protect \citeauthoryear {%
Ruelle%
}{%
Ruelle%
}{%
{\protect \APACyear {1984}}%
}]{%
Ruelle1984}
\APACinsertmetastar {%
Ruelle1984}%
\begin{APACrefauthors}%
Ruelle, D.%
\end{APACrefauthors}%
\unskip\
\newblock
\APACrefYearMonthDay{1984}{}{}.
\newblock
{\BBOQ}\APACrefatitle {Characteristic exponents for a viscous fluid subjected
  to time dependent forces} {Characteristic exponents for a viscous fluid
  subjected to time dependent forces}.{\BBCQ}
\newblock
\APACjournalVolNumPages{Commun. Math. Phys.}{93}{}{285--300}.
\PrintBackRefs{\CurrentBib}

\bibitem [\protect \citeauthoryear {%
Schemm%
, Rivière%
, Ciasto%
\BCBL {}\ \BBA {} Li%
}{%
Schemm%
\ \protect \BOthers {.}}{%
{\protect \APACyear {2018}}%
}]{%
Schemmetal2018}
\APACinsertmetastar {%
Schemmetal2018}%
\begin{APACrefauthors}%
Schemm, S.%
, Rivière, G.%
, Ciasto, L\BPBI M.%
\BCBL {}\ \BBA {} Li, C.%
\end{APACrefauthors}%
\unskip\
\newblock
\APACrefYearMonthDay{2018}{}{}.
\newblock
{\BBOQ}\APACrefatitle {Extratropical cyclogenesis changes in connection with
  tropospheric {ENSO teleconnections to the North Atlantic: Role} of stationary
  and transient waves} {Extratropical cyclogenesis changes in connection with
  tropospheric {ENSO teleconnections to the North Atlantic: Role} of stationary
  and transient waves}.{\BBCQ}
\newblock
\APACjournalVolNumPages{{J. Atmos. Sci.}}{75}{}{3943--3964}.
\newblock
\begin{APACrefDOI} \doi{10.1175/JAS-D-17-0340.1} \end{APACrefDOI}
\PrintBackRefs{\CurrentBib}

\bibitem [\protect \citeauthoryear {%
Sell%
}{%
Sell%
}{%
{\protect \APACyear {1971}}%
}]{%
Sell.1971}
\APACinsertmetastar {%
Sell.1971}%
\begin{APACrefauthors}%
Sell, G\BPBI R.%
\end{APACrefauthors}%
\unskip\
\newblock
\APACrefYear{1971}.
\newblock
\APACrefbtitle {{Topological Dynamics and Ordinary Differential Equations}}
  {{Topological Dynamics and Ordinary Differential Equations}}.
\newblock
\APACaddressPublisher{}{Van Nostrand Reinhold}.
\PrintBackRefs{\CurrentBib}

\bibitem [\protect \citeauthoryear {%
Smith%
, Scaife%
, Eade%
\BCBL {}\ \BBA {} et al%
}{%
Smith%
\ \protect \BOthers {.}}{%
{\protect \APACyear {2020}}%
}]{%
Smithetal2020}
\APACinsertmetastar {%
Smithetal2020}%
\begin{APACrefauthors}%
Smith, D\BPBI M.%
, Scaife, A\BPBI A.%
, Eade, R.%
\BCBL {}\ \BBA {} et al.%
\end{APACrefauthors}%
\unskip\
\newblock
\APACrefYearMonthDay{2020}{}{}.
\newblock
{\BBOQ}\APACrefatitle {{North Atlantic} climate far more predictable than
  models imply} {{North Atlantic} climate far more predictable than models
  imply}.{\BBCQ}
\newblock
\APACjournalVolNumPages{Nature}{583}{}{796-800}.
\newblock
\begin{APACrefDOI} \doi{10.1038/s41586-020-2525-0} \end{APACrefDOI}
\PrintBackRefs{\CurrentBib}

\bibitem [\protect \citeauthoryear {%
T\'el%
\ \protect \BOthers {.}}{%
T\'el%
\ \protect \BOthers {.}}{%
{\protect \APACyear {2020}}%
}]{%
Teletal2020}
\APACinsertmetastar {%
Teletal2020}%
\begin{APACrefauthors}%
T\'el, T.%
, B\'odai, T.%
, Dr\'otos, G.%
, Haszpra, T.%
, Herein, M.%
, Kaszás, B.%
\BCBL {}\ \BBA {} Vincze, M.%
\end{APACrefauthors}%
\unskip\
\newblock
\APACrefYearMonthDay{2020}{}{}.
\newblock
{\BBOQ}\APACrefatitle {The Theory of Parallel Climate Realizations} {The theory
  of parallel climate realizations}.{\BBCQ}
\newblock
\APACjournalVolNumPages{J. Stat. Phys.}{179}{}{1496--1530}.
\newblock
\begin{APACrefDOI} \doi{10.1007/s10955-019-02445-7} \end{APACrefDOI}
\PrintBackRefs{\CurrentBib}

\bibitem [\protect \citeauthoryear {%
Timmermann%
\ \BBA {} Jin%
}{%
Timmermann%
\ \BBA {} Jin%
}{%
{\protect \APACyear {2002}}%
}]{%
Timm.Jin.2002}
\APACinsertmetastar {%
Timm.Jin.2002}%
\begin{APACrefauthors}%
Timmermann, A.%
\BCBT {}\ \BBA {} Jin, F\BHBI F.%
\end{APACrefauthors}%
\unskip\
\newblock
\APACrefYearMonthDay{2002}{}{}.
\newblock
{\BBOQ}\APACrefatitle {A nonlinear mechanism for decadal {El Ni{\~n}o amplitude
  changes}} {A nonlinear mechanism for decadal {El Ni{\~n}o amplitude
  changes}}.{\BBCQ}
\newblock
\APACjournalVolNumPages{Geophysical Research Letters}{29}{1}{{3-1--3-4}}.
\newblock
\begin{APACrefDOI} \doi{10.1029/2001GL013369} \end{APACrefDOI}
\PrintBackRefs{\CurrentBib}

\bibitem [\protect \citeauthoryear {%
Timmermann%
, Jin%
\BCBL {}\ \BBA {} Abshagen%
}{%
Timmermann%
\ \protect \BOthers {.}}{%
{\protect \APACyear {2003}}%
}]{%
Timmermannetal2003}
\APACinsertmetastar {%
Timmermannetal2003}%
\begin{APACrefauthors}%
Timmermann, A.%
, Jin, F\BHBI F.%
\BCBL {}\ \BBA {} Abshagen, J.%
\end{APACrefauthors}%
\unskip\
\newblock
\APACrefYearMonthDay{2003}{}{}.
\newblock
{\BBOQ}\APACrefatitle {A nonlinear theory for {El Ni\~no} bursting} {A
  nonlinear theory for {El Ni\~no} bursting}.{\BBCQ}
\newblock
\APACjournalVolNumPages{{J. Atmos. Sci.}}{60}{}{152-165}.
\PrintBackRefs{\CurrentBib}

\bibitem [\protect \citeauthoryear {%
Tondeur%
, Carrassi%
, Vannitsem%
\BCBL {}\ \BBA {} Bocquet%
}{%
Tondeur%
\ \protect \BOthers {.}}{%
{\protect \APACyear {2020}}%
}]{%
Tondeuretal2020}
\APACinsertmetastar {%
Tondeuretal2020}%
\begin{APACrefauthors}%
Tondeur, M.%
, Carrassi, A.%
, Vannitsem, S.%
\BCBL {}\ \BBA {} Bocquet, M.%
\end{APACrefauthors}%
\unskip\
\newblock
\APACrefYearMonthDay{2020}{}{}.
\newblock
{\BBOQ}\APACrefatitle {On temporal scale separation in coupled data
  assimilation with the ensemble {Kalman} filter} {On temporal scale separation
  in coupled data assimilation with the ensemble {Kalman} filter}.{\BBCQ}
\newblock
\APACjournalVolNumPages{J Stat Phys}{179}{}{1161--1185}.
\PrintBackRefs{\CurrentBib}

\bibitem [\protect \citeauthoryear {%
Tziperman%
, Stone%
, Cane%
\BCBL {}\ \BBA {} Jarosh%
}{%
Tziperman%
\ \protect \BOthers {.}}{%
{\protect \APACyear {1994}}%
}]{%
Tzip.ea.1994}
\APACinsertmetastar {%
Tzip.ea.1994}%
\begin{APACrefauthors}%
Tziperman, E.%
, Stone, L.%
, Cane, M\BPBI A.%
\BCBL {}\ \BBA {} Jarosh, H.%
\end{APACrefauthors}%
\unskip\
\newblock
\APACrefYearMonthDay{1994}{}{}.
\newblock
{\BBOQ}\APACrefatitle {{El Ni\~no chaos: overlapping of resonances between the
  seasonal cycle and the Pacific ocean-atmosphere oscillator}} {{El Ni\~no
  chaos: overlapping of resonances between the seasonal cycle and the Pacific
  ocean-atmosphere oscillator}}.{\BBCQ}
\newblock
\APACjournalVolNumPages{Science}{264}{}{72--74}.
\PrintBackRefs{\CurrentBib}

\bibitem [\protect \citeauthoryear {%
Vannitsem%
}{%
Vannitsem%
}{%
{\protect \APACyear {2017}}%
}]{%
Vannitsem2017}
\APACinsertmetastar {%
Vannitsem2017}%
\begin{APACrefauthors}%
Vannitsem, S.%
\end{APACrefauthors}%
\unskip\
\newblock
\APACrefYearMonthDay{2017}{}{}.
\newblock
{\BBOQ}\APACrefatitle {Predictability of large-scale atmospheric motions:
  {Lyapunov} exponents and error dynamics} {Predictability of large-scale
  atmospheric motions: {Lyapunov} exponents and error dynamics}.{\BBCQ}
\newblock
\APACjournalVolNumPages{Chaos}{27}{}{32101}.
\newblock
\begin{APACrefDOI} \doi{10.1063/1.4979042} \end{APACrefDOI}
\PrintBackRefs{\CurrentBib}

\bibitem [\protect \citeauthoryear {%
Vannitsem%
, Demaeyer%
, {De Cruz}%
\BCBL {}\ \BBA {} Ghil%
}{%
Vannitsem%
\ \protect \BOthers {.}}{%
{\protect \APACyear {2015}}%
}]{%
Vannitsemetal2015}
\APACinsertmetastar {%
Vannitsemetal2015}%
\begin{APACrefauthors}%
Vannitsem, S.%
, Demaeyer, J.%
, {De Cruz}, L.%
\BCBL {}\ \BBA {} Ghil, M.%
\end{APACrefauthors}%
\unskip\
\newblock
\APACrefYearMonthDay{2015}{}{}.
\newblock
{\BBOQ}\APACrefatitle {Low-frequency variability and heat transport in a
  low-order nonlinear coupled ocean-atmosphere model} {Low-frequency
  variability and heat transport in a low-order nonlinear coupled
  ocean-atmosphere model}.{\BBCQ}
\newblock
\APACjournalVolNumPages{Physica D: Nonlinear Phenomena}{309}{}{71--85}.
\newblock
\begin{APACrefDOI} \doi{10.1016/j.physd.2015.07.006} \end{APACrefDOI}
\PrintBackRefs{\CurrentBib}

\bibitem [\protect \citeauthoryear {%
Vannitsem%
\ \BBA {} Ghil%
}{%
Vannitsem%
\ \BBA {} Ghil%
}{%
{\protect \APACyear {2017}}%
}]{%
VannitsemGhil2017}
\APACinsertmetastar {%
VannitsemGhil2017}%
\begin{APACrefauthors}%
Vannitsem, S.%
\BCBT {}\ \BBA {} Ghil, M.%
\end{APACrefauthors}%
\unskip\
\newblock
\APACrefYearMonthDay{2017}{}{}.
\newblock
{\BBOQ}\APACrefatitle {{Evidence of coupling in ocean-atmosphere dynamics over
  the North Atlantic}} {{Evidence of coupling in ocean-atmosphere dynamics over
  the North Atlantic}}.{\BBCQ}
\newblock
\APACjournalVolNumPages{Geophys. Res. Lett.}{44}{4}{2016--2026}.
\PrintBackRefs{\CurrentBib}

\bibitem [\protect \citeauthoryear {%
Vannitsem%
\ \BBA {} Lucarini%
}{%
Vannitsem%
\ \BBA {} Lucarini%
}{%
{\protect \APACyear {2016}}%
}]{%
VannitsemLucarini2016}
\APACinsertmetastar {%
VannitsemLucarini2016}%
\begin{APACrefauthors}%
Vannitsem, S.%
\BCBT {}\ \BBA {} Lucarini, V.%
\end{APACrefauthors}%
\unskip\
\newblock
\APACrefYearMonthDay{2016}{}{}.
\newblock
{\BBOQ}\APACrefatitle {Statistical and dynamical properties of covariant
  {Lyapunov} vectors in a coupled atmosphere-ocean model -- multiscale effects,
  geometric degeneracy, and error dynamics} {Statistical and dynamical
  properties of covariant {Lyapunov} vectors in a coupled atmosphere-ocean
  model -- multiscale effects, geometric degeneracy, and error
  dynamics}.{\BBCQ}
\newblock
\APACjournalVolNumPages{J. Phys. A}{49}{22}{224001}.
\PrintBackRefs{\CurrentBib}

\bibitem [\protect \citeauthoryear {%
Varadi%
, Ghil%
\BCBL {}\ \BBA {} Kaula%
}{%
Varadi%
\ \protect \BOthers {.}}{%
{\protect \APACyear {1999}}%
}]{%
Varadi.ea.1999}
\APACinsertmetastar {%
Varadi.ea.1999}%
\begin{APACrefauthors}%
Varadi, F.%
, Ghil, M.%
\BCBL {}\ \BBA {} Kaula, W\BPBI M.%
\end{APACrefauthors}%
\unskip\
\newblock
\APACrefYearMonthDay{1999}{}{}.
\newblock
{\BBOQ}\APACrefatitle {{Jupiter, Saturn, and the edge of chaos}} {{Jupiter,
  Saturn, and the edge of chaos}}.{\BBCQ}
\newblock
\APACjournalVolNumPages{Icarus}{139}{2}{286--294}.
\PrintBackRefs{\CurrentBib}

\bibitem [\protect \citeauthoryear {%
Wilson%
}{%
Wilson%
}{%
{\protect \APACyear {1985}}%
}]{%
Wilson.1985}
\APACinsertmetastar {%
Wilson.1985}%
\begin{APACrefauthors}%
Wilson, C.%
\end{APACrefauthors}%
\unskip\
\newblock
\APACrefYearMonthDay{1985}{}{}.
\newblock
{\BBOQ}\APACrefatitle {{The great inequality of Jupiter and Saturn: from Kepler
  to Laplace}} {{The great inequality of Jupiter and Saturn: from Kepler to
  Laplace}}.{\BBCQ}
\newblock
\APACjournalVolNumPages{Archive for history of exact
  sciences}{33}{1-3}{15--290}.
\PrintBackRefs{\CurrentBib}

\end{thebibliography}

\end{document}